\documentclass[3p,10pt]{elsarticle}
\usepackage{graphics}
\usepackage{graphicx}
\usepackage{epsfig}

\usepackage{amssymb}
\usepackage{amsthm}
\usepackage{amsmath}

\usepackage{empheq}
\usepackage{epstopdf}
\usepackage{changepage}
\usepackage{rotating}
\usepackage{adjustbox}
\usepackage{color}
\usepackage{xcolor}
\usepackage{dsfont}
\usepackage{float}
\usepackage{lineno}
\usepackage{resizegather}
\usepackage{multicol}
\usepackage{booktabs} 
\usepackage{colortbl} 
\usepackage{multirow} 
\usepackage{longtable,tabu} 
\usepackage{hhline}   
\usepackage{anyfontsize} 
\usepackage{everysel}
\usepackage{psfrag}
\usepackage{bbm}
\usepackage{bm}
\usepackage{pgf,tikz}
\usetikzlibrary{arrows}
\usepackage{url,microtype}
\usepackage{relsize} 

\usepackage{hyperref}

\renewcommand{\S}{\mathbf{S}} 
\newcommand{\Q}{\mathbf{Q}} 
 
\renewcommand{\u}{\mathbf{u}} 
\newcommand{\F}{\mathbf{F}}

\newcommand{\w}{\mathbf{w}} 
\newcommand{\q}{\mathbf{q}} 
\newcommand{\x}{\mathbf{x}} 
\renewcommand{\d}{\, d}

\renewcommand{\AA}{{\boldsymbol{A}}}

\newcommand{\vv}{{\boldsymbol{v}}}

\newcommand{\alpharho}{\bar{\rho}}
\newcommand{\alphap}{\bar{p}}
\newcommand{\alphasigma}{\bar{\sigma}}

\newcommand{\pd}{\partial}
\newcommand{\calE}{\mathcal{E}}

\newcommand{\JJ}{{\boldsymbol{J}}}
\newcommand{\QQ}{{\mathbf{Q}}}

\newcommand{\PP}{{\mathbf{P}}}

\newcommand{\Id}{{\boldsymbol{I}}}

\newcommand{\mbf}[1]{\mathbf{#1}}

\newcommand{\xbin}{\mbf{x}_{\mbf{b}_i}^n}    
\newcommand{\xbinp}{\mbf{x}_{\mbf{b}_i}^{n+1}}    
\newcommand{\xbixn}{{x}_{{b}_i}^n}    
\newcommand{\xbiyn}{{y}_{{b}_i}^n}

\renewcommand{\epsilon}{\varepsilon}
\renewcommand{\phi}{\varphi}
\newcommand{\phitilde}{\tilde{\phi}}

\journal{Frontiers in Physics}
\date{}

\begin{document}

\begin{frontmatter}

\title{High order ADER schemes for continuum mechanics}

\author[LAM]{S. Busto}
\ead{saray.busto@unitn.it}

\author[LAM]{S. Chiocchetti}
\ead{simone.chiocchetti@unitn.it}

\author[LAM]{M. Dumbser\corref{cor1}}
\ead{michael.dumbser@unitn.it}

\author[LAM]{E. Gaburro}
\ead{elena.gaburro@unitn.it}

\author[LAM]{I. Peshkov}
\ead{ilya.peshkov@unitn.it}

\cortext[cor1]{Corresponding author}

\address[LAM]{Laboratory of Applied Mathematics, DICAM, University of Trento, Via Mesiano 77, 38123 Trento, Italy}

\begin{abstract}
In this paper we first review the development of high order ADER finite volume and ADER discontinuous Galerkin schemes on fixed and moving meshes, since their introduction in 1999 by Toro et al.
We show the modern variant of ADER based on a space-time predictor-corrector formulation in the context of ADER discontinuous Galerkin schemes with a posteriori subcell finite volume limiter on fixed and moving grids, as well as on space-time adaptive Cartesian AMR meshes.  
We then present and discuss the unified symmetric hyperbolic and thermodynamically compatible (SHTC) 
formulation of continuum mechanics developed by Godunov, Peshkov and Romenski (GPR model), which allows to describe fluid and solid mechanics in one single and unified first order hyperbolic system. In order to deal with free surface and moving boundary problems, a simple diffuse interface approach is employed, which is compatible with Eulerian schemes on fixed grids as well as direct Arbitrary-Lagrangian-Eulerian methods on moving meshes.    
We show some examples of moving boundary problems in fluid and solid mechanics. 
\end{abstract}


\end{frontmatter}

\section{Introduction and review of the ADER approach}
\label{sec.Intro}

The development of high order numerical schemes for hyperbolic conservation laws has been one of the major challenges of numerical analysis for the last decades. \cite{GodunovRS} proved that for the linear advection equation no monotone linear schemes of second or higher order of accuracy can be constructed. Therefore, even if physical viscosity is considered, a linear high order scheme will present spurious oscillations near discontinuities, as it can be seen, for instance for the Lax-Wendroff scheme, \cite{lax}. A first idea to circumvent this theorem has been proposed in \cite{Kolgan:1972}, where limited slopes are employed to produce a non-linear scheme of second order of accuracy in space. Since then, many high order numerical methods have been developed like the Total Variation Disminishing methods (TVD) and Flux limiter methods (see, for instance, \cite{hartentvd,swebytvd,shu2,leer5,vanLeer:1985a,ToroBook}). Despite these methodologies  being already well established at the end of the last century, their major drawback was that they just provided global second order of accuracy and reduced locally to first order in the vicinity of smooth extrema. 

More advanced non-linear methods for advection dominated problems involve the family of ENO and WENO schemes, see \cite{HartenENO,harten,shu1}. In particular, the method of \cite{harten} is a fully discrete high order scheme that can be re-interpreted in terms of the solution of a generalized Riemann problem (GRP), see \cite{CastroToro}. Moreover, it can be seen as a generalization of the MUSCL-Hancock method of van Leer, see \cite{vanLeer:1985a,ToroBook,Berthon2006}. 

\begin{figure}[!b]
	\centering
	\includegraphics[width=0.5\linewidth]{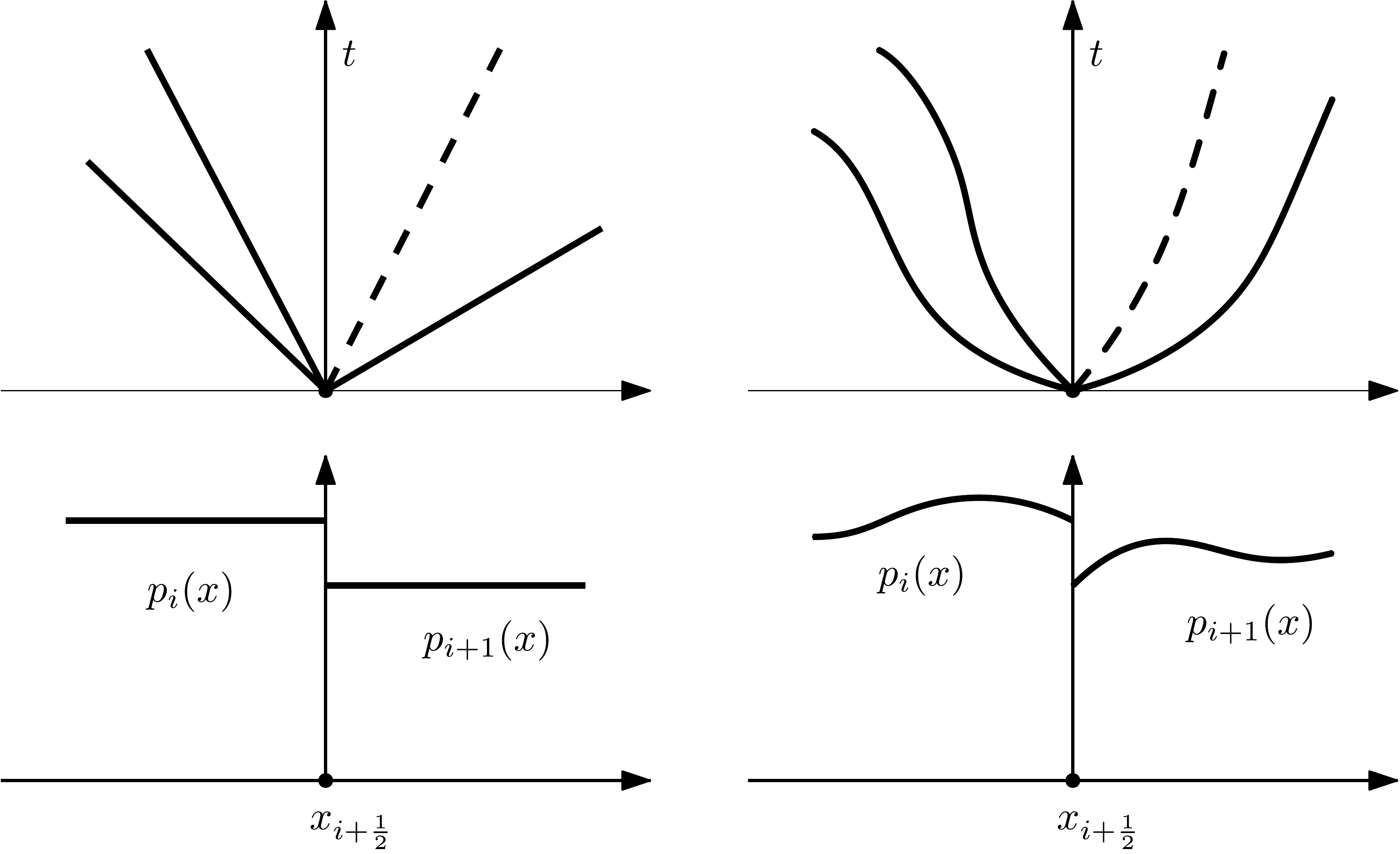}
	\caption{Classical piece-wise reconstruction polynomials used in the ADER approach, $p_{i}(x)$ and $p_{i+1}(x)${\color{black},} 
		and the structure of the Riemann problem solution at the intercell boundary $x_{i+\frac{1}{2}}$. Left: classical piece-wise constant data. Right: piece-wise smooth reconstruction.}
	\label{fig:ader1dplot}
\end{figure}

Following the idea of solving a generalized Riemann problem (GRP), see also \cite{Artzi,LeFloch:1991a,BenArtzi:2006a,HanLi}, the ADER approach (Arbitrary high order DErivative Riemann problem) has been first put forward for the linear advection equation with constant coefficients by \cite{mill,toro1}. 
The first step of the methodology involves piece-wise polynomial data reconstruction, where a nonlinear ENO reconstruction is applied in order to avoid spurious oscillations of the numerical solution. Then, a GRP is defined at each cell interface.
Classically, the initial condition for the GRP was given as piece-wise linear polynomials and second order schemes could be obtained by constructing a space-time integral of the solution in an appropriate control volume \cite{Toro1989,billett2}, or following a MUSCL approach, \cite{leer2,Colella1985}. 
An alternative methodology proposed in  \cite{lit:BenArtzi-Falcovitz-book:03} consists in 
expressing the solution of the GRP as a Taylor series expansion in time. The ADER approach obtains 
the high order time derivatives of the GRP solution at the cell interface via the 
Cauchy-Kovalevskaya procedure, which replaces time derivatives by spatial derivatives using 
repeated differentiation of the differential form of the PDE. The spatial derivatives, which may 
also jump at the interface, are defined via the solution of \textit{linearized} Riemann problems 
for the derivatives, where linearization is carried out about the Godunov state obtained from the 
classical Riemann problem between the boundary extrapolated values at the interface. In 
Figure~\ref{fig:ader1dplot}, the classical piece-wise constant polynomials are plotted against a 
high order reconstruction and the similarity solutions for both cases are sketched. Finally, these 
similarity solutions are used to construct the numerical flux. The resulting schemes are arbitrary 
high order accurate in both space and time, in the sense that they have no theoretical accuracy 
barrier.

Since their introduction in \cite{toro1,mill}, many extensions of the ADER methodology have been 
proposed. Regarding 2D linear PDEs, one may refer to \cite{schwartzkopff} and their simplification for the particular case of structured grids in \cite{schwartzkopff-dumbser-munz}. Moreover, non linear systems have been initially addressed in
\cite{toro4,toro3}. Further applications of ADER on non-Cartesian meshes have been presented in \cite{Kaser2003,Kaser05,DumbserKaeser07,CastroToro}. 
One should also mention the development of ADER schemes in the framework of discontinuous Galerkin (DG) 
finite element methods, see \cite{QiuDumbserShu,dumbser_jsc,Gassner2011a}. 
One of the main advantages of using DG is that the reconstruction step of classical ADER 
finite volume (ADER-FV) schemes can be skipped, since the discrete solution is already given by high order 
piecewise polynomials that can be directly evolved during each time step. Furthermore, ADER-DG schemes avoid the 
use of classical Runge-Kutta time stepping and thus provide efficient communication-avoiding schemes for parallel 
computing, see \cite{ADERGRMHD} and allow for simple and natural time-accurate local time stepping (LTS), see 
\cite{LTS}. 

An important step forward in the development of more general ADER schemes was achieved in \cite{DumbserEnauxToro}, where a new class of ADER-FV methods has been introduced.  
The main contribution of this paper consists in the introduction of a new element-local space-time DG predictor, which allows at the same time the treatment of stiff source terms, as well as the replacement of the cumbersome Cauchy-Kovalevskaya procedure. 
First, a high order WENO method is employed to compute a polynomial reconstruction of the data inside each spatial element; then, an element-local weak formulation of the conservation law is considered in space-time and the predictor is applied to construct the time evolution of the WENO polynomials within each cell. Note that, in this step, the integration by parts is performed only in time, which differs from global space-time DG schemes \cite{spacetimedg1,spacetimedg2}, which are globally implicit. Finally, the cell averages are updated with an explicit fully discrete one-step scheme, considering the integral form of the equations. As a result, the proposed methodology maintains arbitrary high order of accuracy, while avoiding the issues related to the use of a Taylor series expansion in time. As already mentioned above, it naturally provides an  approach for the treatment of stiff source terms (for further details on this topic, see \cite{HidalgoDumbser} and references therein). 

The above methodology can also be applied in the discontiuous Galerkin framework as presented in 
\cite{Dumbser2008}, where, a unified $P_NP_M$ framework for arbitrary high order one-step 
finite volume and DG schemes has been introduced. For other reconstruction-based DG schemes, see e.g. 
\cite{luo1,luo2}. Afterwards, the methodology has been extended to solve 
a wide variety of different PDE systems, such as the resistive relativistic MHD equations, \cite{DumbserZanotti};  
non conservative hyperbolic systems found in geophysical flows, \cite{ADERNC} in which a well-balanced and path-conservative version of the scheme has been developed; compressible multi-phase flows \cite{USFORCE2}, the compressible Navier-Stokes equations, \cite{DumbserNSE}; the compressible Euler equations and divergence-free schemes for MHD,  \cite{BalsaraMultiDRS,BalsaraDivB2015}, where ADER schemes were used in combination with genuinely 
multidimensional Riemann solvers. The last extensions concern the special and general relativistic MHD equations, see
\cite{Zanotti2015d,ADERGRMHD}, as well as the Einstein field equations of general relativity \cite{ADERCCZ4,dumbser2020glm}. 

Later, ADER schemes have been extended to adaptive mesh refinement on Cartesian grids (AMR), in combination with 
time accurate local time stepping (LTS). This technique has initially been introduced in 
\cite{AMR3DCL,AMR3DNC} for conservative and non-conservative hyperbolic systems, respectively.  
Moreover, the schemes of the ADER family were the first high order methods to be applied for the numerical 
solution of the unified first order hyperbolic formulation of continuum mechanics by Godunov, Peshkov and 
Romenski \cite{GodRom1972,PeshRom2014,GodRom2003}, see  
\cite{GPRmodel,GPRmodelMHD,DFTBW2018}. In the rest of this paper, we will refer to the Godunov-Peshkov-Romenski model of continuum mechanics as GPR model. 

The ADER approach has also been extended to the direct Arbitrary-Lagrangian-Eulerian framework (ALE), where the mesh moves with an arbitrary velocity, taken as close as possible to the local fluid velocity. Initially developed for one space dimension, it has been soon extended to the case of the two and three dimensional Euler equations on unstructured meshes, \cite{Lagrange2D,Lagrange3D}, including the discretization of non-conservative products.  
Further works in this area involve the use of local timestepping techniques, \cite{ALELTS1D,ALELTS2D}; coupling with multidimensional HLL Riemann solvers, \cite{LagrangeMDRS}; solution of magnetohydrodynamics problems (MHD), \cite{bonazzoli2014high,LagrangeMHD}; development of a quadrature-free approach to increase the computational efficiency of the overall method, \cite{LagrangeQF}; use of curvilinear unstructured meshes, \cite{LagrangeISO}; or extension to solve the GPR model, \cite{BoscheriDumbser2016,HypoHyper2}. 
Furthermore, in \cite{GaburroDumbserSWE} a novel algorithm to deal with moving nonconforming polygonal grids has been presented. The methodology reduces the typical mesh distortion arising in shear flows and provides high quality elements even for long-time simulations. An exactly well-balanced path-conservative version of this approach for the Euler equations with gravity can be found in  \cite{GaburroDumbserEuler}.
Still in the ALE framework, within this article, we will present new results for the family of ADER-FV and ADER-DG schemes on moving unstructured Voronoi meshes \cite{Springel}, as recently introduced in \cite{gaburroReview,GBCKSD2019}. 

It is well known that when dealing with high order schemes special care must be paid to the 
limiting methodology employed. In most of the previous referenced papers classical \textit{a 
	priori} limiters have been used, such as WENO reconstruction. Nevertheless, some alternative contributions to this 
topic can be found in the series of papers  
\cite{DGLimiter1,ADER_MOOD_14,DGLimiter2,DGLimiter3,Zanotti2015d,ALEMQF,ALEMOOD2,ADERDGVisc,Tavelli2019,GBCKSD2019}, where 
a novel \textit{a posteriori} sub-cell FV limiter of high order DG schemes, based on the MOOD paradigm of \cite{CDL1,CDL2,CDL3}, has been employed. 

Besides the references given above, which focus on the development of the ADER methodology with a local space-time Galerkin predictor, many recent papers have been devoted to the development of other families of ADER schemes, like the classical ADER finite volume methods. Without pretending to be exhaustive, we may refer to 
\cite{castro2009,Hidalgo2009,TDMS2009,ToroBook,MCT2012,MT2014,TM2014,TM2015,TCL2015,Bustophd2018,MLLOT2016,BTVC2016,CTMBK2016,BFTVC2018,DMT2019} 
and references therein.

In this paper, as a promising application of the family of ADER schemes, we solve a diffuse interface formulation 
of the GPR model of continuum mechanics. In comparison with existing continuum mechanics 
models, the novel feature of the GPR model is in that it incorporates the two main branches of continuum mechanics, fluid and solid mechanics, in one single unified PDE system. Recall that traditionally 
fluid and solid mechanics are described by PDE systems of different types, i.e. parabolic (viscous 
fluids) and hyperbolic (linear elasticity and hyperelasticity), which imposes many theoretical and 
technical difficulties if one wishes to model natural and industrial processes involving 
co-existence of the fluid and solid states such as in fluid-structure interaction (FSI) problems, 
modeling of general solid-fluid transition such as in 
melting and solidification processes, e.g. additive manufacturing, see for example 
\cite{Francois2017}, flows of granular media \cite{Forterre2013}, viscoplastic flows, e.g. debris flows, avalanches, mantle convection, flows of many industrial Bingham-type fluids, see 	\cite{FrigaardReview2014}. 
Due to the unified treatment of fluids and solids, the GPR model thus has a great potential for simplifying the 	modeling process and code development for solving the aforementioned problems. Yet, before to be applied to practical problems, the GPR model may require a coupling with an 	interface tracking/capturing technique for the modeling of moving material boundaries such as in 	free surface flows or solid body motion. In particular, in this paper, we couple the GPR model with 	a simple diffuse interface approach, see \cite{Tavelli2019,DIM2D,DIMWB,kemm2020simple}. For example, very 	interesting computational results 	with similar diffuse interface approaches and level set techniques for compressible 	multi-material flows have been obtained for example in 	\cite{Gavrilyuk2008,FavrieGavrilyukSaurel,FavrGavr2012,NdanouFavrieGavrilyuk,Iollo2017,Michael2018a,Jackson2019,Barton2019}.	Finally, we demonstrate that the ADER family of schemes is capable to resolve the GPR model in both 	solid and fluid regimes.

The paper is organized as follows. In Section \ref{sec.scheme} we present the family of ADER finite volume and ADER discontinuous Galerkin finite element schemes on fixed Cartesian and moving polygonal meshes in two space dimensions. 
Next, in Section \ref{sec.Model} we introduce the diffuse interface formulation of the GPR model. In Section \ref{sec.results} we show some computational results obtained with different kinds of ADER schemes (ADER-FV and ADER-DG) 
on different mesh topologies, including moving unstructured Voronoi meshes, as well as fixed and adaptive Cartesian grids. 
The paper is rounded off by some concluding remarks and an outlook to future work in Section \ref{sec.concl}.

\section{ADER finite volume and discontinuous Galerkin schemes}
\label{sec.scheme} 

The numerical method adopted in this paper is the variant of the arbitrary high-order accurate 
ADER approach based on the space-time predictor-corrector formalism, which we have briefly reviewed in the previous Section~\ref{sec.Intro}. It easily applies to the context of finite volume (FV) and discontinous Galerkin (DG) methods, using either space-time adaptive Cartesian grids (AMR), see \cite{Peano1,Peano2,AMR3DCL,DGLimiter2,ADERDGVisc,ADERGRMHD} and references therein, or unstructured meshes, and both on fixed Eulerian domains or in a moving Arbitrary-Lagrangian-Eulerian (ALE) framework, see \cite{ALELTS2D,LagrangeMHD, Lagrange2D,Lagrange3D, ALEDG, ALEMQF, gaburro2018Thesis, GBCKSD2019} and references therein.

Here, we briefly describe the key features of our numerical scheme, keeping the notation as general as possible, and referring to the literature for further details.
We start by introducing the general form of our governing PDE system and a moving unstructured discretization 
of two-dimensional domains (Sections~\ref{ssec.PDEgeneral} and~\ref{ssec.domain}); next, in Section~\ref{ssec.data} we describe the data representation of the discrete solution. 
Then, we explain how to obtain high order of accuracy in \textit{space}: 
this is available by construction in the DG case, and obtained via some variants of the well known WENO procedure (\cite{JiangShu1996,balsarashu,DumbserKaeser07,DumbserKaeser06b,ZhangShu3D,shu2016high}) for the FV approach. 
Finally, we focus on the predictor-corrector version of the ADER scheme that allows to achieve arbitrary 
high order of accuracy in space and \textit{time}.  
Since it is out of the scope of this paper to recall all the details, 
a general overview is given in Sections~\ref{ssec.predictor} and~\ref{ssec.corrector}, 
and an inedited proof of the convergence of the predictor for a nonlinear
conservation law is presented in Section~\ref{ssec.PredictorProof}. 

We would like to emphasize that, besides this novel convergence proof, 
other progress has been introduced within this work. Indeed, up to our knowledge, it is 
the first time that: (i) the ADER approach is used to solve a diffuse 
interface formulation of the GPR model that addresses the free surface problem in both solid and fluid  
mechanics context (previously, a similar formulation was used only in the solid dynamics context 
\cite{Hank2017,Hank2016a,Ndanou2015}); (ii) 
non-conservative products 
are taken into account in the high order direct ALE scheme of \cite{GBCKSD2019}, where they have 
to be integrated also on degenerate space--time control volumes (see 
Section\,\ref{ssec.PredictorSliver}).

\medskip
\subsection{Governing PDE system}
\label{ssec.PDEgeneral}

In this paper we consider high order fully-discrete schemes for nonlinear systems of hyperbolic PDE with non-conservative products and algebraic source terms of the form  
\begin{equation}
\label{eqn.pde}
\frac{\partial \mathbf{Q} }{\partial t} + \nabla \cdot \mathbf{F}\left( \mathbf{Q} \right) 
+ \mathbf{B}(\Q) \cdot \nabla \mathbf{Q}  = \mathbf{S}(\mathbf{Q}), 
\end{equation}
where $\Q = \Q(\x,t) \in \Omega_{Q} \subset \mathbb{R}^m$ is the state vector, $t \in \mathbb{R}_0^+$ is the time, $\mathbf{x} \in \Omega \subset \mathbb{R}^d$ is the spatial coordinate, $d$ is the number of space dimensions,
$\Omega_{Q}$ is the so-called state space or phase space, $\F(\Q)$ is the nonlinear flux tensor, $\mathbf{B}(\Q) \cdot \nabla \mathbf{Q} $ is a non-conservative product and $\S(\Q)$ is 
a purely algebraic source term. Introducing the system matrix  $\mathbf{A}(\Q) = \partial \F / \partial \Q + \mathbf{B}(\Q)$ 
the above system can also be written in quasi-linear form as 
\begin{equation} \begin{aligned}
\label{eq.quasi-linear}
\frac{\partial \mathbf{Q} }{\partial t}   
+ \mathbf{A}(\Q) \cdot \nabla \mathbf{Q}  = \mathbf{S}(\mathbf{Q}).
\end{aligned} \end{equation}
The system is said to be hyperbolic if for all $\mathbf{n} \neq 0$ and for all $\Q \in \Omega_Q$ the matrix $\mathbf{A}(\Q) \cdot \mathbf{n}$ has $m$ real eigenvalues and a full set of $m$ linearly independent right eigenvectors. The system~\eqref{eqn.pde} needs to be provided with an initial condition $\Q(\x,0) = \Q_0(\x)$ and appropriate boundary conditions on $\partial \Omega$. 

In this paper we focus on a particular, but very general, example of a first-order system~\eqref{eqn.pde} describing   elastic and visco-plastic heat-conducting media; it will be discussed in Section~\ref{sec.Model}.

\medskip
\subsection{Domain discretization} 
\label{ssec.domain}

In the general ALE case, we consider a moving two-dimensional $(d=2)$ domain $\Omega(t)$ 
and we cover it using an unstructured mesh made of $N_P$ non overlapping polygons $P_i, i=1, \dots N_P$. 
The mesh is first built at time $t=0$ and then it is rearranged at each time step~$t^n$: 
elements and nodes are moved following the local fluid velocity and when necessary, 
in order to prevent mesh distortion, also the mesh topology (i.e. the shape of the elements and their connectivities) is changed.

Given a polygon $P_i^n$ we denote by $\mathcal{V}(P_i^n) = \{v_{i_1}^n, \dots, v_{i_j}^n,$ $ \dots, v_{i_{N_{V_i}^n}}^n \}$ the set of its $N_{V_i}^n$ Voronoi neighbors (the neighbors that share with $P_i^n$ at least a vertex), and
by $\mathcal{E}(P_i^n) = \{e_{i_1}^n, \dots, e_{i_j}^n,$$ \dots,$$ e_{i_{N_{V_i}^n}}^n \}$ the set of its $N_{V_i}^n$ edges, and 
by $\mathcal{D}(P_i^n) = \{d_{i_1}^n, \dots, d_{i_j}^n,$$ \dots,$$ d_{i_{N_{V_i}^n}}^n \}$ the set of its $N_{V_i}^n$ vertexes, 
consistently ordered counterclockwise.
Finally, the barycenter of $P_i^n$ is noted as $\xbin = (\xbixn,\xbiyn)$. 
When necessary, by connecting $\xbin$ with each vertex of $\mathcal{D}(P_i)$ 
we can subdivide a polygon $P_i^n$ in $N_{V_i}^n$ subtriangles denoted as 
$\mathcal{T}(P_i^n) = \{T_{i_1}^n, \dots, T_{i_j}^n, \dots, T_{i_{N_{V_i}^n}}^n \}$.

The coordinates of each node at time $t^n$ are denoted by $\x_k^n$, and $\overline{\mathbf{V}}^n_k$ 
represents the velocity at which it is supposed to move, 
so that its new coordinates at time $t^{n+1}$ are given from the following relation
\begin{equation} \begin{aligned}
\x^{n+1}_k = \x^n_k + \Delta t \overline{\mathbf{V}}^n_k.
\end{aligned} \end{equation}
More details on how to obtain $\overline{\mathbf{V}}$ 
can be found in \cite{LagrangeMHD, Lagrange3D, ALEDG} for what concerns classical direct ALE schemes 
on conforming unstructured grids, in \cite{GaburroDumbserSWE, GaburroDumbserEuler} for nonconforming unstructured grids, 
in \cite{LagrangeISO} for curvilinear meshes, 
and we refer in particular to Section 2.4 and 2.5 of \cite{GBCKSD2019} for what concerns moving 
unstructured polygonal grids allowing for topology changes, which indeed is the ALE case considered in this paper (see case B below). 
Moreover, working in the ALE framework, we are allowed to take $\overline{\mathbf{V}}=\mbf{0}$, 
i.e. we can also work in a fixed \textit{Eulerian} system where the initial mesh is never modified. 

In particular, in this paper we will consider the following two situations for our domain discretization: 
\begin{itemize}
	\item[A.] A fixed Cartesian mesh made of $N_P$ quadrilaterals elements, which is not moved during the simulation, 
	but which can be successively refined, with a general space-tree-type data structure that allows element-by-element refinement with a general refinement factor $\mathfrak{r} \geq 2$, 
	in order to increase the resolution in the areas of interest, as can be seen in Figure~\ref{fig:amrgridrefinement} (for the details on the refinement procedure we refer to \cite{AMR3DCL,ADERGRMHD}). 
	To ease the description of the numerical method, we will associate to each quadrilateral element $P_{i}^{n}$, a set of indices that refer to its Cartesian coordinates, $\left\lbrace j,k \right\rbrace$, such that
	$P_{jk}^{n}:=P_{i}^{n}=[x_{j-\frac{1}{2}},x_{j+\frac{1}{2}}]\times [y_{k-\frac{1}{2}},y_{k+\frac{1}{2}}]$,
	$\Delta x_{j} = x_{j+\frac{1}{2}}-x_{j-\frac{1}{2}}$, $\Delta y_{k} = y_{k+\frac{1}{2}}-y_{k-\frac{1}{2}}$.
	\begin{figure}
		\centering
		\includegraphics[width=0.25\linewidth]{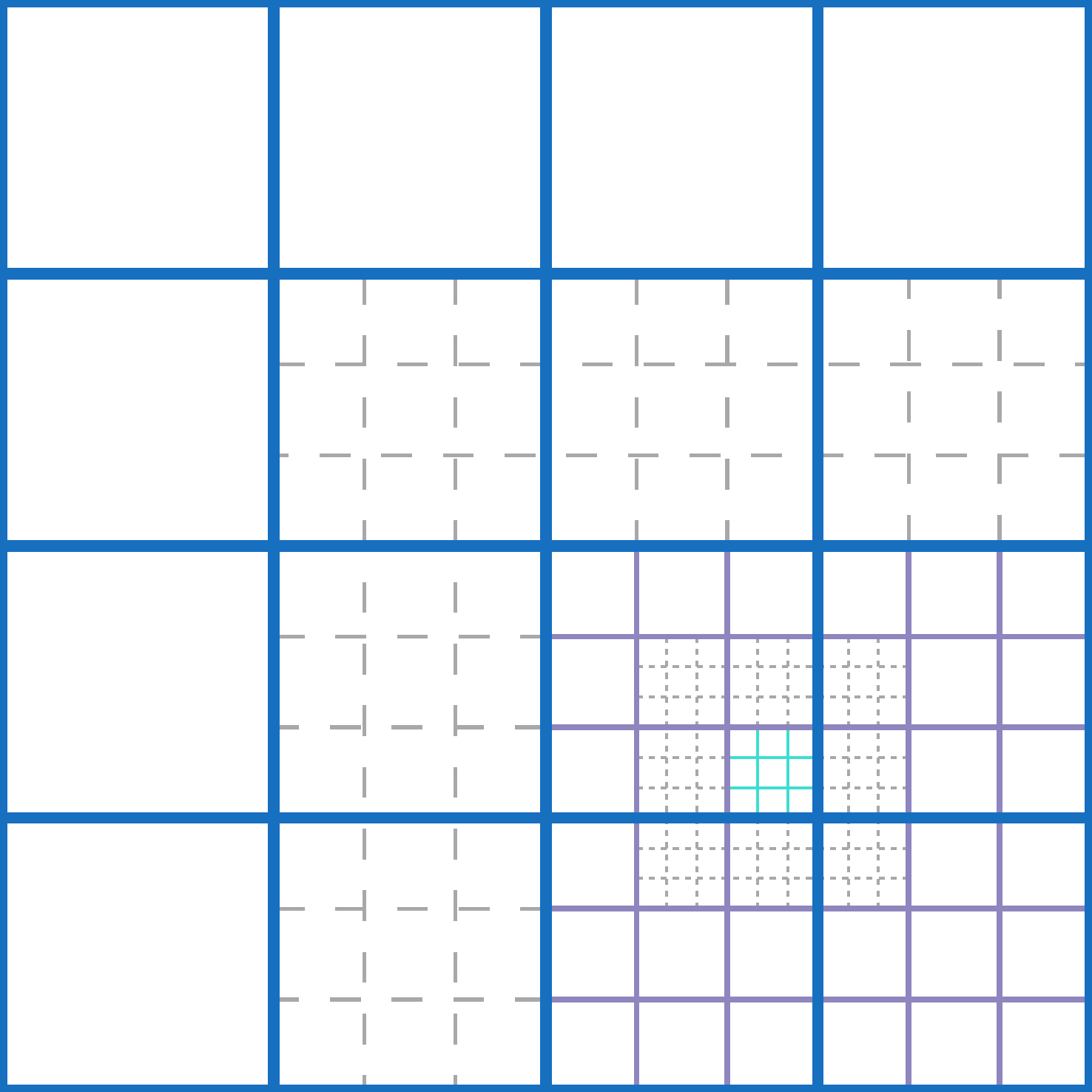}
		\caption{Sketch of the mesh refinement structure of three AMR levels with refinement factor $\mathfrak{r}=3$. Solid lines indicate active cells, whereas the dashed ones are the virtual cells allowing interpolation between the coarse and the refined mesh, needed in the case of high order WENO reconstruction.}
		\label{fig:amrgridrefinement}
	\end{figure}

	\item[B.] 
	A moving polygonal grid as the one described in \cite{GBCKSD2019} that 
	i) moves with the fluid flow in order to reduce the numerical dissipation associated with transport terms and 
	ii) also allows for topology changes at any time step in order to maintain always a high quality of the moving mesh; 
	in this case we remark that our method is also able to deal with degenerate space time control volumes at arbitrary high order of accuracy.
\end{itemize}

\medskip
\subsubsection{Space-time connectivity}
\label{ssec.SpaceTimeConnectivity}

\begin{figure}
	{\includegraphics{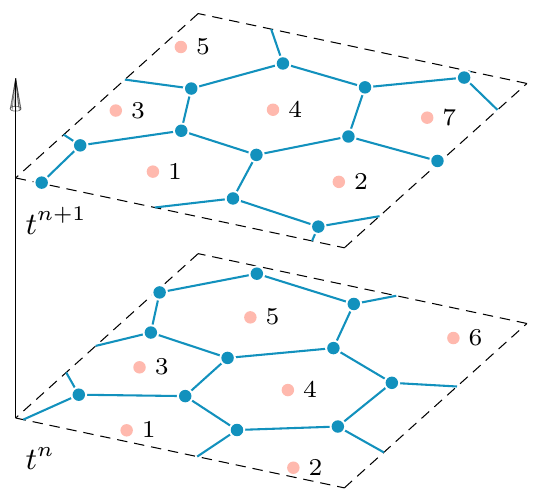}}%
	{\includegraphics{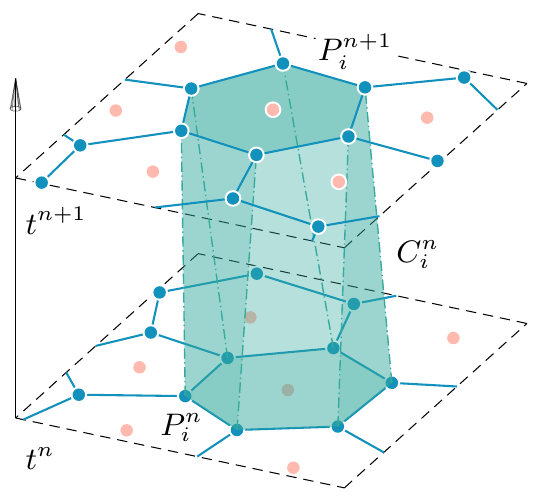}}%
	{\includegraphics{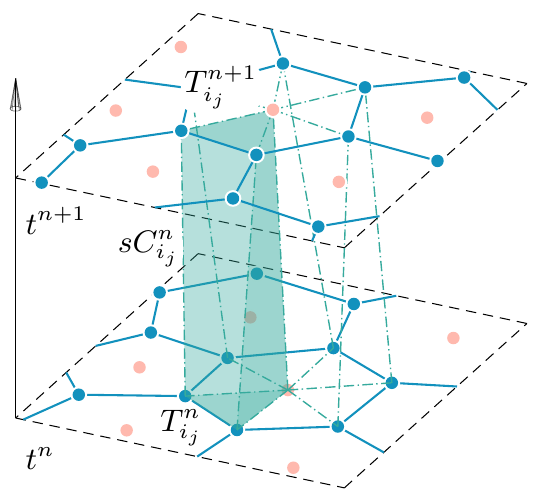}}%
	\caption{Space time connectivity. Left: The tessellation at time $t^n$ and time $t^{n+1}$. 
		Middle: $P_i^n$ is connected with $P_i^{n+1}$ to construct the space--time control volume $C_i^n$. 
		Right: The sub-triangle $T_{i_j}^n$ is connected with $T_{i_j}^{n+1}$ to construct the sub--space--time control volume $sC_{i_j}^n$. }
	\label{fig.ControlVolume_noTopChange}
\end{figure}

To better understand the context of moving meshes we refer the reader to Figure~\ref{fig.ControlVolume_noTopChange}: 
note that the tessellation at time $t^n$ has been evolved resulting in a slightly different tessellation at time $t^{n+1}$; 
for each element $P_i^n$ the new vertex coordinates $\x_k^{n+1}$, $ k = 1, \dots, N_{V_i}^n$, are connected to the old coordinates $\x_k^{n+1}$ via straight line segments, yielding the 
multidimensional \textit{space-time control volume} $C_i^n$, that involves $N_{V_i}^{n,st}+2$ space-time sub-surfaces.
Specifically, the space-time volume $C_i^n$ is bounded on
the bottom and on the top by the element configuration at the current time level $P^n_i$ and at the new time level $P^{n+1}_i$, 
respectively, while it is closed with a total number of $N_{V_i}^{n,st}$ lateral space--time surfaces $\partial C_{i_j}^n, j=1, \dots, N_{V_i}^{n,st}$ that are given by the evolution of each edge $e_{i_j}^n$ of element $P_i^n$ within the
time step $\Delta t = t^{n+1} - t^n$. 
\textit{A priori}, $\partial C_{i_j}^n$ are not parallel to the time direction: 
thus to be treated numerically they can be mapped to a reference square by using a set of of bilinear basis functions (see \cite{Lagrange2D}).
To resume, the space-time volume $C_i^n$ is bounded by its surface $\partial C^n_i$ which is given by
\begin{equation}
\partial C^n_i = \left( \bigcup \limits_{j} \partial C^n_{i_j} \right) 
\,\, \cup \,\, P_i^{n} \,\, \cup \,\, P_i^{n+1}.
\label{dCi}
\end{equation}

Note that in the fixed Cartesian case, $C_i^n$ reduces to a right parallelepiped with four 
lateral space--time surfaces $\partial C_{i_j}^n$ parallel to the time-direction, 
so many simplifications are possible.

We close this part by emphasizing that the family of direct ALE schemes proposed in this work, 
based on the ADER predictor-corrector approach, 
is based on the integration of the governing equation~\eqref{eqn.pde} 
\textit{in space and in time} directly over these \textit{space--time} control volumes, 
see Section~\ref{ssec.corrector}. Note that this procedure, which is more evident when $C_i^n$ is an oblique prism, is also hidden when $C_i^n$ is just a right parallelepiped.

\medskip
\subsection{Data representation}
\label{ssec.data}

The conserved variables $\Q$ in~\eqref{eqn.pde} are discretized in each polygon $P_i^n$ at the current time $t^n$ via piecewise polynomials of arbitrary high order $N$,
denoted by $\mathbf{u}_h^n(\x,t^n)$ and defined as
\begin{equation}
\mathbf{u}_h^n(\x,t^n) = \sum \limits_{\ell=0}^{\mathcal{N}-1} \phi_\ell(\x,t^n) \, \hat{\mathbf{u}}^{n}_{\ell,i} 
= \phi_\ell(\x,t^n) \, \hat{\mathbf{u}}^{n}_{\ell,i} , \quad \x \in P_i^n,
\label{eqn.uh}
\end{equation}
where in the last equality we have employed the classical tensor index notation based on the Einstein summation convention, which implies summation over two equal indices. The functions $\phi_\ell(\x,t^n)$ can be either: 

\begin{itemize}
	\item[i.] \textit{Nodal} spatial basis functions given by a set of Lagrange interpolation polynomials of maximum degree
	$N$ with the property
	\begin{equation} \begin{aligned}
	\label{eq.nodalGeneralBasis}
	\phi_\ell(\x_{\text{GL}}^m) = \left\{ \begin{array}{rl} 1 & \text{if}\;\;
		\ell=m; \\0 & \text{otherwise}; \end{array}\right.\hspace{0.7cm}
	\ell,m=1,\ldots,(N+1)^{\text{d}},
	\end{aligned} \end{equation}
	where $\{\x_{\text{GL}}^m\}$ are the set of the Gauss-Legendre (GL)
	quadrature points on $P_i^n$ (see \cite{stroud} for the multidimensional case). 
	
	\noindent In particular, when employing these basis functions on a Cartesian grid,
	each quadrilateral $P_i^n$ is easily mapped to a reference square, 
	we only need the tensor product of the GL quadrature points in the unit interval $[0,1]$, 
	and the $\phi_\ell$ are simply generated by multiplying one-dimensional nodal basis functions, i.e 
	\begin{equation} \begin{aligned}
	\label{eq.nodalBasisAMR}
	\phi_\ell(\x,t^n) = \phi_{\ell_1}\left(\xi(x)\right)\phi_{\ell_2}\left(\eta(y)\right)
	\end{aligned} \end{equation}
	with $\phi_{\ell_i}$ satisfying~\eqref{eq.nodalGeneralBasis} with $d=1$,
	and $x= x_{j-\frac{1}{2}} + \xi \Delta x_{j}$, $y= y_{k-\frac{1}{2}}+ \eta \Delta y_{k}$ being the set of reference coordinates related to $P_i^n$.
	In this case, the total number of GL quadrature points per polygon, 
	as well as the total number of basis functions  $\{ \phi_\ell\}$ and expansion coefficients 
	$\hat{\mathbf{u}}^{n}_{\ell,i}$,  the so-called degrees of freedom (DOF), is 
	$\mathcal{N} = (N+1)^d$. These basis functions are used on Cartesian grids, i.e. for Case A.

	\item[ii.] \textit{Modal} spatial basis functions written through a Taylor series of degree $N$ in the variables $\mathbf{x}=(x,y)$ 
	directly defined on the \textit{physical element} $P_i^n$, expanded about its current barycenter $\xbin$ and normalized by its current characteristic length $h_i$
	\begin{equation} 
	\label{eq.Dubiner_phi_spatial}
	\phi_\ell(\x,t^n) |_{P_i^n} = \frac{(x - \xbixn)^{p_\ell}}{p_\ell! \, h_i^{p_\ell}} \, \frac{(y - \xbiyn)^{q_\ell}}{q_\ell! \, h_i^{q_\ell}}, \qquad 
	\ell = 0, \dots, \mathcal{N}-1, \quad \ 0 \leq p_\ell + q_\ell \leq N,
	\end{equation} 
	$h_i$ being the radius of the circumcircle of $P_i^n$.
	In this case the total number $\mathcal{N}$ of DOF $\hat{\mathbf{u}}^{n}_{l}$ is 
	$
	\mathcal{N} = \frac{1}{d!} \prod \limits_{m=1}^{d} (N+m).
	\label{eqn.nDOF}
	$
	We employ this kind of basis functions in the moving unstructured polygonal Case B.
\end{itemize}

The discontinuous finite element data representation~\eqref{eqn.uh} leads naturally 
to discontinuous Galerkin (DG) schemes if $N>0$, but also to finite volume (FV) schemes in the case $N=0$. This indeed means that for $N=0$ we have $\phi_\ell(\x) = 1$, with $\ell=0$ and~\eqref{eqn.uh} reduces to the classical piecewise constant data that are typical of finite volume methods.  
In the case $N>0$ (DG) the form given by~\eqref{eqn.uh} already provides a spatially high order accurate data representation with accuracy $N+1$, where instead for the case $N=0$ (FV), if we are interested in increasing the spatial order of accuracy, up to $M+1$ for examle, we need to perform a spatial \textit{reconstruction}.
With this notation, our method falls within the more general class of $P_NP_M$ schemes introduced in \cite{Dumbser2008} for fixed unstructured meshes.

\medskip
\subsection{Data reconstruction}

In this section we focus on the reconstruction procedure needed in the finite volume context ($N=0$, $M>0$) 
in order to obtain order of accuracy $M+1$ in space starting from the piecewise constant values of $\mathbf{u}_h^n(\x,t^n)$ in $P_i^n$ and its neighbors, 
i.e. in order to obtain a high order polynomial of degree $M$ representing our solution in each $P_i^n$
\begin{equation} \begin{aligned}
&\mathbf{w}_h^n(\x,t^n) = \sum \limits_{\ell=0}^{\mathcal{M}-1} \psi_\ell(\x,t^n) \, \hat{\mathbf{w}}^{n}_{\ell,i} 
= \psi_\ell(\x,t^n) \, \hat{\mathbf{w}}^{n}_{\ell,i},\x \in P_i^n, 
\label{eqn.wh}
\end{aligned} \end{equation}
where the $\psi_\ell$ functions simply coincide with the $\phi_{\ell}$ basis functions of~\eqref{eqn.uh}.
Our reconstruction procedures are based on the WENO algorithm in its \textit{polynomial} formulation as presented in \cite{DumbserEnauxToro,DumbserKaeser07,DumbserKaeser06b,MixedWENO2D,MixedWENO3D,LPR:2001,dumbser2017central,SCR:CWENOquadtree},
and not based on the original version of WENO proposed in \cite{JiangShu1996,balsarashu,HuShuVortex1999,ZhangShu3D} which provides only \textit{point values}. 
For each $P_i^n$, the basic idea consists in i) selecting a central stencil of elements
$\mathcal{S}_i^0$ with a total number of 
\begin{equation} \begin{aligned}
\label{eq.StencilTotElements}
n_e= f \cdot \frac{1}{d!} \prod \limits_{m=1}^{d} (M+m)
\end{aligned} \end{equation}
elements, containing the cell $P_i^n$ itself, its first layer of Voronoi neighbors $\mathcal{V}(P_i^n)$
and filled by recursively adding neighbors of those elements that have been already included in the stencil,
and in ii) using the cell-average values of the elements of $\mathcal{S}_i^0$ to reconstruct a polynomial of degree $M$ by imposing the integral conservation criterion, i.e by requiring that its average on each cell match the known cell average. 
If $f>1$ (which occurs in the unstructured case, where we take $f=1.5$), this of course leads to an overdetermined linear system, which is solved using a constrained least-squares technique (CLSQ)~\cite{DumbserKaeser06b}, i.e. the reconstructed polynomial 
has exactly the cell average $\hat{\mathbf{u}}^{n}_{0,i}$ on the polygon $P_i^n$ and matches all the other cell averages of the remaining stencil elements in the least-square sense. 

However, as well known thanks to the Godunov theorem (\cite{GodunovRS}), the use of only one central stencil (which is indeed a linear procedure) would introduce oscillations in the presence of shock waves or other discontinuities.  
So, in order to make the reconstruction procedure nonlinear, we will compute the final reconstruction polynomial as a \textit{nonlinear combination} or \textit{more} than only one reconstruction polynomial, each one defined on a different reconstruction stencil $\mathcal{S}_i^s$. 

We refer to the cited literature for further details, and here we just highlight the main 
characteristics of the two reconstruction procedures adopted in this work.

\medskip
\subsubsection*{Case A: Cartesian mesh.}
\begin{figure}[!b]
	\centering
	\includegraphics[width=0.7\linewidth]{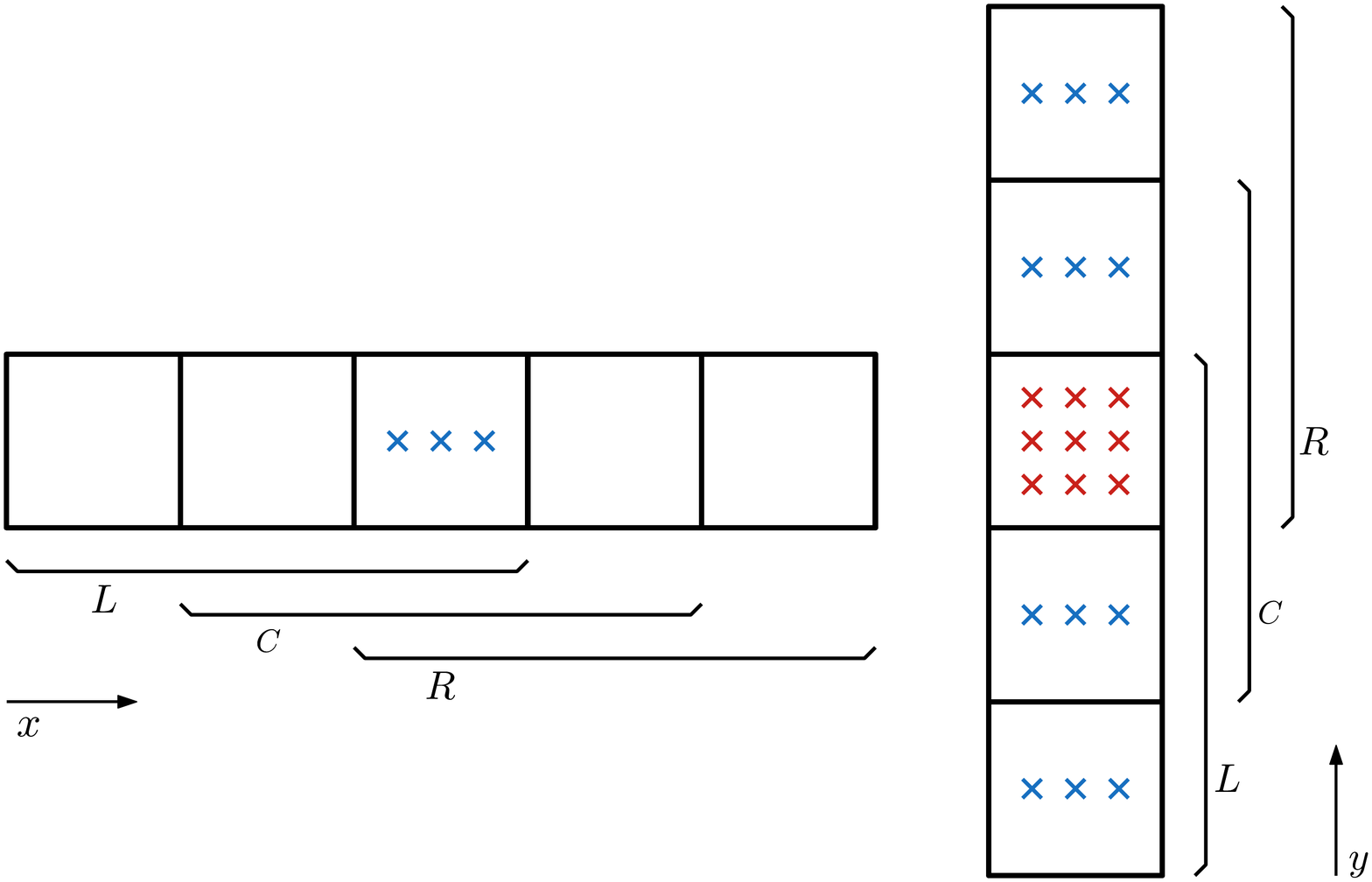}
	\caption{Reconstruction stencils for a fixed Cartesian mesh with $M=2$, where $L$, $C$ and $R$ denote the left--sided, central and right--sided stencils, respectively. Left: reconstruction on $x$ direction. Right: reconstruction on $y$ direction.}
	\label{fig:stencilxy}
\end{figure}

In Case A, of a fixed Cartesian mesh, we employ the polynomial WENO procedure given in \cite{AMR3DCL}, which is implemented in a dimension by dimension fashion. For each cell, we define its related sets of one-dimensional reconstruction stencils as
\begin{equation}
\mathcal{S}^{s,x}_{i}=\displaystyle{\bigcup_{m=j-L}^{j+R}} P_{mk}^{n}, \quad \mathcal{S}^{s,y}_{i}=\displaystyle{\bigcup_{m=k-L}^{k+R}} P_{jm}^{n},
\end{equation}
where $L=\lbrace M,s\rbrace$ and $R=\lbrace M,s\rbrace$ denote the order and stencil dependent 
spatial extension of the stencil to the left and to the right.
For odd order schemes we consider three stencils, one central, one fully left--sided, and one 
fully right--sided stencil in each space dimension (see Figure~\ref{fig:stencilxy} for a graphical 
interpretation for $M=2$), while for even order schemes we have four stencils, two of which are central, 
while the remaining two are again given by the fully left--sided and fully right--sided in each 
space dimension. In both cases the total amount of elements in each stencil is always $n_e = M+1$, the order of the scheme. 

Focusing on the reconstruction procedure on the $x$ direction, given a element $P_{i}^{n}$, we start by expressing the first coordinate of the reconstruction polynomial at each stencil in terms of one dimensional basis functions,
\begin{equation} \begin{aligned}
\mathbf{w}^{s,x}_{h}(x,t^n)  = \sum_{\ell_{1}=0}^{M}  \psi_{\ell_{1}}\left(\xi \right) \hat{\mathbf{w}}^{n,s}_{jk,\ell_{1}}=\psi_{\ell_{1}}\left(\xi \right) \hat{\mathbf{w}}^{n,s}_{jk,\ell_{1}}.
\end{aligned} \end{equation}
Then, we integrate on the stencil elements obtaining an algebraic system on the polynomial coefficients:
\begin{equation} \begin{aligned}
\frac{1}{\Delta x_{m}}  \int_{x_{m-\frac{1}{2}}}^{x_{m+\frac{1}{2}}} \psi_{\ell_{1}}\left(\xi ( x )  \right)  \hat{\mathbf{w}}^{n,s}_{jk,\ell_{1}} dx =  \bar{\mathbf{u}}^{n}_{mk},  \quad \forall P_{mk}^{n}\in \mathcal{S}^{s,x}_{i}
\end{aligned} \end{equation}
with $\bar{\mathbf{u}}^{n}_{mk}$ the average value obtained by integrating the solution at the previous time step on the cell $P_{mk}$. 
Once the coefficients, and thus the polynomials, related to all the stencils are obtained, 
we compute a reconstruction polynomial in the $x$ direction as the data-dependent nonlinear combination of these,
\begin{equation} \begin{aligned}
\mathbf{w}^{x}_{h}\left(x,t^{n}\right) = \psi_{\ell_{1}}\left(\xi \right)\hat{\mathbf{w}}^{n}_{jk,\ell_1},\quad  \hat{\mathbf{w}}^{n}_{jk,\ell_1} = \sum_{s=1}^{n_{s}}  \omega_{s} \hat{\mathbf{w}}^{n, s}_{jk,\ell_1},
\end{aligned} \end{equation}
where $n_{s}$ is the number of stencils, $n_{s}=3$ if $M=\dot{2}$ and $n_{s}=4$ otherwise; 
and $\omega_{s}$ denote the nonlinear weights (see \cite{AMR3DCL} for further details). 

To complete the reconstruction polynomial, we now repeat the above procedure in the $y$ direction for each degree of freedom 
$\hat{\mathbf{w}}^{n}_{jk,\ell_1}$. First, we write the reconstruction polynomial in terms of the basis functions,
\begin{equation} \begin{aligned}
\mathbf{w}^{s,y}_{h}(x,y,t^n)  =  \psi_{\ell_{1}}\left(\xi \right) \psi_{\ell_{2}}\left(\eta \right)  \hat{\mathbf{w}}^{n,s}_{jk,\ell_{1}\ell_{2}}.
\end{aligned} \end{equation}
Then, we solve the algebraic system
\begin{equation} \begin{aligned}
\frac{1}{\Delta y_{m}}  \int_{y_{m-\frac{1}{2}}}^{y_{m+\frac{1}{2}}} \psi_{\ell_{2}}\left(\eta\left( y\right)  \right)  \hat{\mathbf{w}}^{n,s}_{jk,\ell_{1}\ell_{2}} dy =  \hat{\mathbf{w}}^{n}_{jm,\ell_{1}},  \quad \forall P_{jm}^{n}\in \mathcal{S}^{s,y}_{i}
\end{aligned} \end{equation}
and calculate 
\begin{equation} \begin{aligned}
\hat{\mathbf{w}}^{n}_{jk,\ell_1\ell_{2}} = \sum_{s=1}^{n_{s}}  \omega_{s} \hat{\mathbf{w}}^{n, s}_{jk,\ell_1\ell_{2}}.
\end{aligned} \end{equation}
Finally, we get the WENO reconstruction polynomial
\begin{equation} \begin{aligned}
\mathbf{w}_{h}^{n}\left(\mathbf{x},t^{n}\right) = \psi_{\ell_{1}}\left(\xi \right) \psi_{\ell_{2}}\left(\eta \right) \hat{\mathbf{w}}^{n}_{jk,\ell_1 \ell_2}.
\end{aligned} \end{equation}
In order to enforce bounds on the WENO reconstruction polynomial, such as the condition $0 \leq \alpha \leq 1$ on the volume fraction function of for example \eqref{eqn.alpha}, we \textit{rescale} the reconstruction coefficients $\hat{\mathbf{w}}^{n}_{jk,\ell_1 \ell_2}$ around the cell average as follows: 
\begin{equation}
\hat{\mathbf{w}}^{*}_{jk,\ell_1 \ell_2} = \bar{\mathbf{u}}^{*}_{jk} + 
\varphi_{jk} \left( \hat{\mathbf{w}}^{n}_{jk,\ell_1 \ell_2} - \bar{\mathbf{u}}^{*}_{jk} \right),  
\end{equation} 
where the scaling factor $\varphi_{jk}$ is computed via the Barth and Jespersen limiter (see \cite{BarthJespersen})  applied to 
the volume fraction function $\alpha$ in all Gauss-Legendre and Gauss-Lobatto quadrature nodes, i.e. 
$\varphi_{jk} = \min(\varphi_{jk,p})$ is the global minimum in each element, with the nodal limiter values given by  
\begin{equation}
\varphi_{jk,p} = \left\{ \begin{array}{ccc} 
\min \left( 1, \frac{\alpha_{\max} - \bar{\alpha}}{\alpha_p - \bar{\alpha}} \right), 
& \textnormal{ if } & \alpha_p - \bar{\alpha} > 0, \\ 
\min \left( 1, \frac{\alpha_{\min} - \bar{\alpha}}{\alpha_p - \bar{\alpha}} \right), 
& \textnormal{ if } & \alpha_p - \bar{\alpha} > 0, \\ 
1,                                & \textnormal{ if } & \alpha_p - \bar{\alpha} = 0. 
\end{array} \right. 
\end{equation}  
Here $\alpha_{\max} = 1 - \varepsilon \leq 1$ is the upper bound of the volume fraction function and $\alpha_{\min}=\varepsilon \geq 0$ is its lower bound; $\bar{\alpha}$ denotes the cell average of $\alpha$ and 
$\alpha_p$ denotes the node value of $\alpha$ in the quadrature point $\mathbf{x}_p$ under consideration. 
As already mentioned above, this strategy is inspired from the Barth and Jespersen limiter \cite{BarthJespersen}, 
but also from the new bound-preserving polynomial approximation introduced in \cite{despres2017polynomials, campos2019projection}. Since the physical solution of $\alpha$ must satisfy $0 \leq \alpha \leq 1$, the above bound
preserving limiter does \textit{not} reduce the formal order of accuracy of the reconstruction, as proven in 
\cite{despres2017polynomials}. 

\medskip
\subsubsection*{Case B: moving polygonal mesh.}

\begin{figure}
	\centering
	\includegraphics{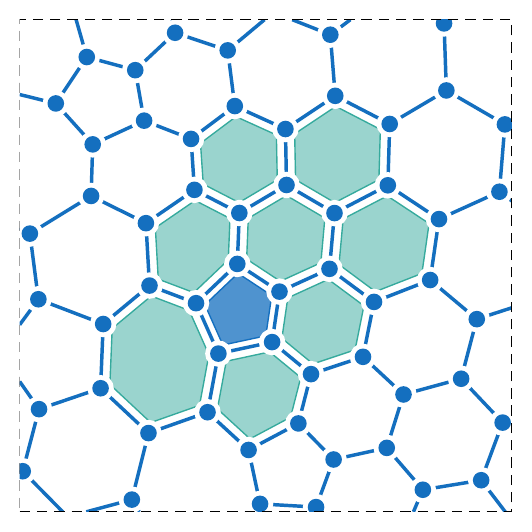}%
	\includegraphics{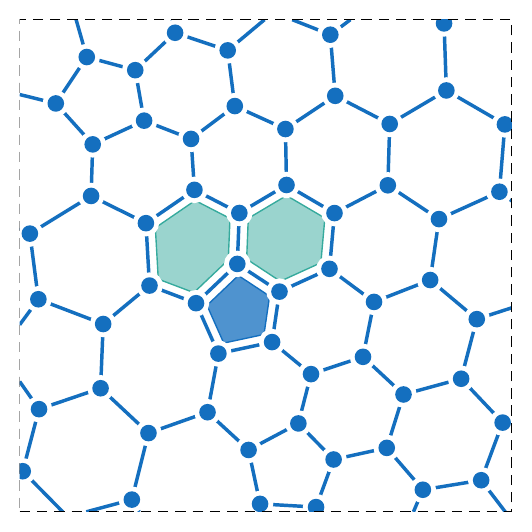}%
	\includegraphics{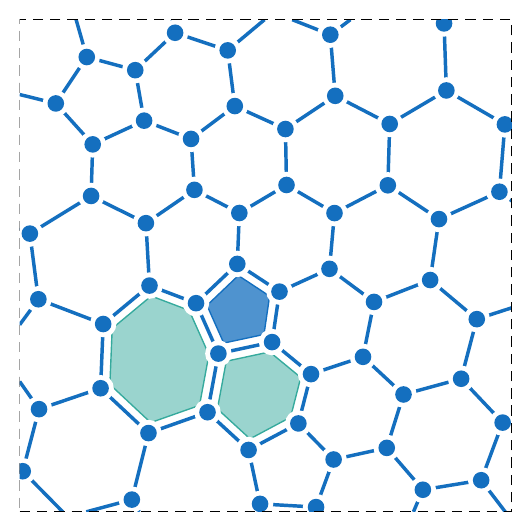}%
	\caption{Stencils for the CWENO reconstruction of order three ($M=2$) with $f=1.5$ for a pentagonal element $P_i^n$. Left: central stencil made of the element itself $P_i^n$ (in violet) and $n_e-1 = 8$ of its neighbors (in blue). In the other panels we report two of the $N_{V_i}^n =5$ sectorial stencils containing the element itself and two consecutive neighbors belonging to~$\mathcal{V}(P_i^n)$.}
	\label{fig.cweno_stencils}
\end{figure}

In Case B of our moving and topology changing polygonal mesh we adopt a CWENO reconstruction algorithm, first introduced in \cite{LPR:99,LPR:00,LPR:02,SCR:CWENOquadtree}, 
and which can be cast in the general framework described in \cite{cravero2018cweno}. 
We closely follow the work outlined in \cite{dumbser2017central, DGCWENO} for unstructured triangular and tetrahedral meshes, and extended it to moving polygonal grids in \cite{GBCKSD2019}.

We emphasize that the main advantages of such a procedure is that only one stencil (the central one) is required to contain the total amount of elements stated in~\eqref{eq.StencilTotElements} and only this one is used to construct a polynomial of degree $M$; the other ones are used to compute polynomials of lower degree.
In particular, we consider $N_{V_i}^n$ stencils $S_i^s$, each of them containing exactly $\hat{n}_e=(d+1)$ cells, i.e. the central cell $P^n_i$ and two consecutive neighbors belonging to~$\mathcal{V}(P_i^n)$.
Refer to Figure~\ref{fig.cweno_stencils} for a graphical description of the stencils.
For each stencil $S_i^s$ we compute a linear polynomial by solving a simple reconstruction system which is not overdetermined. 
According to the above mentioned literature, the reconstructed polynomial obtained via a nonlinear combination of the polynomial of degree $M$, computed over $S_0^s$, and of the $N_{V_i}^n$ linear polynomials, computed over  $S_i^s$, maintains the order of convergence of the method and avoids unwanted spurious oscillations.
In particular, in the case of moving meshes with topology changes, where the set of neighbors may change at any time step, the use of smaller so-called sectorial stencils significantly speeds up computations.

\smallskip
For the sake of uniform notation, in the DG case, i.e. when $N>0$ and $M=N$, we trivially impose that 
the reconstruction polynomial is given by the DG polynomial, i.e.  
$\mathbf{w}_h^n(\x,t^n)=\mathbf{u}_h^n(\x,t^n)$, which automatically implies
that in the case $N=M$ the reconstruction operator is simply the identity.

\medskip
\subsection{Space-time predictor step}
\label{ssec.predictor}

In this section we focus on the key feature, the element-local \textit{space-time predictor} step, of our ADER FV-DG schemes: 
this part of the algorithm (the \textit{predictor}) produces a high order approximation in both space and time of $\Q$ in all $P_i^n$. This allows to obtain a fully discrete one-step scheme that is uniformly high order accurate in both space and time.  

The predictor step consists in a completely \textit{local} procedure which solves the 
governing PDE~\eqref{eqn.pde} \textit{in the small}, see~\cite{harten}, inside each space-time element $C_i^n$,
and it only considers the geometry of volume $C_i^n$, the initial data $\w_h^n$ on $P_i^n$ and 
the governing equations~\eqref{eqn.pde}, without taking into account any interaction 
between $C_i^n$ and its neighbors. Because of this absence of communications, we refer to it as \textit{local}.
The procedure finally provides, for each $C_i^n$, a space-time polynomial data representation $\q_h^n$, which 
serves as a predictor solution, only valid inside $C_i^n$, to be used for evaluating the numerical fluxes, 
the non-conservative products and the algebraic source terms when integrating the PDE in the 
final \textit{corrector} step (see Section~\ref{ssec.corrector}) of the ADER scheme. 

The predictor $\q_h^n$ is a polynomial of degree $M$, which takes the following form 
\begin{equation} \begin{aligned}
\q_h^n(\x, t) = \sum_{\ell=0}^{\mathcal{Q}-1} \theta_\ell (\x, t) \hat{\q}_\ell^n, \qquad (\x,t) \in C_i^n, 
\label{eqn.qh}
\end{aligned} \end{equation}
where $\theta_\ell (\x, t)$ can be either
\begin{itemize}
	\item[i.] For fixed and adaptive Cartesian grids (Case A), \textit{nodal} space-time basis functions of degree $M$ given by the product of one-dimensional nodal basis functions verifying~\eqref{eq.nodalGeneralBasis} (with $d=1$),
	\begin{equation} \begin{aligned}
	\label{eq.nodalBasisTimeAMRpredictor}
	\theta_\ell(x,y,t) = \phi_{\ell_1}\left(\xi(x)\right)\phi_{\ell_2}\left(\eta(y)\right)\phi_{\ell_3}\left(\tau(t)\right),
	\end{aligned} \end{equation}
	two of them mapped to the unit interval $[0,1]$ as in~\eqref{eq.nodalBasisAMR} and with the time coordinate mapped to the reference time $\tau \in [0,1]$ via $t = t^n + \tau \Delta t$.
	In this case, the total number of GL quadrature points per cell, as well as the total number of DOF is $\mathcal{Q} = (M+1)^{d+1}$, see also Figure~\ref{fig:degreesoffreedom_AMR}.
	\begin{figure}[!b]
		\centering
		\includegraphics[width=0.4\linewidth]{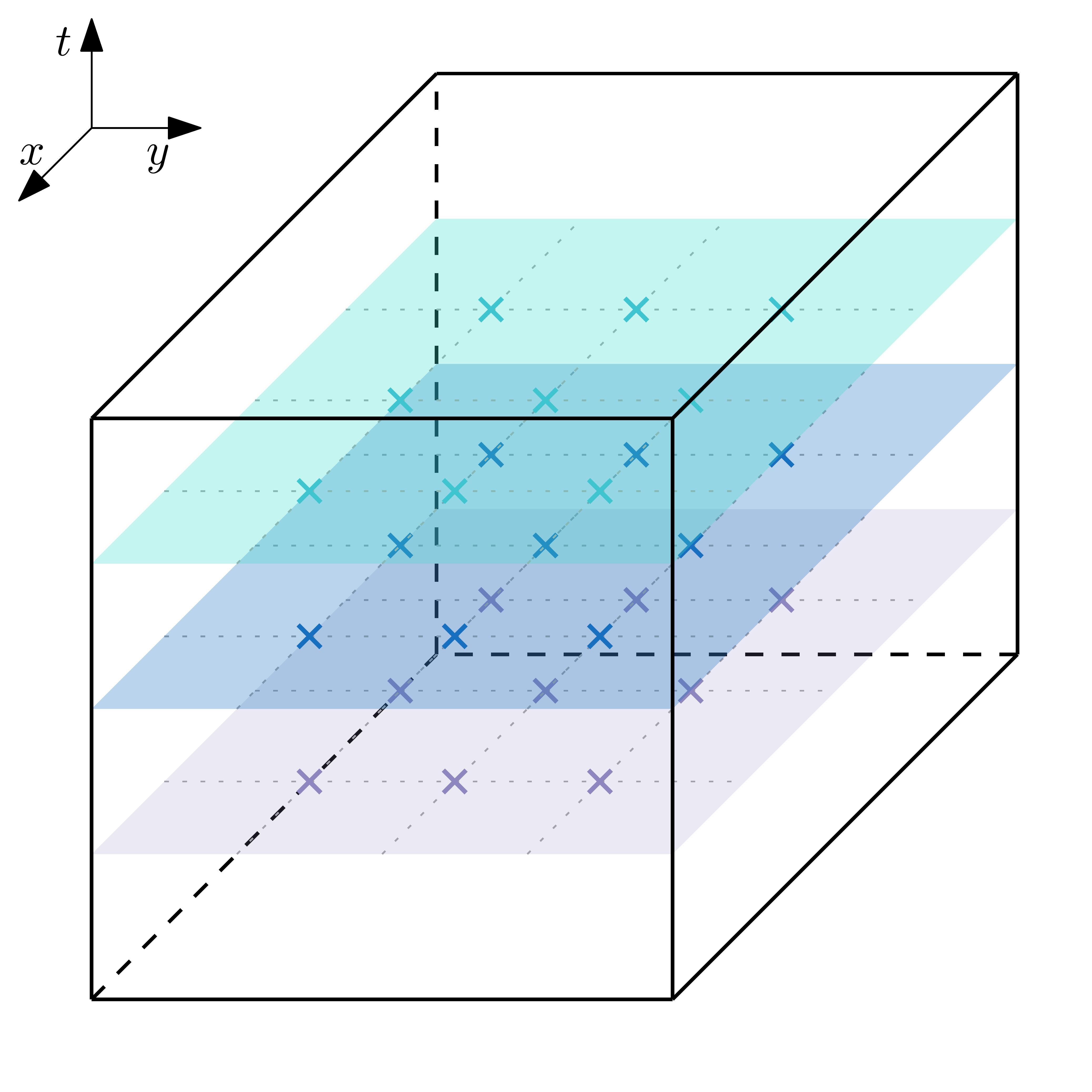}
		\caption{Quadrature points on a space-time element, $C_{i}^{n}$, of a fixed Cartesian mesh with $M=2$.}
		\label{fig:degreesoffreedom_AMR}
	\end{figure}

	\item[ii.] For our moving polygonal meshes (Case B), \textit{modal} space time basis functions of degree $M$ in $d+1$ dimensions ($d$ space dimensions plus time) are used, which read
	\begin{equation}
	\label{eq.Dubiner_phi}
	\theta_\ell(x,y,t)|_{C_i^n} = \frac{(x - \xbixn)^{p_\ell}}{{p_\ell}! \, h_i^{p_\ell}} \, \frac{(y - \xbiyn)^{q_\ell}}{{q_\ell}! \, h_i^{q_\ell}}
	\, \frac{(t - t^n)^{q_\ell}}{{q_\ell}! \, h_i^{q_\ell}}, 
	\qquad \ell = 0, \dots, \mathcal{Q}, 
	\quad 0 \leq p_\ell + q_\ell + r_\ell \leq M, \\
	\end{equation}  
	with the total number of DOF $\ \mathcal{Q} = \frac{1}{(d+1)!} \prod \limits_{m=1}^{d+1} (M+m)$, see also Figure~\ref{fig.quadraturePoints}.
	\begin{figure}
		\centering
		{\includegraphics{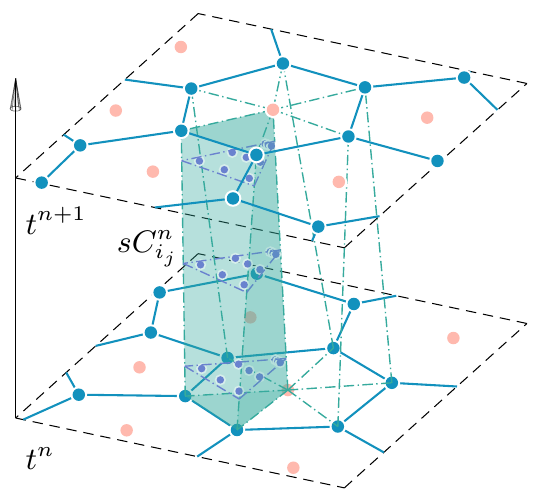}}%
		{\includegraphics{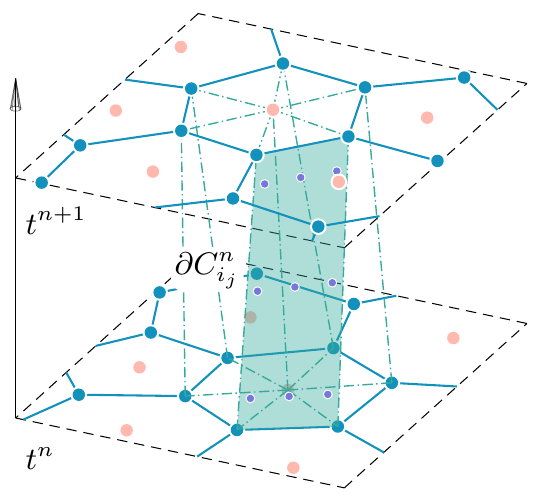}}%
		{\includegraphics{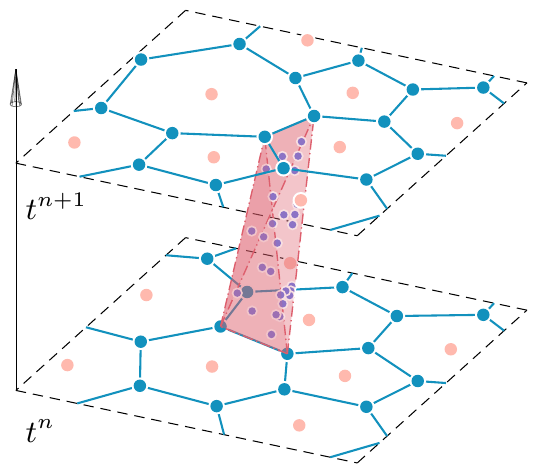}}
		\caption{Space--time quadrature points for third order methods, i.e. $M=2$, on a moving polygonal mesh with topology changes. 
			Left: quadrature points for the volume integrals and the space--time predictor. Middle: quadrature points for the surface integrals, i.e. for flux computation. Right: quadrature points for the volume integrals and the space--time predictor of a sliver element.}
		\label{fig.quadraturePoints}
	\end{figure}
\end{itemize}

Now, multiplying our PDE system~\eqref{eqn.pde} with a test function $\theta_k$ and integrating
over the space-time control volume $C_i^n$ (see Section~\ref{ssec.SpaceTimeConnectivity}), 
we obtain the following weak form of the governing PDE, where both the test and the
basis functions are \textit{time dependent}
\begin{equation} \begin{aligned}
\label{eqn.pde.st1}
\int_{C_i^n} \theta_k(\x,t) \frac{\partial \q_h^n}{\partial t} \d\x \d t
+\int_{C_i^n} \theta_k(\x,t) \left( \nabla \cdot \F(\q_h^n) + \mathbf{B}(\q_h^n) \cdot \nabla \q_h^n  \right) \d\x \d t
= \int_{C_i^n} \theta_k(\x,t) \mathbf{S}(\q_h^n) \d\x \d t\,.
\end{aligned} \end{equation}
Since we are only interested in an element local predictor solution,
i.e. we do not need to consider the interactions with the neighbors,
we do not yet take into account the jumps of $\q_h^n$ across the space--time lateral surfaces, 
because this will be done in the final corrector step (Section~\ref{ssec.corrector}). 

Instead, we insert the known discrete solution $\w_h^n(\x,t^n)$ at time $t^n$ in order to 
introduce a weak initial condition for solving our PDE; note that $\w_h^n(\x,t^n)$ 
uses information coming from the past only (following an \textit{upwinding approach}) 
in such a way that the causality principle is correctly respected. 
To this purpose, the first term is integrated by parts in time. 
This leads to
\begin{equation} \begin{aligned}
\label{eqn.pde.st2}
\int_{P_i^{n+1}} &  \theta_k(\x,t^{n+1}) \q_h^n(\x,t^{n+1}) \d\x 
- \int_{P_i^{n}}  \theta_k(\x,t^{n}) \w_h^n(\x,t^{n}) \d\x 
- \int_{C_i^{n}} \frac{\partial }{\partial t} \theta_k(\x,t) \q_h^n(\x,t) \d\x \d t\\
& + \ \int_{C_i^{n} \backslash \partial C_i^n}   \theta_k(\x,t) \nabla \cdot \F(\q_h^n)  \d\x \d t =  
\int_{C_i^{n} \backslash \partial C_i^n}  \theta_k(\x,t)
\left( \mathbf{S}(\q_h^n) - \mathbf{B}(\q_h^n) \cdot \nabla \q_h^n \right) \d\x \d t. 
\end{aligned} \end{equation}
Equation~\eqref{eqn.pde.st2} results in an element-local nonlinear system for the unknown 
degrees of freedom $\hat{\q}_\ell^n$ of the space-time polynomials $\q_h^n$. 
The solution of~\eqref{eqn.pde.st2} can be found via a simple and fast converging 
fixed point iteration (a discrete Picard iteration) as detailed e.g. in \cite{Dumbser2008,HidalgoDumbser}. For linear homogeneous 
systems, the discrete Picard iteration converges in a finite number of at most $N+1$ steps, 
since the involved iteration matrix is nilpotent, see \cite{Jackson}. 
Moreover a proof of the convergence of this procedure in the case of a nonlinear homogeneous conservation 
law in 1D is given in next Section~\ref{ssec.PredictorProof}.

\medskip
\subsubsection*{Simplification in the case of a fixed Cartesian mesh}
\label{ssec.cartesianpredictor}
The space-time predictor step formerly presented can be simplified in the case of a 
Cartesian mesh with nodal basis functions resulting in a more efficient algorithm.
Under these assumptions the governing PDE~\eqref{eqn.pde}, can be rewritten as
\begin{equation}
\frac{\partial \mathbf{Q}}{\partial \tau} + \frac{\partial \mathbf{f}^{\star}}{\partial \xi} + \frac{\partial \mathbf{g}^{\star}}{\partial \eta} + \mathbf{B}_{1}^{\star}  \frac{\partial \mathbf{Q}}{\partial \xi} + \mathbf{B}_{2}^{\star}  \frac{\partial \mathbf{Q}}{\partial \eta}  = \mathbf{S}^{\star}
\end{equation}
with 
\begin{equation} \begin{aligned}
\mathbf{f}^{\star} = \frac{\Delta t}{\Delta x_{j}}\mathbf{f}, \quad \mathbf{g}^{\star} = \frac{\Delta t}{\Delta y_{k}}\mathbf{g}, \quad \mathbf{B}_{1}^{\star}  = \frac{\Delta t}{\Delta x_{j}} \mathbf{B}_{1},\quad
\mathbf{B}^{\star}_{2} = \frac{\Delta t}{\Delta y_{k}}\mathbf{B}_{2},\quad \mathbf{B} = \left[\mathbf{B}_{1}, \mathbf{B}_{2}\right], \quad \mathbf{S}^{\star} = \Delta t \mathbf{S}.
\end{aligned} \end{equation}
Next, we multiply each term by a test function $\theta_{k}$ and we integrate over the 
reference space-time control volume $\mathcal{I}_{0}=\left[0,1\right]^{3}$
\begin{equation} \begin{aligned}
\int_{0}^{1} \int_{0}^{1} \int_{0}^{1} & \theta_{k} \left(   \frac{\partial \mathbf{Q}}{\partial \tau} + \frac{\partial \mathbf{f}^{\star}\left( \mathbf{Q} \right)}{\partial \xi} + \frac{\partial \mathbf{g}^{\star}\left( \mathbf{Q} \right)}{\partial \eta}  \right)   d\xi d\eta d\tau \\ &= 
\int_{0}^{1} \int_{0}^{1} \int_{0}^{1} \theta_{k} \left( \mathbf{S}^{\star}\left( \mathbf{Q} \right) - \mathbf{B}_{1}^{\star}\left( \mathbf{Q} \right)  \frac{\partial \mathbf{Q}}{\partial \xi} - \mathbf{B}_{2}^{\star}\left( \mathbf{Q} \right)  \frac{\partial \mathbf{Q}}{\partial \eta}  \right)  d\xi d\eta d\tau .
\end{aligned} \end{equation}
Now, by substituting the discrete space-time predictor solution $\mathbf{q}_{h}^{n}$ with its expansion on the nodal basis and after integrating by parts in time, we obtain
\begin{equation} \begin{aligned}
\label{eqn.pde.st2.cartesian}
\int_{0}^{1}  &\int_{0}^{1} \int_{0}^{1}   \theta_{k}\left(\xi,\eta,1\right) \theta_{\ell}\left(\xi,\eta,1\right)  \hat{\mathbf{q}}_{\ell}^{n}  d\xi d\eta d\tau+  
\int_{0}^{1} \int_{0}^{1} \int_{0}^{1}  \frac{\partial  \theta_{k}  \left(\xi,\eta,\tau\right)}{\partial \tau}   \theta_{\ell}\left(\xi,\eta,\tau\right)  \hat{\mathbf{q}}_{\ell}^{n} d\xi d\eta d\tau
\\
& =
\int_{0}^{1}  \int_{0}^{1} \int_{0}^{1}  \theta_{k}\left(\xi,\eta,0 \right)   \mathbf{w}_{h}^{n} \left(\xi,\eta,t^{n} \right)  d\xi d\eta d\tau
-\int_{0}^{1} \int_{0}^{1} \int_{0}^{1} \theta_{k} \left( \frac{\partial \mathbf{f}^{\star}\left( \mathbf{q}_{h}^{n} \right) }{\partial \xi} + \frac{\partial \mathbf{g}^{\star}\left( \mathbf{q}_{h}^{n} \right)}{\partial \eta}  \right) d\xi d\eta d\tau \\
& +
\int_{0}^{1} \int_{0}^{1} \times \int_{0}^{1} \theta_{k} \left( \mathbf{S}^{\star}\left( \mathbf{q}_{h}^{n}\right) - \mathbf{B}_{1}^{\star}\left( \mathbf{q}_{h}^{n} \right)  \frac{\partial \mathbf{q}_{h}^{n}}{\partial \xi} - \mathbf{B}_{2}^{\star}\left( \mathbf{q}_{h}^{n}\right)  \frac{\partial \mathbf{q}_{h}^{n}}{\partial \eta}  \right)  d\xi d\eta d\tau .
\end{aligned} \end{equation}
To recover the value of the unknown degrees of freedom $ \hat{\mathbf{q}}_{\ell}^{n}$, it is sufficient to solve the above equation locally for each element. One important advantage of using the nodal Gauss-Legendre basis is that the terms in~\eqref{eqn.pde.st2.cartesian} can be evaluated in a \textit{dimension-by-dimension} fashion.

\medskip
\subsubsection*{Space-time predictor for sliver space--time elements}
\label{ssec.PredictorSliver}

\begin{figure}
	\centering
	{\includegraphics[width=0.33\linewidth]{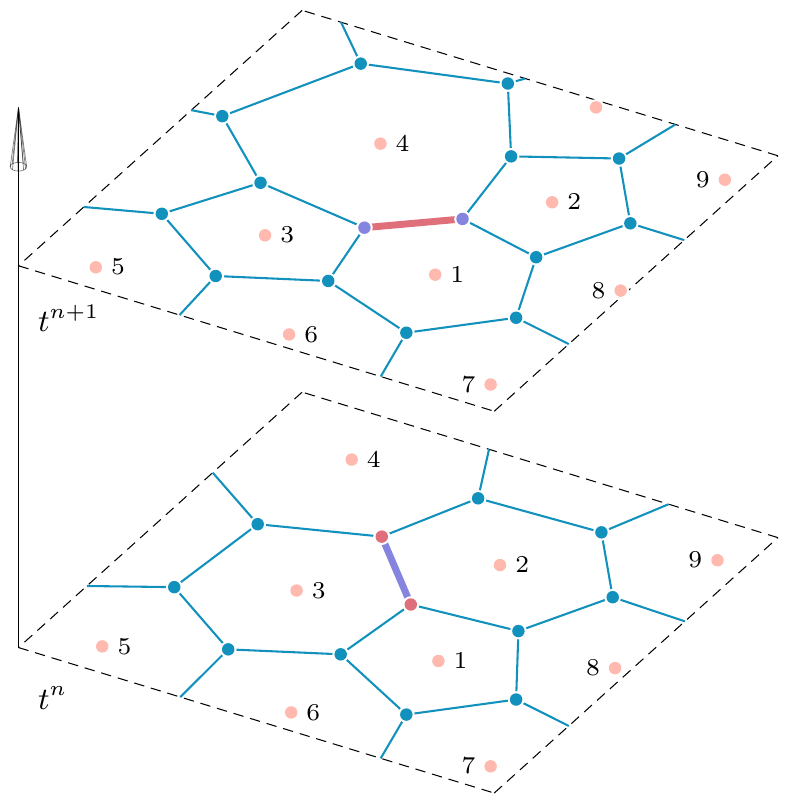}\label{sfig.Crazy_a}}%
	{\includegraphics[width=0.33\linewidth]{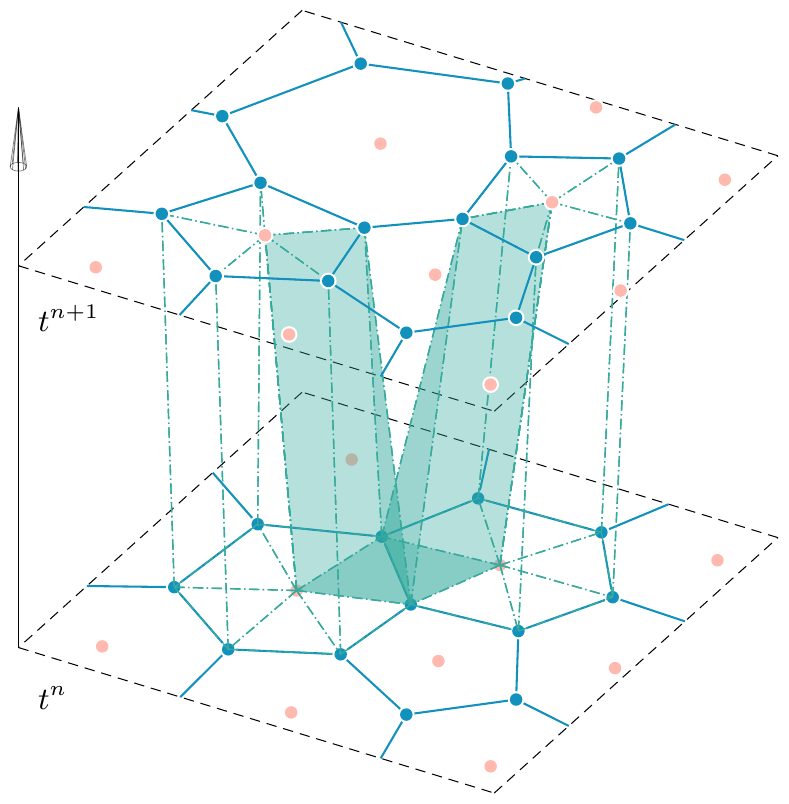}\label{sfig.Crazy_b}}%
	{\includegraphics[width=0.33\linewidth]{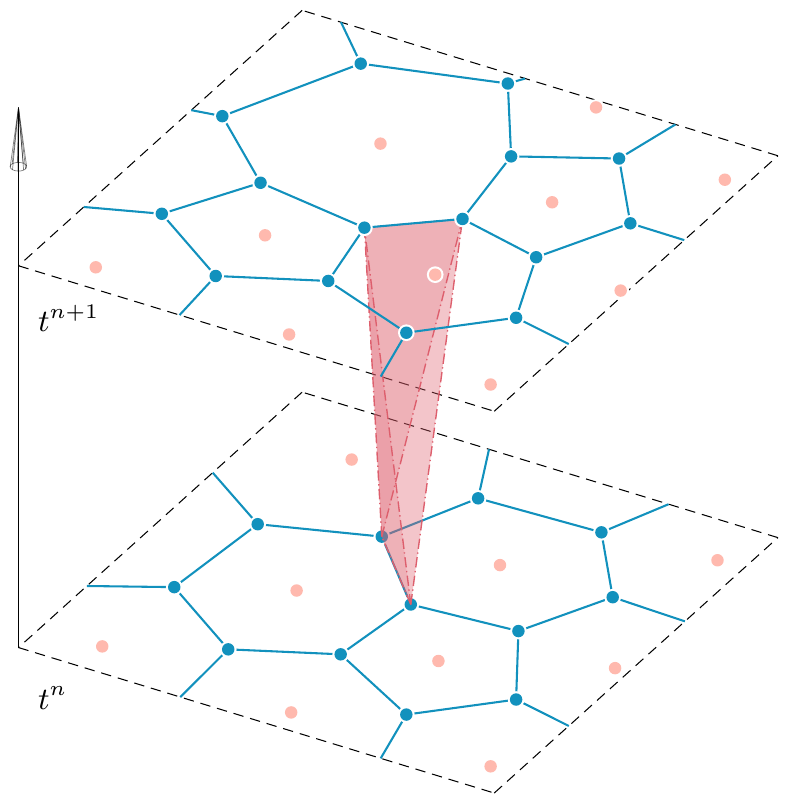}\label{sfig.Crazy_d}}%
	\caption{Space time connectivity \textit{with} topology changes and sliver element. Left: at time $t^n$ the polygons $P_2^n$ and $P_3^n$ are neighbors and share the highlighted edge, instead at time $t^{n+1}$ they do not touch each other; the opposite situation occurs for polygons $P_1^n$ and $P_4^n$. This change of topology causes the appearance of degenerate elements of different types (refer to \cite{GBCKSD2019} for all the details). 
		In particular, so-called \textit{space--time sliver elements} (right) need to be taken into account when considering the space--time framework, so the predictor and the corrector step have to be a adapted to their special features.
		Sliver elements (right) are indeed completely new control volumes which do neither exist at time $t^n$, nor at time $t^{n+1}$, since they coincide with an edge of the tessellation and, as such, have zero areas in space. However, they have a \textit{non-negligible volume} in space--time. The difficulties associated to this kind of element are due to the fact that $\w_h$ is not clearly defined for it at time $t^n$ (thus the predictor has to be modified) and that contributions across it should not be lost at time $t^{n+1}$ in order to guarantee conservation (thus the corrector has to be modified).  } 
	\label{fig.ControlVolume_TopChange_1sliver}
\end{figure}

When a topology change occurs, some space--time sliver elements, as those shown on the right side of Figure~\ref{fig.ControlVolume_TopChange_1sliver}, are  originated (see \cite{GBCKSD2019}), and the predictor procedure over them needs particular care.
The problem connected with sliver elements is the fact that 
their bottom face, which consists only in a line segment, is degenerate, 
hence the spatial integral over $P_i^n$ vanishes, i.e. there is no possibility to introduce 
an initial condition for the local Cauchy problem at time $t^n$ into their predictor.
Thus, in order to couple however~\eqref{eqn.pde.st1} with some known data from the past, we will end up with a formula different from~\eqref{eqn.pde.st2}.
We underline that we \textit{first} carry out the space--time predictor for all standard 
elements using, which can be computed independently of each other, and only subsequently we process the remaining space--time sliver elements. 
Then, when considering a sliver, we use the upwinding in time approach on the entire
space--time surface $\partial C_{i}^n$ that closes a sliver control volume, and again respecting 
the causality principle, we take the information to feed the predictor only from the past, i.e. only 
from those space--time neighbors $C_j^n$ whose common surface $\partial C_{ij}^n$ exhibit
a \textit{negative} time component of the outward pointing space--time normal vector ($ \tilde{\mbf{n}}_t < 0$). 
In this way, we can introduce information from the past into the space--time sliver elements.

As a consequence, the predictor solution $\q_{h}^{n}$ is again obtained by means of~\eqref{eqn.pde.st1}, 
but by treating the \textit{entire} $\partial C_{i}^n$ with the upwind in time approach,
i.e. by considering also the jump terms between the still unknown predictor of the slivers (call it $\q_{h}^{n,-}$) 
and the already known predictors of its neighbors (call them $\q_{h}^{n,+}$), 
\begin{equation} \begin{aligned}
\label{eqn.pde.st2.sliver}
& \int_{C_i^{n}}  \theta_k(\x,t) \frac{\partial }{\partial t} \q_h^n(\x,t) \d\x \d t 
- \int_{\partial C_i^{-} }  \theta_k(\x,t^{n}) \left(  (\q_h^{n,+} - \q_h^{n,-}) -  (\mathbf{B} \cdot \tilde{\mbf{n}}) (\q_h^{n,+} - \q_h^{n,-}) \right )  \d\S  \d t \\
& + \int_{C_i^{n} \backslash \partial C_i^n}   \theta_k(\x,t) \nabla \cdot \F(\q_h^n)  \d\x \d t 
= \int_{C_i^{n} \backslash \partial C_i^n}  \theta_k(\x,t) \left( \mathbf{S}(\q_h^n) - \mathbf{B}(\q_h^n) \cdot \nabla \q_h^n \right) \d\x \d t,
\end{aligned} \end{equation}
where $\partial C_i^- = \partial C_i^{n} \text{ with } \tilde{\mbf{n}}_t < 0$ is the part of the space-time boundary that has a negative time component of the space-time normal vector. 
Note that here we have taken into account also the jump of the nonconservative terms, 
and that these contributions have been added entirely (i.e. not only half of them, as in~\eqref{eqn.fluxPC}).
Indeed, in~\eqref{eqn.fluxPC} half of the jump contribution goes to one element, while 
the other half goes to the neighboring element; 
here instead, since the interaction between neighbors is only computed from the side of the sliver element, 
the entire jump contributes to the predictor in the sliver element.

\subsection{Convergence proof of the predictor step for a nonlinear conservation law}
\label{ssec.PredictorProof} 
In this section, the convergence proof of the predictor for a nonlinear conservation law is given. 
The proof is provided, for simplicity, in the case of a fixed mesh in one space dimension, following the 
nomenclature already employed in Section~\ref{ssec.cartesianpredictor}, but it still holds in higher dimensions.  
Let us consider a general hyperbolic system of conservation laws of the form
\begin{equation}
\frac{\partial \Q}{\partial t} + \frac{\partial \mathbf{f}}{\partial x} = 0.
\label{eqn.nonlinear.scalar} 
\end{equation}
Then, the corresponding space-time DG predictor used in the ADER-DG framework reads 	
\begin{equation} 
\int \limits_0^1 \int \limits_0^1 \theta_k \frac{\partial \mathbf{q}_h}{\partial \tau} d\xi d\tau  + 
\frac{\Delta t}{\Delta x} \, 
\int \limits_0^1 \int \limits_0^1 \theta_k \frac{\partial \mathbf{f}_h}{\partial \xi}  d\xi d\tau  
= 0. \label{eq:stDGpredictor}
\end{equation} 	
For convenience, all derivatives and integrals in~\eqref{eq:stDGpredictor} have been transformed to the reference space-time element $[0,1]^2$.  
Moreover, the discrete solution is given by $\mathbf{q}_h = \theta_l(\xi,\tau) \hat{\mathbf{q}}_\ell$, 
and the flux is expanded in the same basis as $\mathbf{f}_h = \theta_\ell(\xi,\tau) \hat{\mathbf{f}}_\ell$. When using a nodal basis, we can compute the degrees of freedom for the flux 
interpolant $\mathbf{f}_h$ simply as $ \hat{\mathbf{f}}_\ell = \mathbf{f} \left( \hat{\mathbf{q}}_\ell \right)$. We also recall that the initial condition given by
the DG scheme at time $t^n$ reads $\mathbf{w}_h = \phi_\ell(\xi) \hat{\mathbf{w}}_\ell$. 
Then, integration of the first term in~\eqref{eq:stDGpredictor} by parts in time yields   
\begin{equation} 
\int \limits_0^1 \theta_k(\xi,1) \mathbf{q}_h  d\xi 
-\int \limits_0^1 \int \limits_0^1  \frac{\partial \theta_k }{\partial \tau} \mathbf{q}_h  d\xi d\tau  + 
\frac{\Delta t}{\Delta x} \, 
\int \limits_0^1 \int \limits_0^1 \theta_k \frac{\partial \mathbf{f}_h}{\partial \xi}  d\xi d\tau  
= \int \limits_0^1 \theta_k(\xi,0) \mathbf{w}_h  d\xi,
\end{equation} 		
and insertion of the definitions of the discrete solution leads to  	
\begin{equation} 
\left( \int \limits_0^1 \theta_k(\xi,1) \theta_l(\xi,1)  d\xi \, 
- \int \limits_0^1 \int \limits_0^1 \frac{\partial \mathbf{\theta}_k}{\partial \tau} \theta_l d\xi d\tau \, \right) \, \hat{\mathbf{q}}_l  + 
\frac{\Delta t}{\Delta x} \, 
\int \limits_0^1 \int \limits_0^1 \theta_k \frac{\partial \mathbf{\theta}_l}{\partial \xi}  d\xi d\tau \, \hat{\mathbf{f}}_l  
= \int \limits_0^1 \theta_k(\xi,0) \phi_l(\xi)  d\xi \, \hat{\mathbf{w}}_l. 
\end{equation} 		
The iterative scheme employed to find the solution for the space-time degrees of freedom $\hat{\mathbf{q}}$, at any Picard iteration $r$, can therefore be rewritten in compact matrix-vector notation as 
\begin{equation} 
\mathbf{K}_1 \hat{\mathbf{q}}^{r+1}  + \frac{\Delta t }{\Delta x} \mathbf{K}_\xi \, \mathbf{f}\left(\hat{\mathbf{q}}^{r+1}  \right) = \mathbf{F}_0 \hat{\mathbf{w}}^{n}
\end{equation} 
with 
\begin{equation}
\mathbf{K}_1 = \int \limits_0^1 \theta_k(\xi,1) \theta_l(\xi,1)  d\xi \, 
- \int \limits_0^1 \int \limits_0^1 \frac{\partial \mathbf{\theta}_k}{\partial \tau} \theta_l d\xi d\tau,
\end{equation} 
\begin{equation}
\mathbf{K}_\xi = \int \limits_0^1 \int \limits_0^1 \theta_k \frac{\partial \mathbf{\theta}_l}{\partial \xi}  d\xi d\tau,
\qquad 
\mathbf{F}_0 = \int \limits_0^1 \theta_k(\xi,0) \phi_l(\xi)  d\xi,
\end{equation} 
where we have dropped the indices to ease the notation.
After inverting $\mathbf{K}_1$ (this matrix is built using the linearly independent basis functions so that it is invertible), 
we obtain the explicit iteration formula 
\begin{equation} 
\hat{\mathbf{q}}^{r+1}  =  \mathbf{K}_1^{-1} \mathbf{F}_0 \hat{\mathbf{w}}^{n} -  \frac{\Delta t }{\Delta x} \mathbf{K}_1^{-1} \mathbf{K}_\xi \, \mathbf{f}\left(\hat{\mathbf{q}}^{r}  \right). \label{eq:explitform} 
\end{equation} 
To prove that the former iterative formula will converge, we introduce the operator
\begin{equation} 
\boldsymbol{\phi} \left( \hat{\mathbf{q}} \right)  =  \mathbf{K}_1^{-1} \mathbf{F}_0 \hat{\mathbf{u}}^{n} -  \frac{\Delta t }{\Delta x} \mathbf{K}_1^{-1} \mathbf{K}_\xi \, \mathbf{f}\left(\hat{\mathbf{q}}   \right),  
\end{equation} 
and the induced matrix norm 
\begin{equation}
\left\| \mathbf{A} \right\| = \sup \limits_{\mathbf{x} \neq 0} \frac{ \left\| \mathbf{A} \mathbf{x} \right\|}{ \left\| \mathbf{x} \right\|} . \label{eq:inducednorm}
\end{equation}
Furthermore, we assume the flux to be Lipschitz continuous with Lipschitz constant 
$L>0$ so that
\begin{equation}
\left\|  \mathbf{f}\left(\hat{\mathbf{p}} \right)  - \mathbf{f}\left(\hat{\mathbf{q}} \right) \right\|  \leq L 
\left\|  \hat{\mathbf{p}}  - \hat{\mathbf{q}}   \right\| .
\end{equation}
We now need to show that the operator $\boldsymbol{\phi}$ is a contraction:  
\begin{eqnarray}
\left\| \boldsymbol{\phi} \left( \hat{\mathbf{q}} \right) - \boldsymbol{\phi} \left( \hat{\mathbf{p}} \right) \right\|
&=& \left\| \mathbf{K}_1^{-1} \mathbf{F}_0 \hat{\mathbf{u}}^{n} - \mathbf{K}_1^{-1} \mathbf{F}_0 \hat{\mathbf{u}}^{n}   
-  \frac{\Delta t }{\Delta x} \mathbf{K}_1^{-1} \mathbf{K}_\xi \, \mathbf{f}\left(\hat{\mathbf{q}}  \right) 
+  \frac{\Delta t }{\Delta x} \mathbf{K}_1^{-1} \mathbf{K}_\xi \, \mathbf{f}\left(\hat{\mathbf{p}}  \right) \right\| \nonumber \\
&= &  \frac{\Delta t }{\Delta x} \left\|   
\mathbf{K}_1^{-1} \mathbf{K}_\xi \, \left( \mathbf{f}\left(\hat{\mathbf{p}} \right) - \mathbf{f}\left(\hat{\mathbf{q}} \right) \right)
\right\| \nonumber \\ 
& \leq &  \frac{\Delta t }{\Delta x} \left\|   
\mathbf{K}_1^{-1} \mathbf{K}_\xi \right\| \,  \left\|  \mathbf{f}\left(\hat{\mathbf{p}} \right) - \mathbf{f}\left(\hat{\mathbf{q}} \right) 
\right\| \nonumber \\ 
& \leq &  L \frac{\Delta t }{\Delta x} \left\|   
\mathbf{K}_1^{-1} \mathbf{K}_\xi \right\| \,  \left\|  \hat{\mathbf{p}}  - \hat{\mathbf{q}}   \right\| \,. 
\end{eqnarray}
The operator is therefore a \textit{contraction} under the CFL-type condition on the time step $\Delta t$
\begin{equation}
0 < L \frac{\Delta t }{\Delta x} \left\|   \mathbf{K}_1^{-1} \mathbf{K}_\xi \right\| < 1, \label{eq:contraction}
\end{equation}
which connects the Lipschitz constant $L$ with the mesh spacing $\Delta x$ and the matrix norm of $\left\|   \mathbf{K}_1^{-1} \mathbf{K}_\xi \right\|$. 
Since the operator is contractive under the above assumptions, the Banach fixed point theorem, \cite{Banach1922}, guarantees convergence of the iterative method. 

In the previous reasoning, we have assumed that the inequality in the right hand side of~\eqref{eq:contraction} be strict. Thus, to conclude the proof, let us assume that the equality holds, this is true if and only if $\left\|   \mathbf{K}_1^{-1} \mathbf{K}_\xi \right\|=0$. 
By taking into account the definition of the induced matrix norm~\eqref{eq:inducednorm}, 
it implies  $\left\| \mathbf{K}_1^{-1} \mathbf{K}_\xi\, \mathbf{x}\right\| = 0$ for 
any $\mathbf{x}$ in the metric space. Thus, $\mathbf{K}_1^{-1} \mathbf{K}_\xi=0$. Direct 
substitution in~\eqref{eq:explitform} gives
\begin{equation} 
\mathbf{K}_1\hat{\mathbf{q}}^{r+1}  =  \mathbf{F}_0 \hat{\mathbf{w}}^{n}, 
\end{equation} 
so that no iterative procedure is done.

Note: The matrix $\mathbf{K}_1^{-1} \mathbf{K}_\xi$ has been proven to be nilpotent and thus all its eigenvalues are zero, see \cite{Jackson}, which guarantees convergence to the exact solution in a finite number of steps for linear homogeneous PDE. 

\medskip
\subsection{Corrector step}
\label{ssec.corrector}

The corrector step is the last step of our \textit{path-conservative} ADER FV-DG scheme, 
where the update of the solution from time $t^n$ up to time $t^{n+1}$ can take place in 
a single step procedure thanks to the use of the predictor $\q_h^n$.

The update formula is recovered starting from the space--time divergence form of the PDE
\begin{equation} \begin{aligned}
\tilde \nabla \cdot \tilde{\F}(\Q) + \tilde{\mbf{B}}(\Q) \cdot \tilde{\nabla} \Q = \S(\Q), \quad  
\tilde{\F} = (\F, \Q), \quad \tilde{\mbf{B}} = (\mbf{B}, \mbf{0}), \ \text{ and } \ 
\tilde \nabla  = \left( \partial_\x, \, \partial_t \right)^T,
\label{PDEdiv}
\end{aligned} \end{equation}
which is multiplied by a set of space--time test functions $\phitilde_k$ and integrated over each space--time control volume $C_i^n$
\begin{equation} \begin{aligned}
\label{eqn.weak1}
\int_{C_i^n} \phitilde_k(\x,t) \left( \nabla \cdot \tilde \F(\Q) + \tilde{\mathbf{B}}(\Q) \cdot \nabla \Q  \right) \d\x \d t
= \int_{C_i^n} \phitilde_k(\x,t) \mathbf{S}(\Q) \d\x \d t\,.
\end{aligned} \end{equation}

Note that the employed test functions $\phitilde_k$ coincide with the $\theta_k$ of~\eqref{eq.nodalBasisTimeAMRpredictor} for the Cartesian Case A.
Instead, for the moving polygonal Case B, they need to be tied to the motion of the barycenter $\x_{\mathbf{b}_i}(t)$ and must be moved together 
with $P_i(t)$ in such a way that at time $t = t^{n}$ they refer to the current barycenter $\xbin$ and 
at time $t = t^{n+1}$ they refer to the new barycenter $\xbinp$, thus they are defined as follows  
\begin{equation} \begin{aligned}
\label{eq.Dubiner_phi_movingspatial}
& \phitilde_\ell(x,y,t)|_{C_i^{n}} = 
\frac{(x - x_{b_i}(t))^{p_\ell}}{{p_\ell}! \, h_i^{p_\ell}} \, \frac{(y - y_{b_i}(t))^{q_\ell}}{{q_\ell}! \, h_i^{q_\ell}}, \quad 
\text{ with } \ \x_{\mbf{b}_i} (t) = \frac{t-t^n}{\Delta t} \xbin + \left(1-\frac{t-t^n}{\Delta t}\right) \xbinp, \\ 
& \ell = 0, \dots, \mathcal{N}, \quad  0 \leq p+q \leq N. 
\end{aligned} \end{equation}
These moving modal basis functions are essential to the moving approach presented in \cite{GBCKSD2019} and used
in this paper. 
They naturally allow for topology changes, without the need of any remapping steps, 
which we want to avoid in a direct ALE formulation.

Now,~\eqref{eqn.weak1} by applying the Gauss theorem to the flux-divergence term and 
by splitting the non-conservative products into their volume and surface contribution, becomes 
\begin{equation} \begin{aligned}
&\int_{P_i^{n+1}} \phitilde_k  \u_h(\x,t^{n+1}) \, d\x   
=  \int_{P_{i}^n} \phitilde_k  \u_h(\x,t^{n})  \, d\x 
- \sum_{j=1}^{N_{V_i}^{n,st}} \int_{\partial C_{ij}^n} \phitilde_k  \mathcal{D}(\q_h^{n,-},\q_h^{n,+}) \cdot \mathbf{\tilde n} \, dS \\
& + \int_{C_i^{n} \backslash \partial C_i^n} \tilde \nabla \phitilde_k \cdot \tilde{\F} (\q_h)\, d\mathbf{x} dt  
+ \int_{C_i^{n} \backslash \partial C_i^n}  \phitilde_k(\x,t) \left( \mathbf{S}(\q_h^n) - \mathbf{B}(\q_h^n) \cdot \nabla \q_h^n \right) \d\x \d t,
\label{eqn.GaussPDE2}
\end{aligned} \end{equation}
where $\Q$ on $P_i^{n+1}$ is represented by the unknown $\u_h^{n+1}$, 
on $P_i^n$ is taken to be the current representation of the conserved variables $\u_h^n$, 
in the interior of $C_i^n$ is given by the predictor $\q_h^{n}$ 
and on the space--time lateral surfaces $\partial C_{ij}^n$ is given by
$\q_h^{n,-}$ and $\q_h^{n,+}$ which are the so-called boundary-extrapolated data, i.e. 
the values assumed respectively by the predictors of the two neighbor elements $C_i^n$ and $C_j^n$ on the shared space--time lateral surface  $\partial C_{ij}^n$.
Furthermore, we have employed a two-point path-conservative numerical flux function of Rusanov-type
\begin{equation} \begin{aligned}
\label{eqn.fluxPC}
\mathcal{D}(\q_h^{n,-},\q_h^{n,+}) \cdot \mathbf{\tilde n} 
=&  \ \frac{1}{2} \left( \tilde{\mbf{F}}(\q_h^{n,+}) + \tilde{\mbf{F}}(\q_h^{n,-}) \right) \cdot \tilde{\mbf{n}} 
- \frac{1}{2} s_{\max} \left( \q_h^{n,+} - \q_h^{n,-} \right)\, \\
& + \frac{1}{2} \left(\int \limits_{0}^{1} \tilde{\mbf{B}} \left(\mbf{\Psi}(\q_h^{n,-},\q_h^{n,+},s) \right)\cdot\mbf{n} \, d\x \right)
\cdot\left(\q_h^{n,+} - \q_h^{n,-}\right),
\end{aligned} \end{equation}
where $s_{\max}$ is the maximum eigenvalue of the ALE Jacobian matrices $\mathbf{A}^{\!\! \mathbf{V}}_{\mathbf{n}}(\q_h^{n,+})$ and $\mathbf{A}^{\!\! \mathbf{V}}_{\mathbf{n}}(\q_h^{n,-})$ 
being
\begin{equation} \begin{aligned}
\label{eq.ALEjacobianMatrix}
\mathbf{A}^{\!\! \mathbf{V}}_{\mathbf{n}}(\Q)=\left(\sqrt{\tilde n_x^2 + \tilde n_y^2}\right)\left[\frac{\partial \mathbf{F}}{\partial \Q} \cdot \mathbf{n}  - 
(\mathbf{V} \cdot \mathbf{n}) \,  \mathbf{I}\right], \qquad    
\mathbf{n} = \frac{(\tilde n_x, \tilde n_y)^T}{\sqrt{\tilde n_x^2 + \tilde n_y^2}},
\end{aligned} \end{equation}
and the path $\mbf{\Psi}=\mbf{\Psi}(\q_h^-,\q_h^+, s)$ is a straight-line segment path
\begin{equation} \begin{aligned}
\mbf{\psi} = \mbf{\psi}(\q_h^-, \q_h^+, s) = \q_h^- + s \left( \q_h^+ - \q_h^- \right)\,,  s \in [0,1]\,, 
\end{aligned} \end{equation}
connecting $\q_h^{n,-}$ and $\q_h^{n,+}$ which allow to treat the jump of the non-conservative 
products following the theory introduced in \cite{DLMtheory,pares2006,Castro2006}, and extended 
to ADER FV-DG schemes of arbitrary high order in \cite{ADERNC,OsherNC}. Despite in this paper 
we only consider the Rusanov flux, the above methodology can be extended to different flux 
functions, adapting to the new flux splitting techniques like the ones presented in \cite{TVC2012}.
Finally, the time step size $\Delta t$ is given by 
\begin{equation} 
\Delta t < \text{CFL}\frac{h_{\text{min}} }{
	\left(2N+1\right)} \frac{1}{|\lambda_{\text{max}}|}, \quad \text{(Case A)}, \qquad 
\Delta t < \textnormal{CFL} \left(\!\! 
\frac{ |P_i^n| }{ (2N+1) \, |\lambda_{\max}| \, \sum_{\partial P_{i_j}^n} |\ell_{i_j}| } 
\!\!\right) \quad \text{(Case B)},  
\label{eq:timestep}
\end{equation} 
where $h_{\text{min}}$ is the minimum characteristic mesh-size, $\ell_{i_j}$ is the length 
of the edge $j$ of $P_i^n$ and $|\lambda_{\max}|$ is the spectral radius of the Jacobian 
of the flux $\mathbf{F}$.
Stability on unstructured meshes is guaranteed by the satisfaction of the 
inequality $\textnormal{CFL} < \frac{1}{d}$, see \cite{Dumbser2008}.

We close this section by remarking that the integration of the governing PDE 
over \textit{closed} space-time volumes $C_i^n$ automatically satisfies the geometric 
conservation law (GCL) for all test functions $\phitilde_k$. This simply follows from the Gauss 
theorem and we refer to~\cite{Lagrange3D} for a complete proof.

\medskip
\subsection{A posteriori subcell finite volume limiter}

Up to now, we have presented a family of FV and DG type schemes which achieves arbitrary high order of accuracy in space and time; 
the main difference between the FV and the DG approach lies in the fact that FV schemes, 
thanks to the WENO-type nonlinear reconstruction procedure, are robust in the presence of shocks and discontinuities, 
while the DG formulation as presented so far, being linear in the sense of Godunov, 
is subject to the appearance of spurious oscillations.
Thus, in order to employ a DG scheme in the context of solving hyperbolic partial differential 
equations, where usually discontinuities are developed,
a technique that is able to limit spurious oscillations (called \textit{limiter}) should be introduced.
Several attempts in that direction can be found in the literature. 
For example, we could recall the \textit{artificial viscosity} 
technique used in \citep{Hartman_02,Persson_06,Feistauer4} which consists in adding a small 
parabolic term in the equation in order to smooth out the discontinuities.

Here, instead, we follow a different approach based on exploiting the respective strengths of FV and DG schemes, 
i.e. the resolution of DG in smooth regions and the robustness of FV across discontinuities.
Thus, we first evolve the solution everywhere by using our DG scheme; 
then, we check \textit{a posteriori}, at the end of each time step, 
if the obtained DG solution in each cell respects or not some criteria 
(as density and pressure positivity, a relaxed discrete maximum principle, specific physical bounds, 
or more elaborate choices as those of \cite{guermond2018second}),
and we mark as \textit{troubled} those cells where the obtained DG solution is marked as not acceptable.
Only for these troubled cells we repeat the time step using, instead of the DG scheme, 
a second order TVD FV method, which always assures a robust solution.

This idea is founded on works as those of \cite{cbs4, QiuShu1, Qiu_2004, balsara2007, Luo_2007, Krivodonova_2007, Zhu_2008, Zhu_13, CDL1,CDL2,CDL3,ADER_MOOD_14,ALEMOOD1,ALEMOOD2}; but in particular, here, we adopt a so-called \textit{subcell} approach aimed at not losing the resolution of the DG scheme when switching to the FV method,
as forwarded in \cite{Sonntag,DGLimiter1,DGLimiter2,DGLimiter3,ALEDG,ADERDGVisc,rannabauer2018ader,DeLaRosaMunzDGMHD,DGCWENO}.
Indeed, at the beginning of the time step we \textit{project} the DG solution $\u_h^n$ of a troubled cell $P_i^{n}$ on a subdivision of it in sub-cells 
$s_{i,\alpha}^n$ obtaining a value for the cell averages on $s_{i,\alpha}^n$ at time $t^n$  
\begin{equation}
\mathbf{v}_{i,\alpha}^n(\x,t^n) = \frac{1}{|s_{i,\alpha}^n|} \int_{s_{i,\alpha}^n} \mathbf{u}_{h}^n(\x,t^n) \, d\x = \frac{1}{|s_{i,\alpha}^n|} \int_{s_{i,\alpha}^n} \phi_\ell(\x) \, d\x \, \hat{\mathbf{u}}^{n}_{l}=\mathcal{P}(\mathbf{u}_h^n) \qquad \forall \alpha.
\label{eqn.vh}
\end{equation}
We evolve the cell averages up to time $t^{n+1}$ using a classical TVD FV scheme, obtaining $\mathbf{v}_{i,\alpha}^{n+1}(\x,t^{n+1})$.
Finally, we recover a DG polynomial representation of the solution at time $t^{n+1}$ over $P_i^{n+1}$ using the values on the sub-grid level $\mathbf{v}_{i,\alpha}^{n+1}$ and by applying a \textit{reconstruction} operator as
\begin{equation}
\int_{S_{i,\alpha}^n} \mathbf{u}_{h}^{n+1}(\x,t^{n+1}) \, d\x = \int_{S_{i,\alpha}^n} \mathbf{v}_{i,\alpha}^{n+1}(\x,t^n) \, d\x =\mathcal{R}(\mathbf{v}_{i,\alpha}^{n+1}(\x,t^n))  \qquad \forall \alpha,
\label{eqn.intRec}
\end{equation}
where the reconstruction is imposed to be \textit{conservative} on the main cell $P_i^{{n+1}}$ yielding the additional linear constraint
\begin{equation}
\int_{P_i^{{n+1}}} \mathbf{u}_{h}(\x,t^{n+1}) \, d\x = \int_{P_i^{{n+1}}} \mathbf{v}_{h}(\x,t^{n+1}) \, d\x.
\label{eqn.LSQ}
\end{equation}
Thus, the limited solution on a troubled cell is \textit{robust} thanks to the use of a 
TVD scheme and \textit{accurate} thanks to the subcell resolution.

For all the details of the \textit{a posteriori} subcell FV limiter used in this work, we refer to \cite{DGLimiter1,ADERGRMHD} for the fixed Cartesian Case A and to \cite{GBCKSD2019} for the moving polygonal Case B. 


\section{A unified first order hyperbolic model of continuum mechanics}
\label{sec.Model}

\medskip
\subsection{Governing PDE system} 

A simplified diffuse interface formulation of the unified continuum fluid and solid mechanics model 
\cite{PeshRom2014,GPRmodel,GPRmodelMHD,HYP2016}, which can be used for modeling moving boundary 
problems of fluids and solids of arbitrary geometry, is given by the following PDE system 
(throughout this paper we make use of the Einstein summation convention over repeated indices) 
\begin{subequations}\label{eqn.GPR}
	\begin{align}
	& \frac{\partial \alpha}{\partial t}+v_k \frac{\partial \alpha}{\partial x_k}=0,  
	\label{eqn.alpha} \\ 
	&\frac{\partial (\alpha \rho)}{\partial t}  + \frac{\partial (\alpha \rho v_k )}{\partial x_k} =0, 
	\label{eqn.mass} \\
	&\frac{\partial (\alpha \rho v_i)}{\partial t}  +\frac{\partial \left(  \alpha \rho v_i v_k + \alpha p \delta_{ik} - \alpha \sigma_{ik}  \right) }{\partial x_k}   = \rho g_i, 
	\label{eqn.momentum} \\ 
	&\frac{\partial A_{ik}}{\partial t} +\frac{\partial (A_{ij}v_j)}{\partial x_k} + 
	v_j \left( \frac{\partial A_{ik}}{\partial x_j} -\frac{\partial A_{ij}}{\partial x_k}  \right) 
	= 
	-\frac{1}{\theta_1(\tau_1)} E_{A_{ik}}, 
	\label{eqn.A} \\  
	& \frac{\partial (\alpha \rho J_i)}{\partial t} + \frac{\partial \left( \alpha \rho J_i v_k + 
		T \delta_{ik}\right) }{\partial x_k}  
	= - \frac{1}{\theta_2(\tau_2)} E_{J_i},  
	\label{eqn.J} \\ 
	& \frac{\partial (\alpha \rho S) }{\partial t} + \frac{\partial \left( \alpha \rho S v_k + E_{J_k} 
		\right)}{\partial x_k} = 
	\frac{\rho}{T} \left( \frac{1}{\theta_1} E_{A_{ik}} E_{A_{ik}} + \frac{1}{\theta_2} E_{J_k} E_{J_k}   \right) \geq 0,  
	\label{eqn.entropy} \\ 
	& \frac{\partial (\alpha \rho {E})}{\partial t}  + \frac{ \partial \left( v_k \alpha \rho {E} + \alpha v_i(p\delta_{ik} - \sigma_{ik}) \right) }{\partial x_k} = \rho g_i v_i.  
	\label{eqn.energy} 
	\end{align}
\end{subequations} 
Here,~\eqref{eqn.alpha} is the evolution equation for the color function $\alpha$ that is needed 
in the diffuse interface approach as introduced in \cite{Tavelli2019} for the description of 
linear elastic solids of arbitrary geometry and as used in \cite{DIM2D,DIMWB} for a simple 
diffuse interface method for the simulation of non-hydrostatic free surface flows. We assume that 
the color function $ \alpha$ equals to 1 in the regions of the computational domain occupied by 
the material and 0 outside these regions. In the computational code, $ \alpha = 1 - \epsilon $ 
inside of the material and $ \alpha = \epsilon $ outside the material. Here, $ \epsilon $ is a 
small parameter $ \epsilon \ll 1 $, see Section\,\ref{sec.results}. Then, inside of the 
diffuse interface, $\alpha$ may take any 
values between 0 and 1 (between $ \epsilon $ and $ 1- \epsilon $ in the computational code). 
Equation~\eqref{eqn.mass} 
is the mass conservation law and $\rho$ is the material density; \eqref{eqn.momentum} is the 
momentum conservation law, where $v_i$ is the velocity field and $g_i$ is the gravity vector; 
\eqref{eqn.A} is the evolution equation 
for distortion field $A_{ik}$ (non-holonomic basis triad, see \cite{PRD-Torsion2019}); 
\eqref{eqn.J} is the evolution equation for the 
specific thermal impulse $J_k$ constituting the heat conduction in the matter via a hyperbolic (non 
Fourier--type) model.  Finally,~\eqref{eqn.entropy} is the entropy balance equation and 
\eqref{eqn.energy} is the energy conservation law.
Other thermodynamic parameters are defined via the total energy potential 
$E=E(\alpha, \rho,S,\vv,\AA,\JJ)$: 
$\Sigma_{ik}=p\delta_{ik} - \sigma_{ik}$ is the total stress tensor ($\delta_{ik}$ is the Kronecker 
delta); $p = \rho^2 E_\rho$ is the thermodynamic pressure;
$\sigma_{ik}=-\rho A_{jk} E_{A_{ji}} $ is the non-isotropic part of the stress tensor, $T = E_S$ is 
the temperature, and the notations such as $E_{\rho} $, $ E_{A_{ik}}$, etc. stand for the partial 
derivatives of the 
energy potential, e.g. $E_{\rho} = \frac{\pd E}{\pd \rho}$, $E_{A_{ik}} = \frac{\pd E}{\pd 
	A_{ik}}$, etc.

The dissipation in the medium includes two relaxation processes: the shear stress relaxation 
characterized by the scalar function $\theta_1(\tau_1) > 0$ depending on the relaxation time 
$\tau_1$ and thermal impulse relaxation characterized by $\theta_2(\tau_2) > 0$ depending on the 
relaxation time $\tau_2$. Both these relaxation processes then contribute to the entropy production 
term (the source on the right hand-side of~\eqref{eqn.entropy}) which is positive because it is 
quadratic in $E_{A_{ik}}$ and $E_{J_k}$.

From the mathematical standpoint, the unification of the model~\eqref{eqn.GPR} consists in the use 
of only first-order hyperbolic equations for both dissipative and non-dissipative processes in 
contrast to the classical continuum mechanics relying on the mixed hyperbolic-parabolic 
formulations such as the famous Navier-Stokes-Fourier equations, for example. From the physical 
standpoint, the 
unification of equations~\eqref{eqn.GPR} consists in treating solid and fluid states of matter from 
the solid-dynamics viewpoint. Indeed, as discussed in \cite{PeshRom2014,GPRmodel,HYP2016}, similarly 
to standard continuum solid-dynamics, the distortion field introduces additional degrees of 
freedom (in comparison to the classical continuum fluid mechanics) which characterizes deformation 
and rotational degrees of freedom of the continuum particles, represented not as 
scaleless mathematical points but characterized by a finite length scale, or equivalently, 
time scale $\tau_1$, e.g. see \cite{HYP2016}. In such a formulation, solid-type behavior 
corresponds to 
relaxation times $\tau_1$ such that $T^\mathsmaller{problem} \ll \tau_1$, 
while the fluid-type behavior corresponds to $\tau_1 \ll T^\mathsmaller{problem}$, 
where $T^\mathsmaller{problem}$ is the characteristic time scale of the problem under 
consideration.

In order to close system \eqref{eqn.GPR}, that is, in order to define pressure $p = \rho^2 E_{\rho}$, stresses
$\sigma_{ik} =-\rho A_{jk}E_{A_{ji}}$, temperature $T = E_{S}$, and the dissipative source terms, 
one needs to provide the energy potential $E$. In this paper, we rely on a rather simple choice 
of $E$, which is, however, enough to deal with Newtonian fluids and simple hyperelastic solids. 
Thus, we assume that the specific total energy can be written as a sum of three 
contributions as 
\begin{equation}
\label{eqn.energy.sum} 
E(\alpha,\rho,S,v_i,A_{ik},J_k) = E_1(\rho,S) + E_2(\alpha,A_{ik}, J_k) + E_3(v_i),   
\end{equation}
with the specific internal energy given by the ideal gas equation of state
\begin{equation}
E_1(\rho,S) = \frac{c_0^2}{\gamma(\gamma-1)}, \ c_0^2 = \gamma\rho^{\gamma-1} e^{S/c_v},
\quad \text{or} \quad 
E_1(\rho,p) = \frac{p}{\rho (\gamma-1)},
\end{equation}
in the case of gases, and given by either the so-called stiffened gas equation of state
\begin{equation}
E_1(\rho,S) = \frac{c_0^2}{\gamma(\gamma-1)}\left (\frac{\rho}{\rho_0}\right )^{\gamma-1} e^{S/c_v} 
+ \frac{\rho_0 c_0^2 -\gamma p_0}{\gamma\rho}
\end{equation}
or the well-known Mie-Gr{\"u}neisen equation of state
\begin{equation}\label{eq:mie_gruneisen_eos}
E_1(\rho,p)= \frac{p-\rho_0 c_0^2 \ f(\nu)}{\rho_0 \Gamma_0}, \quad f(\nu) = 
\frac{(\nu-1)(\nu-\frac12\Gamma_0(\nu-1))}{(\nu - s(\nu-1))^2}, \quad \nu=\frac{\rho}{\rho_0},
\end{equation}
in the case of solids and liquids. Here, $ c_v $ is the specific heat capacity at constant volume, 
$ \gamma $ is the ratio of the specific heats,
$ p_0 $ is the reference (atmospheric) pressure, $ \rho_0 $ is the reference material density, and 
$\Gamma_0$, and $s$ are some material parameters.
The specific energy stored in material deformations and in the thermal impulse is
\begin{equation}\label{E2}
E_2(\alpha,A_{ik},J_k) = \frac{1}{4} \bar{c}_s^2 \mathring{G}_{ij} \mathring{G}_{ij} + \frac{1}{2}
\bar{c}_h^2 J_k
J_k,
\end{equation}
where $ \mathring{G}_{ij} = G_{ij} - \frac{1}{3} G_{kk} \, \delta_{ij} $ is the trace-free part of 
the metric tensor $G_{ij} = A_{ki} A_{kj}$, which is induced by the mapping from
Eulerian coordinates to the current stress-free reference configuration.
The coefficients $ \bar{c}_s(\alpha) $ and $ \bar{c}_h(\alpha) $ in~\eqref{E2} are the
characteristic velocities for
propagation of shear and thermal perturbations accordingly. In the present diffuse interface model,
we choose the following simple linear mixture rule for the computation of the shear sound speed 
and of the heat wave propagation as a function of the volume fraction $\alpha$ 
\begin{equation}
\bar{c}_s(\alpha) = \alpha c_s + (1-\alpha) c_s^g, \qquad \bar{c}_h(\alpha) = \alpha c_h +
(1-\alpha) c_h^g,
\end{equation}
where $c_s$ and $c_h$ are the material parameters inside the continuum and $c_h^g \ll 1$ and $c_s^g \ll 1$ 
are free parameters that can be chosen for the region outside the continuum.  
The specific kinetic energy is contained in the third
contribution to the total energy and reads $ E_3(v_k) = \frac{1}{2} v_i v_i $. 

With the equation of state chosen above, we get the following expressions for the stress tensor, 
the heat flux and the dissipative sources $E_{A_{ik}}$ and $E_{J_k}$ present in the relaxation 
source terms:  
\begin{equation}
\sigma_{ik} = \rho \bar{c}_s^2 G_{ij} \mathring{G}_{jk}, \qquad
q_k = \rho T \bar{c}_h^2 J_k, 
\end{equation} 
\begin{equation}
E_{A_{ik}} = \bar{c}_s^2 A_{ij} \mathring{G}_{jk}, \qquad
E_{J_k} = \bar{c}_h^2 J_k.
\end{equation} 
The functions $\theta_1$ and $\theta_2$ are chosen in such a way that a \textit{constant} viscosity 
and 
heat conduction coefficient are obtained in the stiff relaxation limit, see \cite{GPRmodel} for a  
formal asymptotic analysis, 
\begin{equation}
\theta_1(\tau_1) = \frac{1}{3} \tau_1 \bar{c}_s^2 |\AA|^{\frac{5}{3}}, \qquad \theta_2(\tau_2) =
\tau_2 \bar{c}_h^2\frac{\rho \, T_0}{\rho_0 T}.
\end{equation}
Thus, following the procedure detailed in \cite{GPRmodel}, one can show via formal asymptotic
expansion
that 
in the stiff relaxation limit $\tau_1 \to 0$, $\tau_2 \to 0$, the stress tensor and the heat flux 
reduce to 
\begin{equation}\label{eqn.stiff.limit}
\boldsymbol{\sigma} = -\frac{1}{6} \rho_0 \bar{c}_s^2 \tau_1 \left( \nabla \vv + \nabla
\vv^T
- \frac{2}{3} \left( \nabla \cdot \vv\right) \Id \right)
\end{equation}
and 
\begin{equation} 
\boldsymbol{q} = -  \bar{c}_h^2 \tau_2 \frac{T_0}{\rho_0}\nabla T,
\end{equation} 
that is the effective shear viscosity and effective heat conductivity of model~\eqref{eqn.GPR} are
\begin{equation}
\label{visc.conductivity}
\mu = \frac{1}{6} \rho_0 \tau_1 \bar{c}_h^2, \qquad \kappa = \tau_2 \bar{c}_h^2 \frac{T_0}{\rho_0}
\end{equation}
with $ \rho_0 $ and $ T_0 $ are reference density and temperature, see \cite{GPRmodel}, where also 
an explanation has been provided of how the relaxation times $\tau$ could be obtained experimentally via
ultrasound measurements. 

\medskip
\subsection{Symmetric Godunov form of the model}

It is important to note an interesting structural feature of equations~\eqref{eqn.GPR} that may 
affect future 
developments of the ADER schemes in an attempt to respect such structural properties at the 
discrete level that may help to improve physical consistency of the numerical solution. Thus, as 
many PDE systems studied in some other of our papers 
\cite{GPRmodel,GPRmodelMHD,Rom1998,SHTC-GENERIC-CMAT}, system~\eqref{eqn.GPR} belongs to the class 
of so-called Symmetric Hyperbolic Thermodynamically Compatible (SHTC) PDE systems originally 
studied by Godunov \cite{God1961,God1972MHD} and later by Godunov and Romenski in 
\cite{GodRom1995,Godunov1996,Rom1998,Rom2001}.
Indeed, by simply rescaling the quantities $ \alpharho = \alpha\rho $, $ \alphap = \alpha p = 
\alpharho^2 E_{\alpharho} $, and $ 
\alphasigma_{ik} = \alpha\sigma_{ik} = -\alpharho A_{jk}E_{A_{ji}}$ and replacing the 
non-conservative equation~\eqref{eqn.alpha} by an equivalent (on smooth solutions) conservative 
form~\eqref{eqn.alpha1}, system~\eqref{eqn.GPR} can be written as
\allowdisplaybreaks
\begin{subequations}\label{eqn.GPR1}
	\begin{align}
	& \frac{\partial (\alpha\alpharho)}{\partial t} + \frac{\partial (\alpha\alpharho 
		v_k)}{\partial 
		x_k}=0,  
	\label{eqn.alpha1} \\ 
	&\frac{\partial \alpharho}{\partial t}  + \frac{\partial (\alpharho v_k )}{\partial x_k} 
	=0, 
	\label{eqn.mass1} \\
	&\frac{\partial (\alpharho v_i)}{\partial t}  +\frac{\partial \left(  \alpharho v_i v_k + 
		\alphap \delta_{ik} - \alphasigma_{ik}  \right) }{\partial x_k}   =0, 
	\label{eqn.momentum1} \\ 
	&\frac{\partial A_{ik}}{\partial t} +\frac{\partial (A_{ij}v_j)}{\partial x_k} + 
	v_j \left( \frac{\partial A_{ik}}{\partial x_j} -\frac{\partial A_{ij}}{\partial x_k}  \right) 
	= 
	-\frac{1}{\theta_1} E_{A_{ik}},
	\label{eqn.A1} \\  
	& \frac{\partial (\alpharho J_i)}{\partial t} + \frac{\partial \left( \alpharho J_i v_k+ 
		E_S \delta_{ik}\right) }{\partial x_k}  
	= - \frac{1}{\theta_2} E_{J_i},
	\label{eqn.J1} \\
	& \frac{\partial (\alpharho S) }{\partial t} + \frac{\partial \left( \alpharho S v_k + 
		E_{J_k} \right)}{\partial x_k} = 
	\frac{\alpharho}{\alpha T} \left( \frac{1}{\theta_1} E_{A_{ik}} E_{A_{ik}} + 
	\frac{1}{\theta_2} 
	E_{J_k} 
	E_{J_k}   \right) \geq 0,  
	\label{eqn.entropy1} 
	\end{align}
\end{subequations} 
where we have omitted the energy equation. Now, this system looks exactly as the system studied in 
\cite{GPRmodel}, apart from the additional equation~\eqref{eqn.alpha1} which has the same structure as 
\eqref{eqn.mass1} and does not change the essence. Then, after denoting $ \calE = \alpharho E $ and
introducing new variables $ \PP =
(\varrho_1,\varrho_2,v_i,\alpha_{ik},\Theta_{i},\sigma) $
\begin{equation}\label{new.var}
\varrho_1 = \calE_{\alpha\alpharho}, 
\quad
\varrho_2 = \calE_{\alpharho},
\quad
v_i = \calE_{\alpharho v_i},
\quad
\alpha_{ik} = \calE_{A_{ik}},
\quad
\Theta_{i} = \calE_{\alpharho J_i},
\quad
T = \calE_{\alpharho S},
\end{equation}
which are thermodynamically conjugate to the 
conservative variables $ \QQ = (\alpha\alpharho,\alpharho,\alpharho v_i, A_{ik},\alpharho 
J_i,\alpharho S) $, and a new thermodynamic potential $ L(\PP) = \QQ \cdot
\calE_{\mathsmaller{\QQ}} -
\calE = \QQ\cdot\PP-\calE$, system~\eqref{eqn.GPR1} can be written in a symmetric form
\allowdisplaybreaks
\begin{subequations}\label{eqn.GPR.sym}
	\begin{align}
	&\frac{\pd L_{\varrho_i}}{\pd t}  + \frac{\pd (v_k L)_{\varrho_i}}{\pd x_k} 
	=0, \quad i=1,2,
	\\
	&\frac{\pd L_{v_i}}{\pd t}  + \frac{\pd (v_k L)_{v_i}}{\pd x_k}  + 
	L_{\alpha_{ij}}\frac{\pd\alpha_{kj}}{\pd x_k} - L_{\alpha_{jk}}\frac{\pd \alpha_{jk}}{\pd x_i} 
	= \rho g_i, 
	\\ 
	&\frac{\pd L_{\alpha_{il}}}{\pd t} + \frac{\pd (v_k L)_{\alpha_{il}}}{\pd x_k} + 
	L_{\alpha_{jl}}\frac{\pd v_j}{\pd x_i} - L_{\alpha_{il}}\frac{\pd v_k}{\pd x_k}
	= -\frac{1}{\theta_1} \alpha_{il},
	\\  
	& \frac{\pd L_{\Theta_i}}{\pd t} + \frac{\pd ( v_k L)_{\Theta_i} }{\pd x_k}  + \frac{\pd 
		T}{\pd x_i} = - \frac{1}{\theta_2} \Theta_i,
	\\
	& \frac{\pd L_T
	}{\pd t} + \frac{\pd ( v_k L)_T}{\pd x_k}  + \frac{\pd \Theta_k}{\pd x_k} = 
	\frac{\varrho_2^2}{\varrho_1 T} \left( \frac{1}{\theta_1} \alpha_{ik} \alpha_{ik} + 
	\frac{1}{\theta_2} \Theta_k\Theta_k   \right) \geq 0.
	\end{align}
\end{subequations} 
In this PDE system, the first two terms in each equation form the canonical Godunov form introduced 
in \cite{God1961} which can be immediately written as a quasilinear symmetric form, e.g. see 
\cite{SHTC-GENERIC-CMAT,Rom1998,Rom2001}. The other (non-conservative) terms obviously form a symmetric matrix.
Therefore, the entire system~\eqref{eqn.GPR.sym} can be written in a symmetric quasi-linear form 
and hence, it is a symmetric hyperbolic system if the thermodynamic potential $ L $ is convex.

We note that the understanding of the structural properties of the continuous equations might be 
beneficial for developing of so-called structure-preserving numerical integrators (e.g. symplectic 
integrators). Thus, the energy conservation law~\eqref{eqn.energy} is in fact a consequence of the 
other equations~\eqref{eqn.GPR} or~\eqref{eqn.GPR.sym}, e.g. see \cite{GPRmodel,SHTC-GENERIC-CMAT}, 
and can be viewed as a 
constraint of the system~\eqref{eqn.GPR.sym}. Its non-violation at the discrete level cannot be 
guaranteed by the general purpose ADER family of schemes studied in this paper and hence, usually, 
as well 
as in our implementation, it is included into the set of discretized PDEs instead of the entropy 
equation. In principle, a structure-preserving scheme which satisfies all SHTC properties 
\cite{SHTC-GENERIC-CMAT} of the 
continuous equations at the discrete level should guarantee the automatic satisfaction of the energy 
conservation law, without its explicit discretization. We hope to cover this topic in future 
work.

\section{Numerical results}
\label{sec.results} 

In this section, we present some numerical results in order to illustrate the capabilities and 
potential applicability 
of the proposed numerical approach in nonlinear continuum mechanics. The first three test problems are carried out without
making explicit use of the diffuse interface approach, i.e. setting $\alpha=1$ everywhere in the entire computational domain. The last three test problems illustrate the full potential of the diffuse interface extension of the GPR model 
in the context of moving free boundary problems. Gravity effects are neglected in all test cases, apart from the dambreak
problem shown in Subsection~\ref{sec.dambreak}. Whenever values for $\nu = \mu / \rho_0$ and $c_s$ are provided, the 
corresponding relaxation time $\tau_1$ is computed according to \eqref{visc.conductivity}. 

\medskip
\subsection{Numerical convergence studies in the stiff relaxation limit}  
\label{sec.conv}

In order to verify the high order property of our ADER schemes in both space and time in the stiff relaxation limit, we 
first represent the numerical convergence study that was already carried out in \cite{GPRmodel} 
on a smooth unsteady flow, for which an exact analytical solution is known for the 
compressible Euler equations, i.e. in the stiff relaxation limit 
$\tau_1 \to 0$ and $\tau_2 \to 0$ of the GPR model. The problem setup is the one of the classical isentropic vortex, 
see \cite{HuShuTri}. The initial condition consists in a stationary isentropic vortex, whose 
exact solution can easily be found by solving the compressible Euler equations in cylindrical 
coordinates. Due to the Galilean invariance of the Euler equations and of the GPR model, one can 
then simply superimpose a constant velocity field to this stationary vortex solution in order 
to get an unsteady version of the test problem. The vortex strength is chosen as $\epsilon=5$ 
and the perturbation of entropy $S=\frac{p}{\rho^\gamma}$ is assumed to be zero. For details of the setup, see \cite{HuShuTri,GPRmodel}.  
In this test we set the distortion field initially to $ \AA = \sqrt[3]{\rho} \, \Id$, while the 
heat 
flux vector 
is initialized with $\JJ=0$. As computational domain we choose $\Omega=[0;10]\times[0;10]$ with 
periodic boundary conditions. The reference solution for the GPR model in the stiff relaxation 
limit is given by the exact solution of 
the compressible Euler equatons, which is the time--shifted initial condition $\Q_e(\x,t)=\Q(\x-\vv_c t,0)$, 
where the convective mean velocity is $\vv_c=(1,1)$. We run this benchmark on a mesh sequence until the final time $t=1.0$. 
The physical parameters of the GPR model are chosen as $\gamma = 1.4$, $c_v = 2.5$, $\rho_0=1$, $c_s=0.5$ and $c_h=1$. 
The volume fraction function is set to $\alpha=1$ in the entire computational domain. The resulting 
numerical convergence  rates obtained with ADER-DG schemes using polynomial approximation degrees 
from $N=M=2$ to $N=M=5$ are listed in Table~\ref{tab.conv1}, together with the chosen values for 
the effective viscosity $\mu$ and the effective heat conductivity coefficient $\kappa$. From 
Table~\ref{tab.conv1} one can observe that high order of convergence of the numerical method 
is achieved also in the stiff limit of the governing PDE system. 

\begin{table}[!tp]
	
	\caption{Experimental errors and order of accuracy at time $t=1$ for the density $\rho$ for ADER-DG schemes applied to the GPR model ($c_s=0.5$, $\alpha=1$) in the stiff relaxation limit ($\mu \ll 1, \kappa \ll 1$). The reported errors are floating point numbers that have been obtained for numerical simulations carried out in double precision arithmetics.}
	\begin{center} 
			\renewcommand{\arraystretch}{1.2}
			\begin{tabular}{ccccccc} 
				\hline
				$N_x$ & $\epsilon({L_1})$ & $\epsilon({L_2})$ & $\epsilon({L_\infty})$ & $\mathcal{O}(L_1)$ & $\mathcal{O}(L_2)$ & $\mathcal{O}(L_\infty)$ \\ 
				\hline
				\multicolumn{7}{c}{ADER-DG $P_2P_2$ ($\mu=\kappa=10^{-6}$)}  \\
				\hline
				20	& 9.4367E-03	& 2.2020E-03	& 2.1633E-03	&      &      &		    \\
				40	& 1.9524E-03	& 4.4971E-04	& 4.2688E-04	& 2.27 & 2.29	& 2.34  \\
				60	& 7.5180E-04	& 1.7366E-04	& 1.4796E-04	& 2.35 & 2.35	& 2.61  \\
				80	& 3.7171E-04	& 8.6643E-05	& 7.3988E-05	& 2.45 & 2.42	& 2.41  \\
				\hline
				\multicolumn{7}{c}{ADER-DG $P_3P_3$ ($\mu=\kappa=10^{-6}$)}  \\
				\hline
				10	& 1.7126E-02	& 4.0215E-03	& 3.6125E-03	& 		  &      &        \\
				20	& 6.0405E-04	& 1.7468E-04	& 2.1212E-04	& 4.83	& 4.52 & 	4.09  \\
				30	& 8.3413E-05	& 2.5019E-05	& 2.7576E-05	& 4.88	& 4.79 & 	5.03  \\
				40	& 2.1079E-05	& 6.0168E-06	& 7.6291E-06	& 4.78	& 4.95 & 	4.47  \\
				\hline
				\multicolumn{7}{c}{ADER DG $P_4P_4$ ($\mu = \kappa = 10^{-7}$)}   \\
				\hline
				10	& 1.5539E-03	& 4.5965E-04	& 5.1665E-04	& 		 &        &       \\ 
				20	& 4.3993E-05	& 1.0872E-05	& 1.0222E-05	& 5.14 & 	5.40	& 5.66  \\
				25	& 1.8146E-05	& 4.4276E-06	& 4.1469E-06	& 3.97 & 	4.03	& 4.04  \\
				30	& 8.6060E-06	& 2.1233E-06	& 1.9387E-06	& 4.09 &	4.03	& 4.17  \\
				\hline
				\multicolumn{7}{c}{ADER DG $P_5P_5$ ($\mu = \kappa = 10^{-7}$)}   \\
				\hline
				5	& 1.1638E-02	& 1.1638E-02	& 1.8898E-03	& 		 &       &       \\
				10	& 3.9653E-04	& 9.3717E-05	& 6.5319E-05	& 4.88 & 	6.96 & 	4.85 \\
				15	& 4.4638E-05	& 1.2572E-05	& 1.9056E-05	& 5.39 & 	4.95 & 	3.04 \\
				20	& 9.6136E-06	& 3.0120E-06	& 3.9881E-06	& 5.34 & 	4.97 & 	5.44 \\
				\hline 
			\end{tabular}
	\end{center}
	\label{tab.conv1}
\end{table} 

\medskip
\subsection{Circular explosion problem in a solid}  
\label{sec.ep2d}

\begin{figure}[!p]
	\centering
	\includegraphics[width=0.382\textwidth]{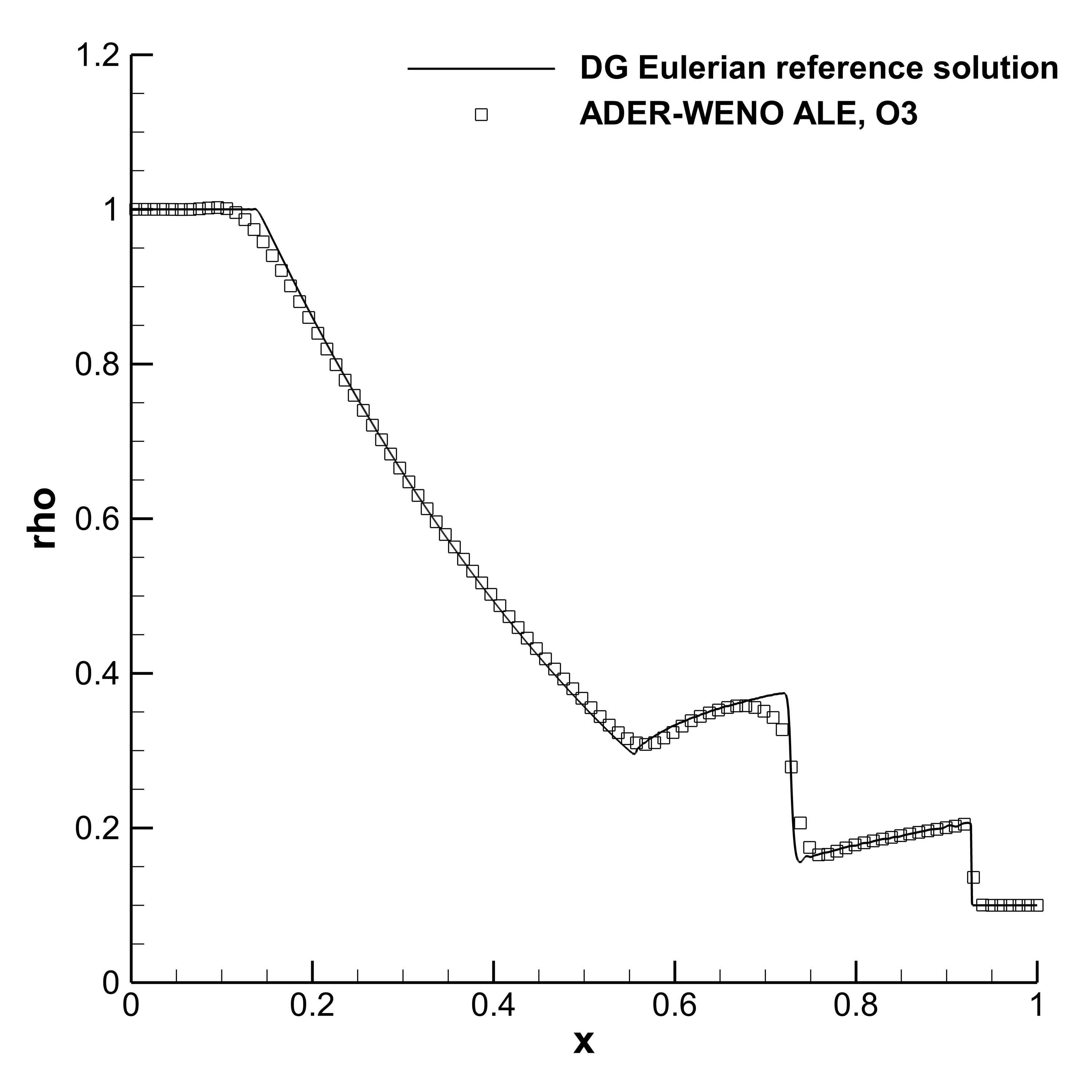}%
	\includegraphics[width=0.382\textwidth]{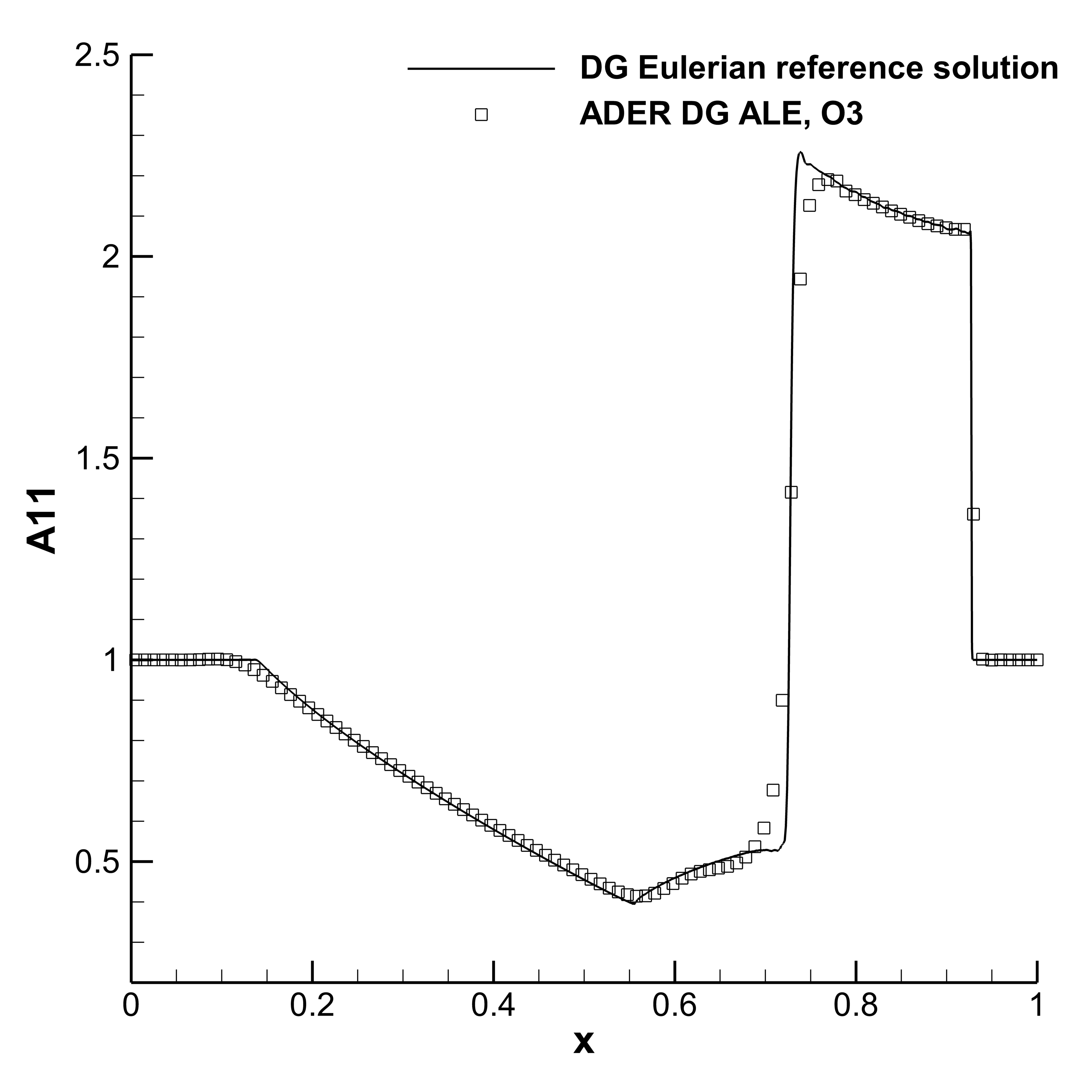}\\
	\includegraphics[width=0.382\textwidth]{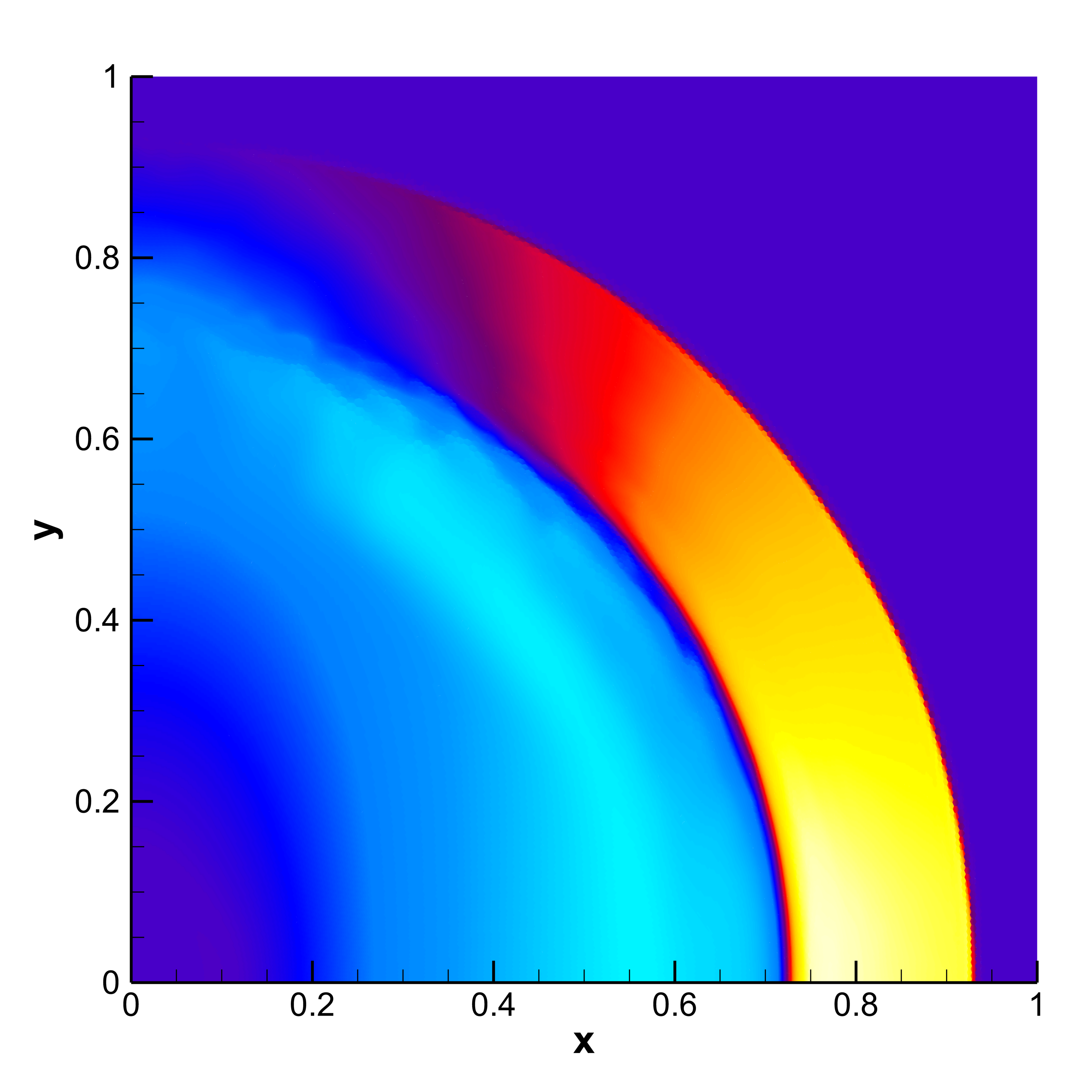}%
	\includegraphics[width=0.382\textwidth]{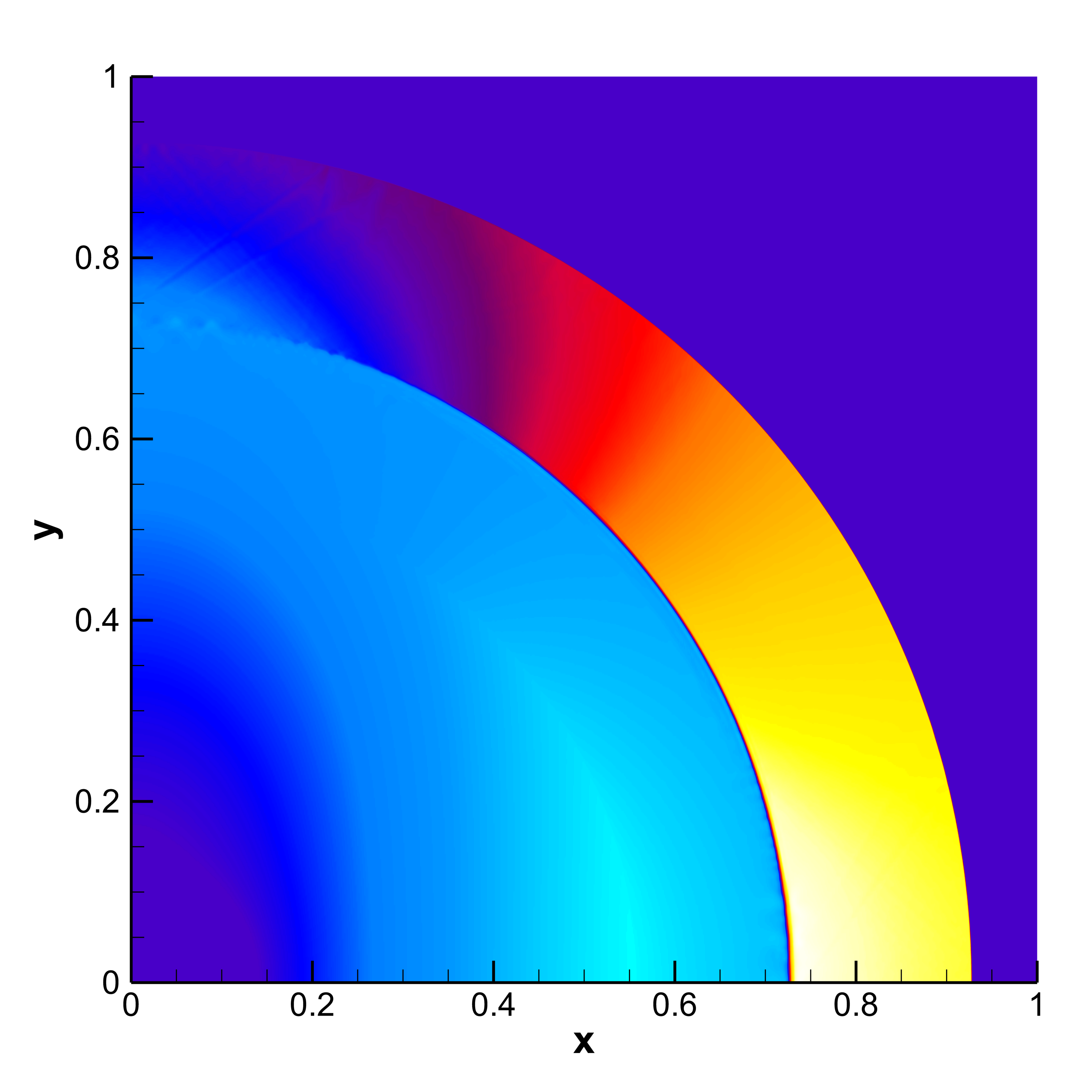}\\
	\includegraphics[width=0.382\textwidth]{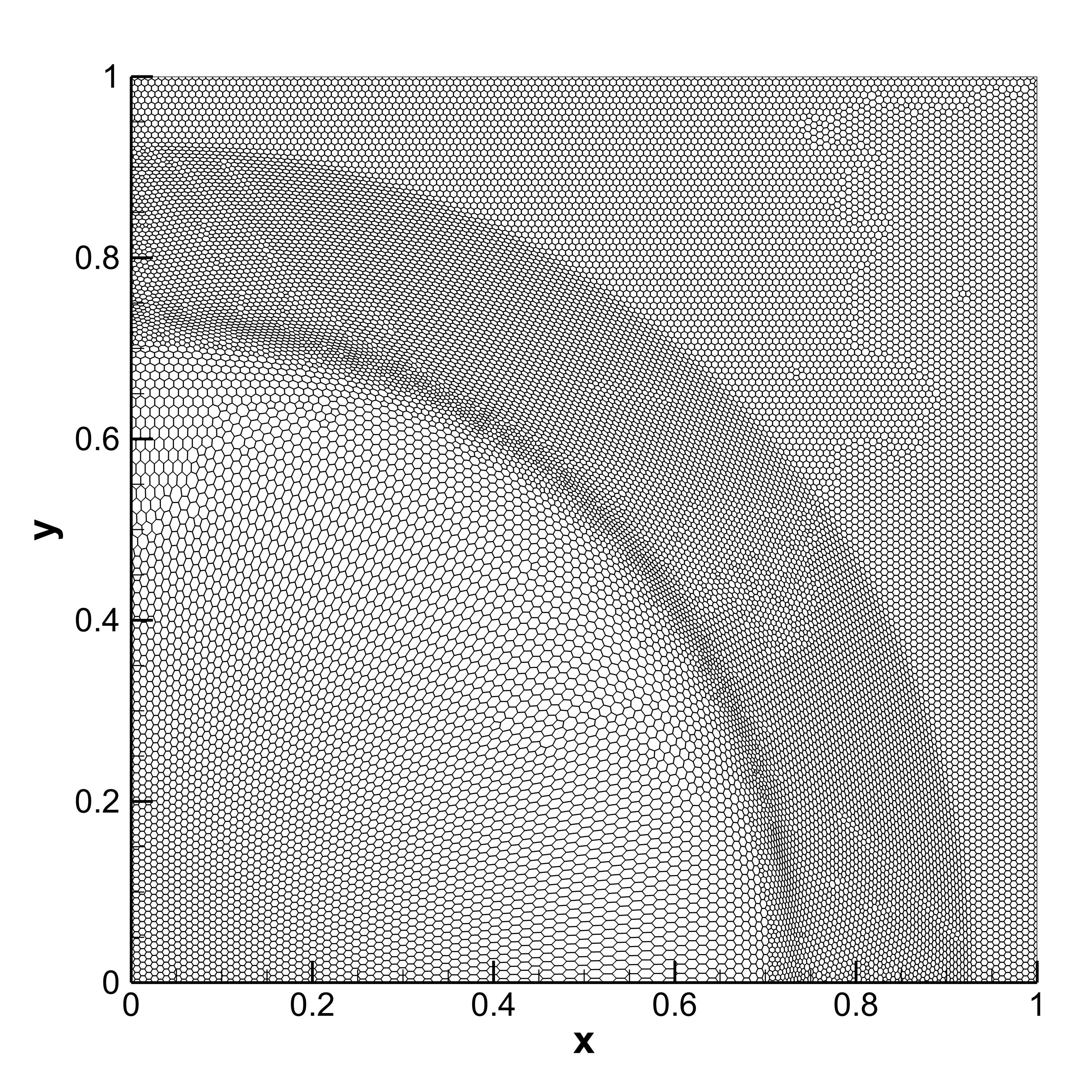}%
	\includegraphics[width=0.382\textwidth]{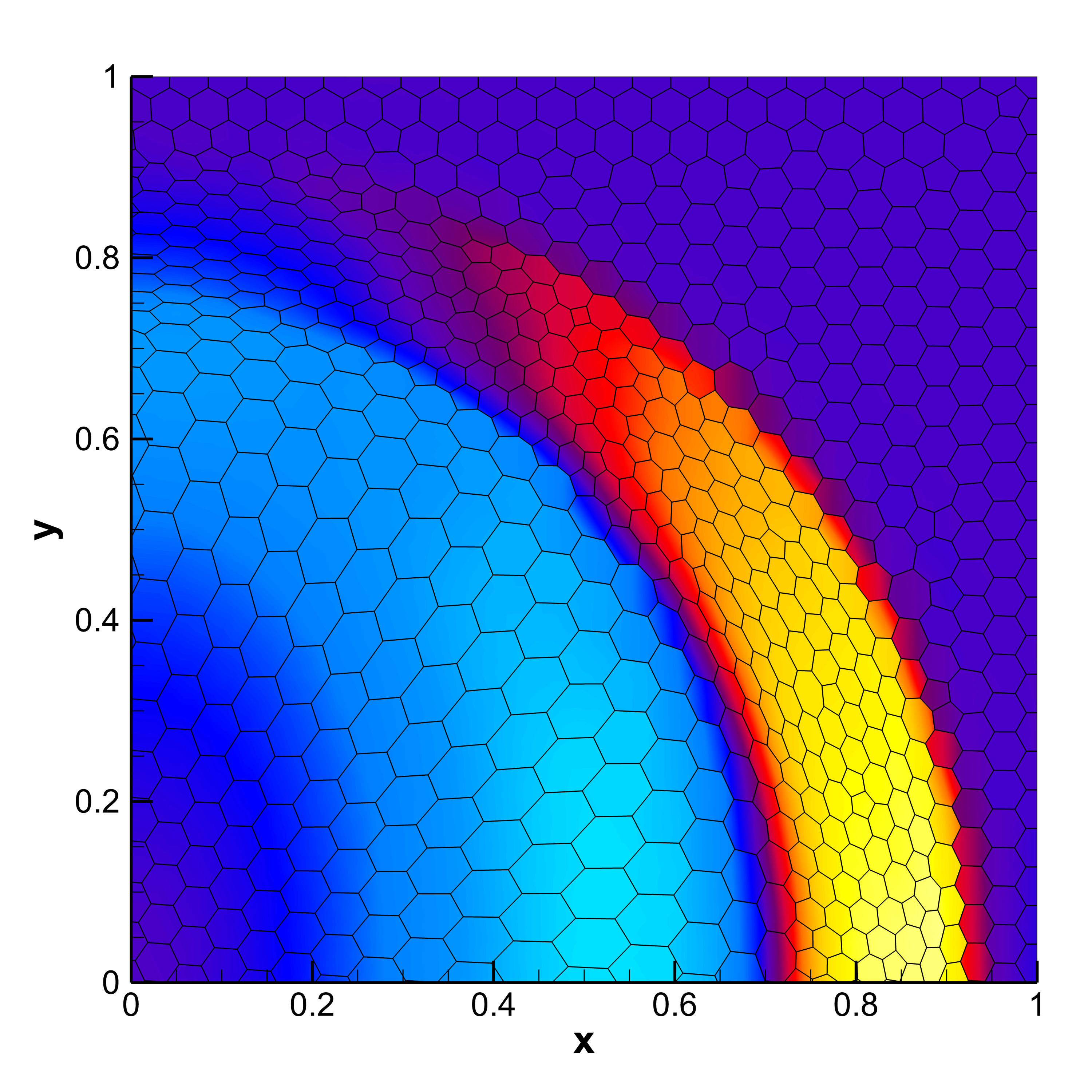}%
	\caption{Simulation results for the explosion problem obtained with a third order ADER-WENO ALE finite volume scheme 
		on a moving Voronoi grid composed of $82\,919$ cells and with
		a fourth order ADER-DG scheme on a Cartesian grid of size $512^2 = 262\,144$ ($4.2\times
		10^6$ DOF). In the top row, two cuts of the solution along the $x$-axis are shown;
		in the middle row, from the left, the solution for $A_{11}$ obtained with the ADER-WENO 
		ALE scheme and with the ADER-DG Eulerian scheme;
		in the bottom row, the Voronoi grid at the final simulation time and the results from the
		ADER-WENO ALE scheme on a coarser grid of $2\,727$ elements.}
	\label{fig.solidexplosion}
\end{figure}

In this Section, we simulate a circular explosion problem in an ideal elastic solid. We compare the results 
obtained with a third order ADER-WENO finite volume scheme on moving 
unstructured Voronoi meshes with possible topology changes, \cite{GBCKSD2019}, with those obtained with a
fourth order ADER discontinuous Galerkin finite element scheme on a very fine uniform Cartesian mesh composed of 
$512 \times 512$ elements, which will be taken as the reference solution for this benchmark. 
The computational domain is $\Omega=[-1,1] \times [-1,1]$ and the final simulation time is $t=0.25$. 
We set $\alpha=1$, $\vv=\mathbf{0}$, $\AA=\Id$ and $\JJ=\mathbf{0}$ in the entire domain.
For $r = \sqrt{x^2+y^2} \leq 0.5$ the initial density and the initial pressure are set to $\rho=1$ and $p=1$, 
while in the rest of the domain we set $\rho=0.1$ and $p=10^{-3}$. The parameters of the GPR model are chosen 
as follows: $c_s=0.2$, $c_h=0$, $\tau_1 \to \infty$ (in order to model an elastic solid). We use the stiffened 
gas equation of state with $\gamma=2$ and $p_0=0$. For the simulation on the moving Voronoi mesh,
we employ
a mesh with $82\,919$ control volumes. The computational results obtained with the unstructured
ADER-WENO ALE scheme and those obtained with the high order Eulerian ADER-DG scheme are presented and compared
with each other in Figure~\ref{fig.solidexplosion}. We can note a very good agreement between the two results.
The high quality of the ADER-WENO finite volume scheme on coarse grids is mainly due to the natural mesh 
refinement around the shock, which is typical for Lagrangian schemes. Furthermore, Lagrangian schemes are well
known to capture material interfaces and contact discontinuities very well, since the mesh is moving with the fluid
and thus numerical dissipation at linear degenerate fields moving with the fluid velocity is significantly lower than
with classical Eulerian schemes.

\medskip
\subsection{Rotor test problem}  
\label{sec.rotor}

A second solid mechanics benchmark consists in the simulation of a plate on which a rotational impulse
is initially impressed, in a circular region centered with respect to the computational domain. This 
\textit{rotor} 
will initially move according to the rotational impulse, while emitting elastic waves 
which ultimately determine the formation of a set of concentric rings with alternating direction of rotation.
The test is analogous to the rotor problem shown in \cite{HypoHyper2}, but with a weakened material 
in order to show stronger motion of the Voronoi grid.

The results of the third order ADER-WENO finite volume method on a moving Voronoi grid with variable
connectivity, composed of $150\,561$ cells, are compared against a reference solution obtained with a fourth order ADER discontinuos Galerkin
scheme on a very fine uniform Cartesian mesh counting $512 \times 512$ elements, for a total of over four million
spatial degrees of freedom. 

\begin{figure}[!bp]
	\centering
	\includegraphics[width=0.5\textwidth]{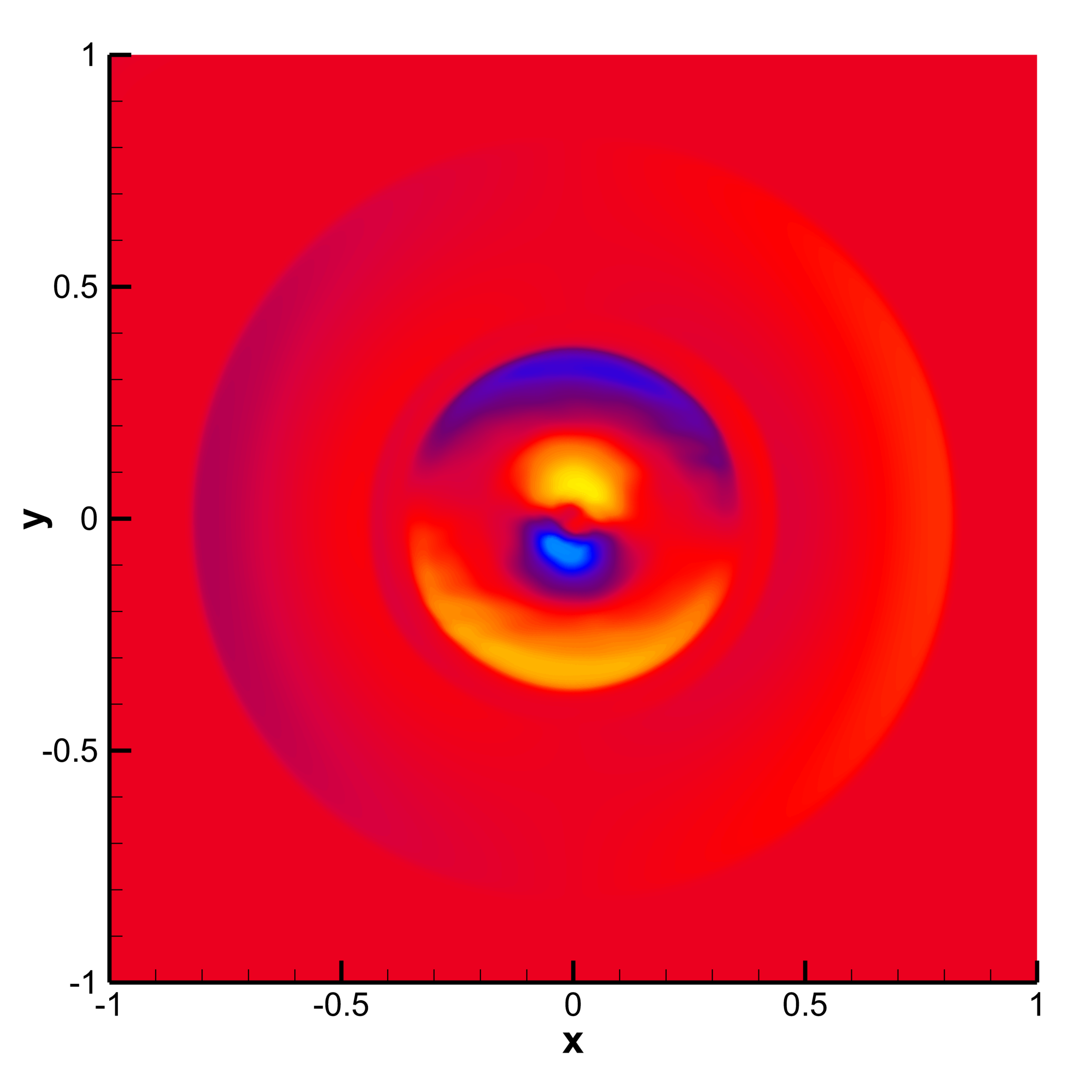}%
	\includegraphics[width=0.5\textwidth]{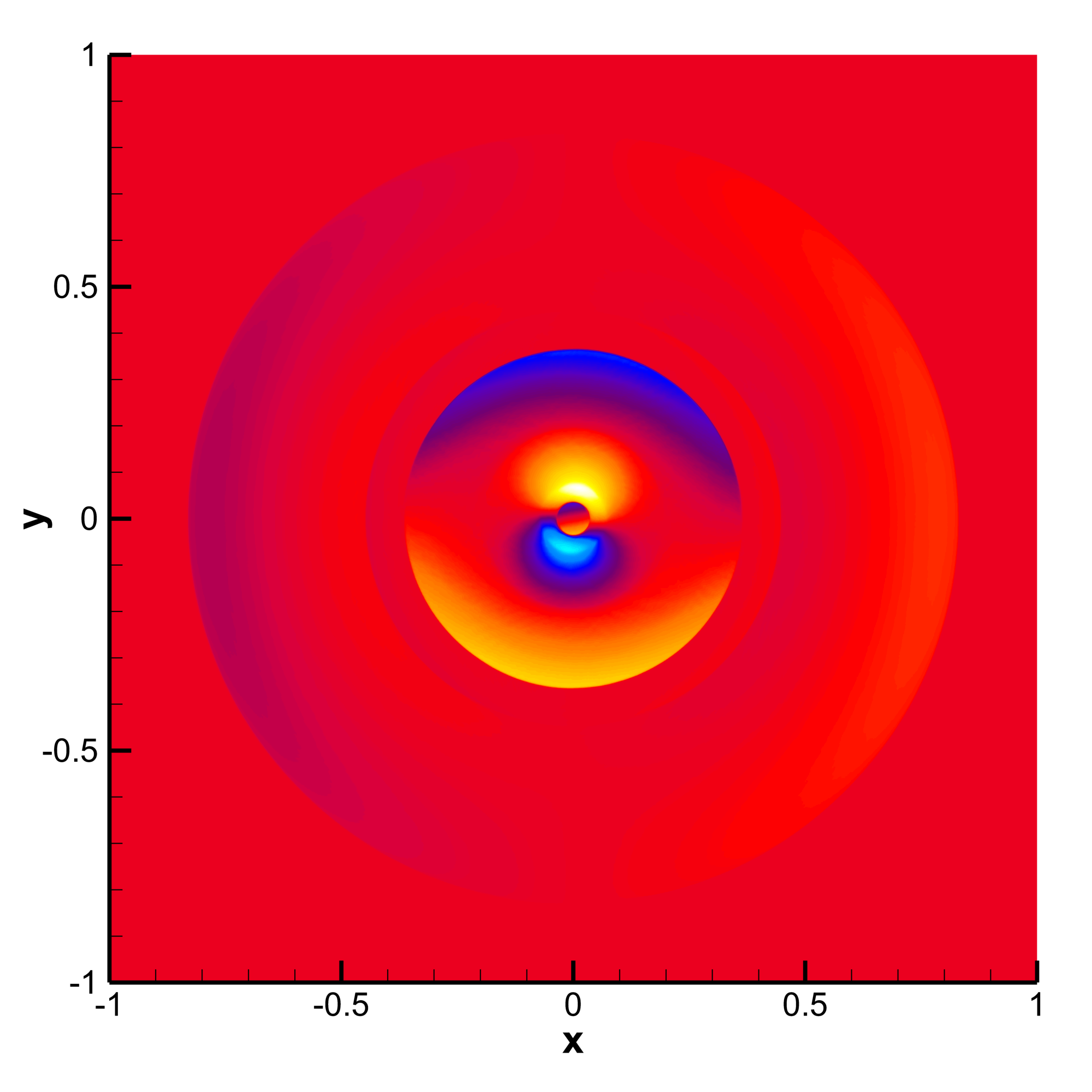}\\
	\includegraphics[width=0.5\textwidth]{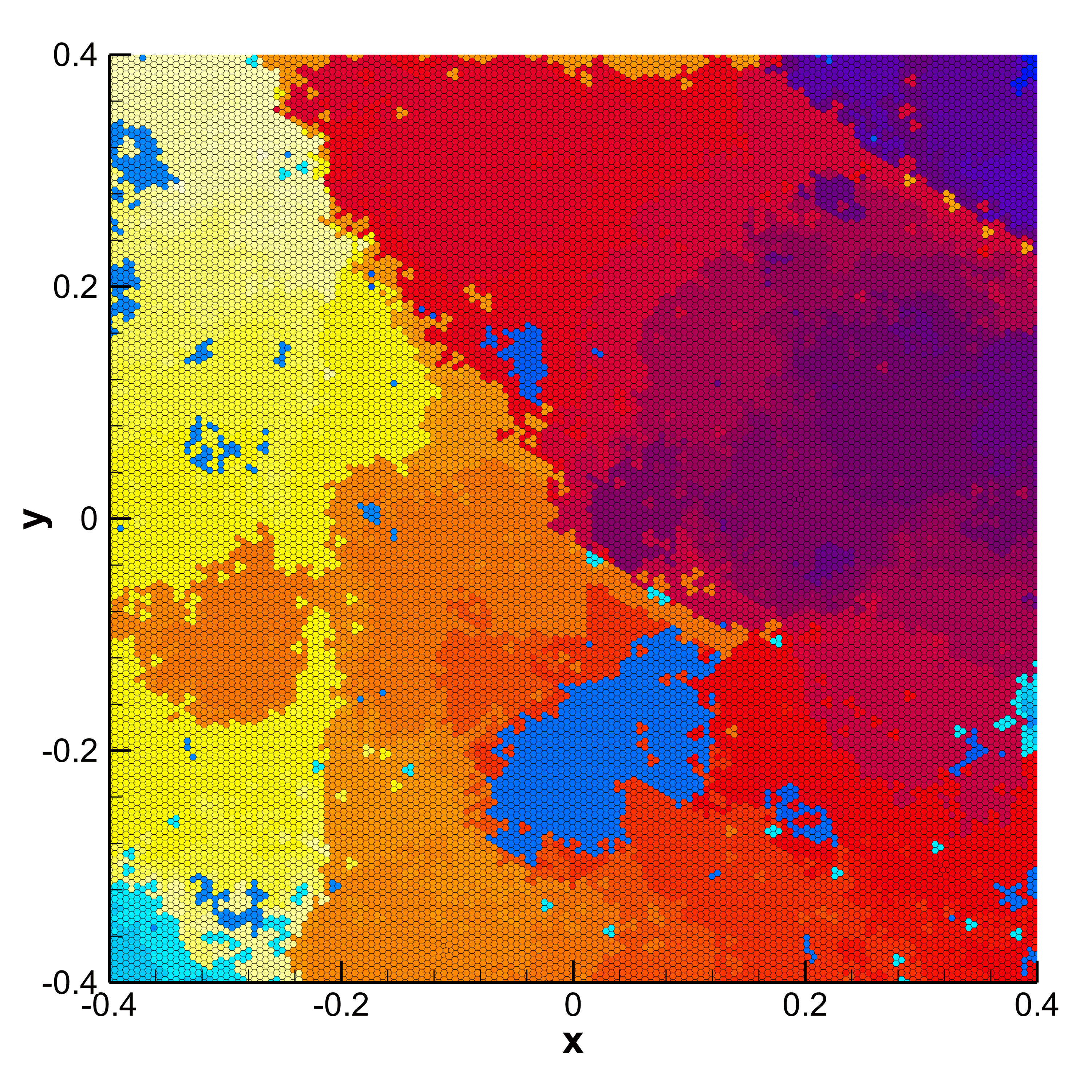}%
	\includegraphics[width=0.5\textwidth]{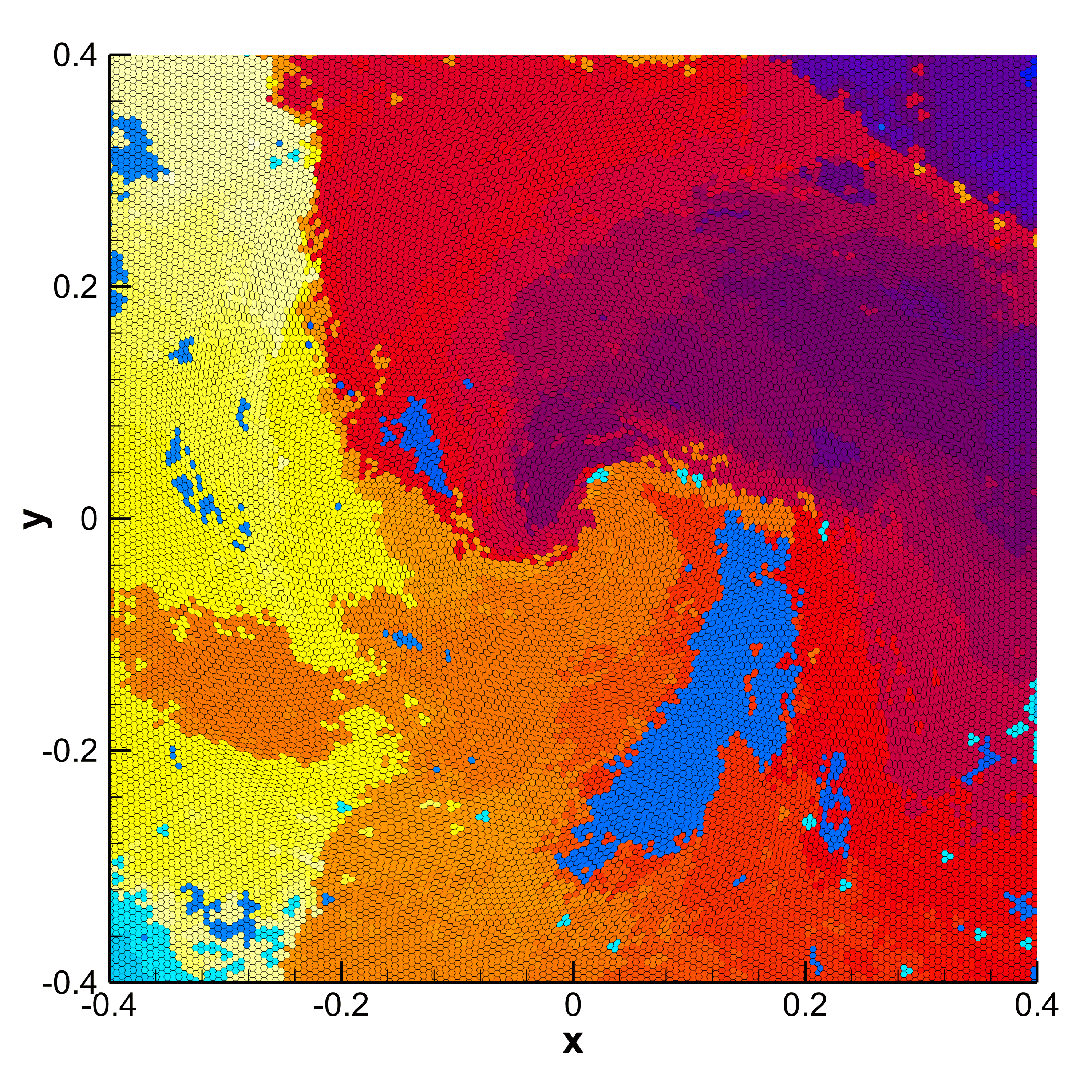}\\
	\caption{Simulation results for the solid rotor problem obtained from 
		a third order ADER-WENO ALE finite volume scheme on a moving Voronoi grid composed of
		$150\,561$ cells and with
		a fourth order ADER-DG scheme on a cartesian grid of size $512^2 = 262\,144$ ($4.2\times
		10^6$ DOF).
		In the top row, the solutions for the $u$ component
		of the velocity field are shown, on the left those obtained with the unstructured ADER-WENO ALE scheme on 
		moving Voronoi meshes and on the right those of the ADER-DG scheme on a fixed Cartesian grid; 
		in the bottom panels the cells are colored according to their mesh numbering to
		show the mesh motion between the beginning of the ALE simulation and the final time.}
	\label{fig.solidrotor}
\end{figure}

The computational domain is the square $\Omega=[-1,1] \times [-1,1]$ and the final simulation time is set to $t = 0.5$.
With exception made for the velocity field, all variables are initially constant throughout the domain. Specifically
we set $\alpha=1$, $\rho=1$, $p=1$, $\AA = \Id$, $\JJ = \mathbf{0}$,
while the velocity field is $\vv = [-y/R, x/R, \ 0]$ 
if $r = \sqrt{x^2+y^2} \leq R$, and $\vv=\mathbf{0}$ otherwise, that is, outside of the circle of
radius $R = 0.2$; 
this way, the initial tangential velocity at $r=R$ is one.
The solid is taken to be elastic ($\tau_1 \to \infty$), heat wave propagation is neglected ($c_h = 0$), and the
characteristic speed of shear waves is $c_s = 0.25$. 
The constitutive law is chosen to be the stiffened-gas EOS with $\gamma = 1.4$ and $p_0 = 0$.
We can see in Figure~\ref{fig.solidrotor} that, although some of the finer features 
are lost (specifically the small central counterclockwise-rotating ring) due to the 
lower resolution of the finite volume method
on a coarser grid, the shear waves travel outwards with the correct velocity 
and the moving Voronoi finite volume simulation can be
said to be in agreement
with the high resolution discontinuous Galerkin results. Also in Figure~\ref{fig.solidrotor}, it is shown
that the central region of the computational grid has undergone significant motion but thanks to the absence
of constraints on the connectivity between elements, the Voronoi control volumes have not been stretched excessively as
would instead happen for a similar moving unstructured grid, but with fixed connectivity.

\medskip

\subsection{Elastic vibrations of a beryllium plate}  
\label{sec.beryllium}

The first benchmark for our new diffuse interface version of the GPR model consists in the purely 
elastic vibrations of a beryllium plate, subject to an initial velocity distribution, see for example 
\cite{Sambasivan_13,Maire_elasto_13,Burton2015,BoscheriDumbser2016,HypoHyper2} 
for a setup of the same test problem in the framework of Lagrangian and ALE schemes.

Unlike in the Lagrangian simulations, the computational domain considered here is \textit{larger} and is set to 
$\Omega=[-5;5]\times [-2.5;2.5]$. The computational grid consists of $512 \times 256$ uniform Cartesian cells
with a characteristic mesh size of about $h=0.02$. We use a third order ADER-WENO finite volume scheme in the entire domain. The initial geometry of the beryllium bar is now simply defined by setting $\alpha(\mathbf{x},0) = 1-\varepsilon$ inside the subdomain $\Omega_b=[-3,3] \times [-0.5,0.5]$, while the solid volume fraction 
function $\alpha$ is set to $\alpha(\x,0)=\varepsilon$ elsewhere, with $\varepsilon = 5 \cdot 10^{-3}$. The initial velocity field inside 
$\Omega_b$ is imposed according to \cite{Burton2015,BoscheriDumbser2016,HypoHyper2}  as  
\begin{equation}
\vv(\mathbf{x}) = \left(0, A \omega\left\{  
C_1 \left( \sinh(\Omega(x+3))+\sin(\Omega(x+3)) \right) - S_1\left( \cosh(\Omega(x+3))+\cos(\Omega(x+3)) \right)
\right\}, 0 \right),
\end{equation}
with 
$\Omega=0.7883401241$, $\omega=0.2359739922$, $A=0.004336850425$, $S_1=57.64552048$ and 
\mbox{$C_1=56.53585154$}, while we simply set $\vv=\mathbf{0}$ outside $\Omega_b$. For this test case we set
$\epsilon = 5 \cdot 10^{-3}$. The distortion field is initially set to $\AA=\Id$.
The material properties of Beryllium in the Mie-Gr\"uneisen equation of state
are taken as follows: $\rho_0=1.845$, $c_0=1.287$, $c_s=0.905$, $\Gamma = 1.11$, and $s_0= 1.124$. 
We furthermore neglect heat conduction and set $c_h=0$
and $\JJ=\mathbf{0}$.

Unlike in Lagrangian schemes, \textit{no boundary conditions} need to be imposed on the surface of the 
bar. We simply use transmissive boundaries on $\partial \Omega$. 
The entire computational domain is initialized with the reference density for beryllium as $\rho(\x,0)=\rho_0$, 
while the pressure is set to $p(\x,0)=0$. The distortion field is initialized with $ \AA=\Id$. 
According to \cite{Burton2015}, the final time is set to $t_f=53.25$ so that it corresponds approximately to 
two complete flexural periods. The simulations are carried out with a third order ADER-WENO scheme on two 
uniform Cartesian meshes composed of $256 \times 128$ and $512 \times 256$ elements, respectively.  

\begin{figure}[!bp]
	\centering
	\begin{tabular}{cc} 
		\includegraphics[width=0.45\textwidth]{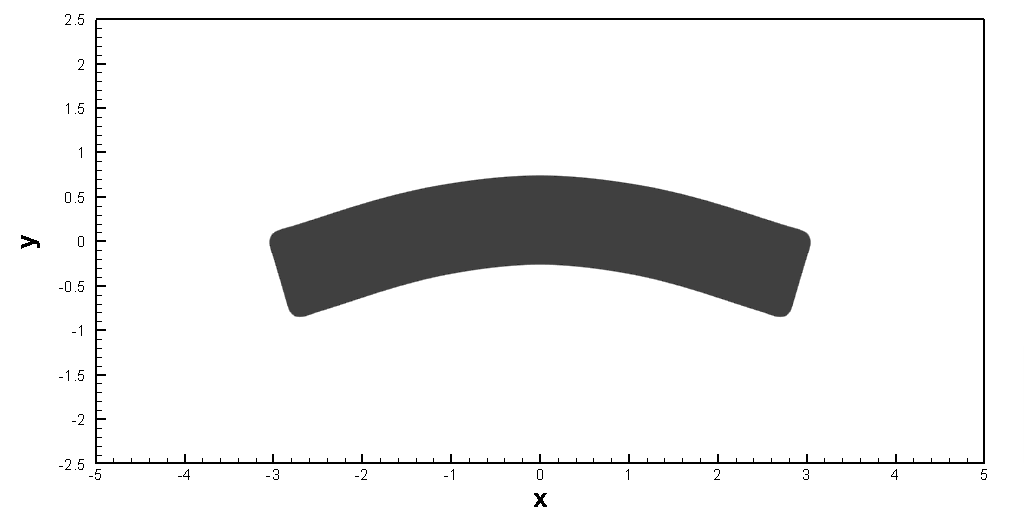}    &  
		\includegraphics[width=0.45\textwidth]{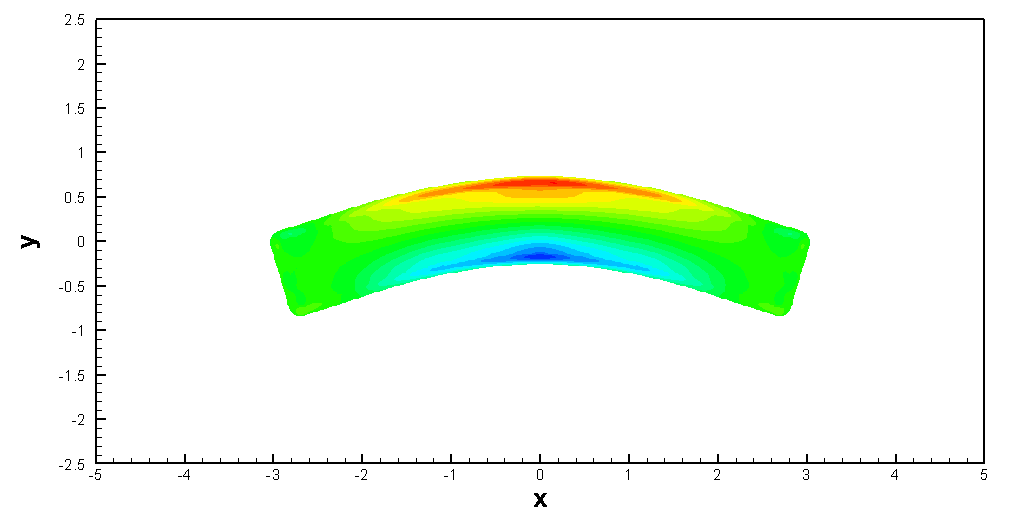}  \\
		\includegraphics[width=0.45\textwidth]{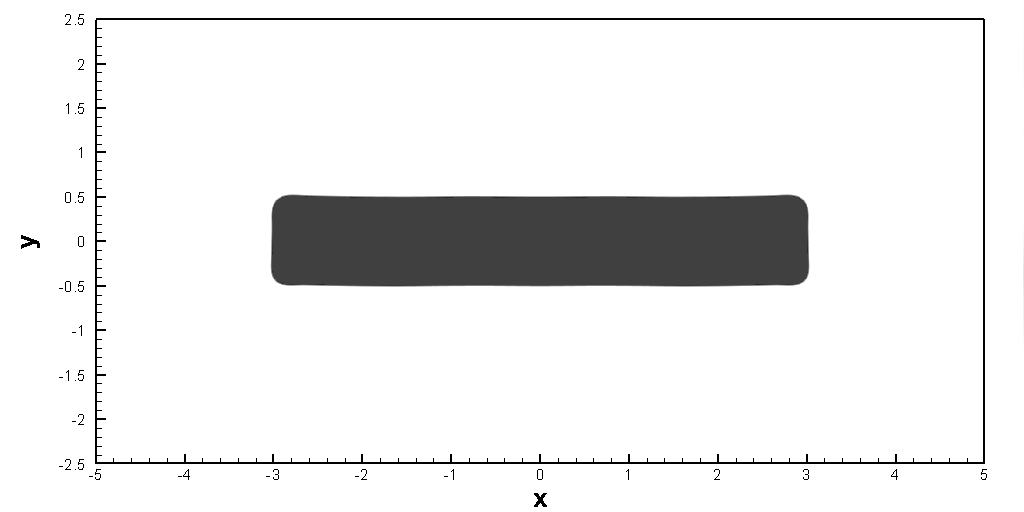}   &  
		\includegraphics[width=0.45\textwidth]{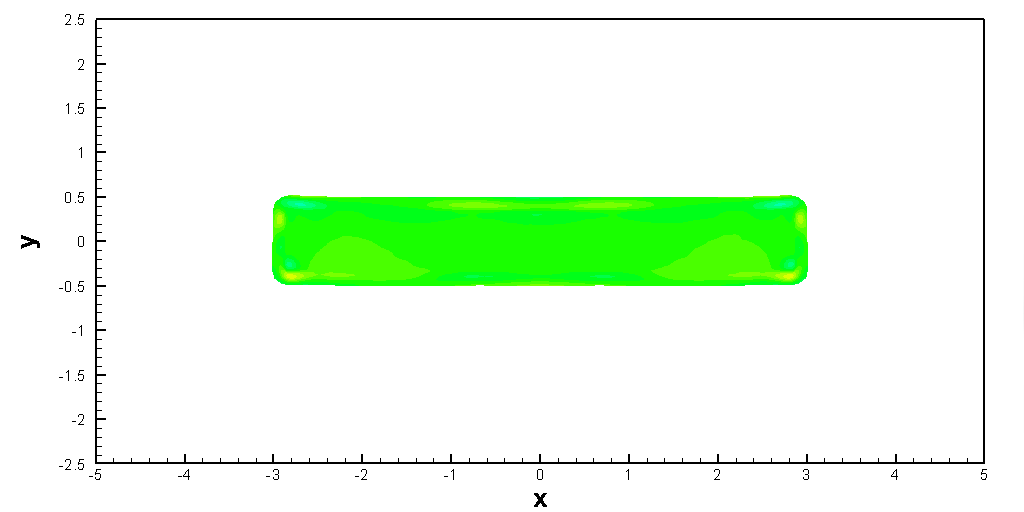} \\
		\includegraphics[width=0.45\textwidth]{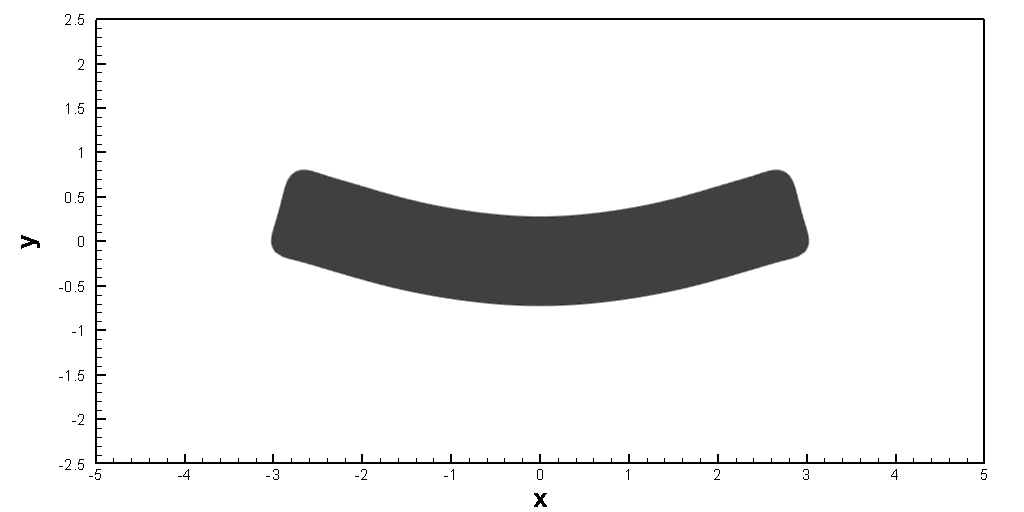}   & 
		\includegraphics[width=0.45\textwidth]{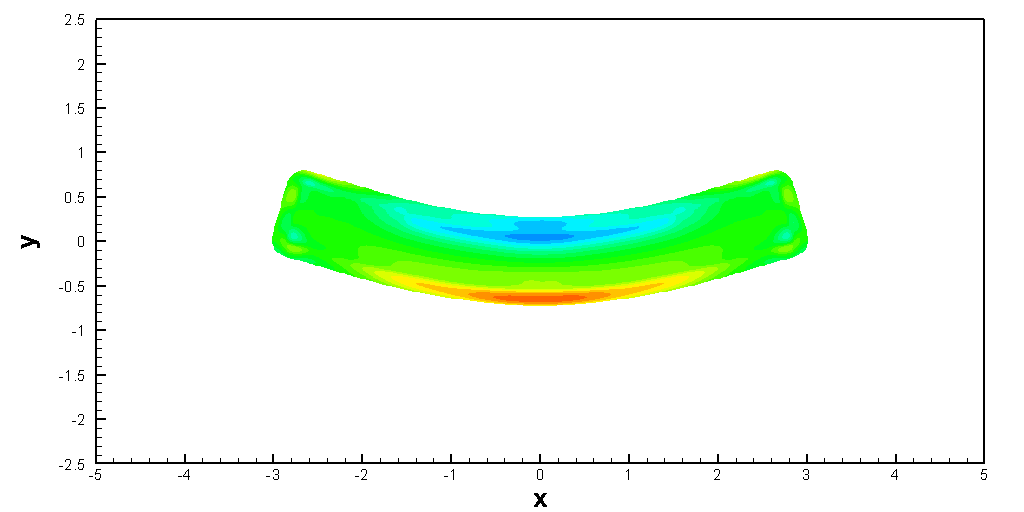} \\
		\includegraphics[width=0.45\textwidth]{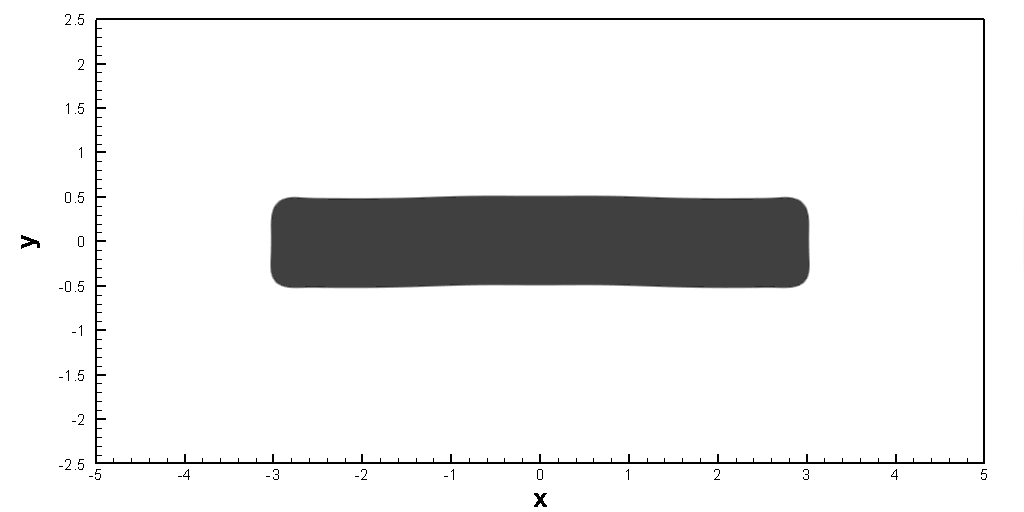}   & 
		\includegraphics[width=0.45\textwidth]{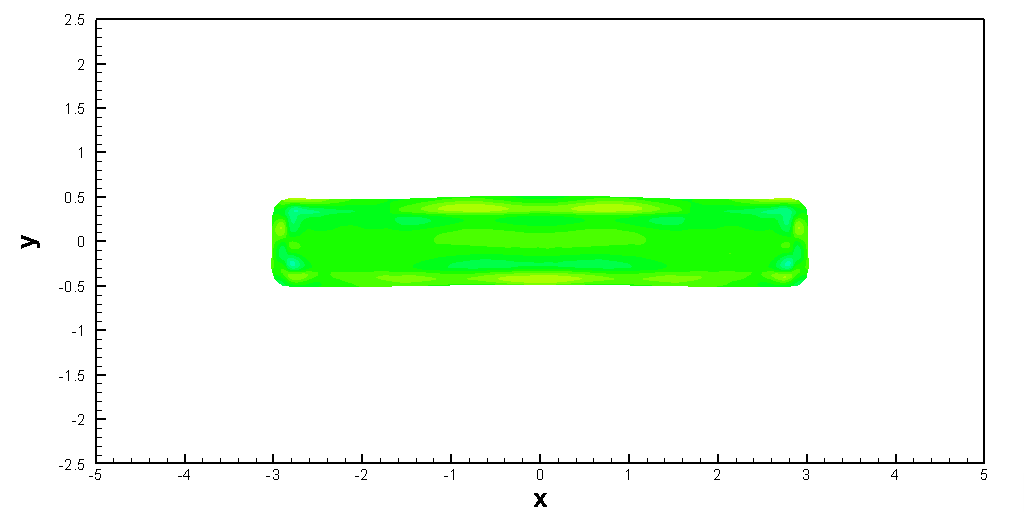}    	
	\end{tabular}    
	\caption{Vibration of an elastic beryllium plate. Temporal evolution of the volume fraction function $\alpha$ (left) and of the pressure field (right) at times 
		$t=5$, $t=14$, $t=23$ and $t=28$, from top to bottom. }
	\label{fig.plate.alphap}
\end{figure}

\begin{figure}[!b]
	\centering
	\includegraphics[width=0.6\textwidth]{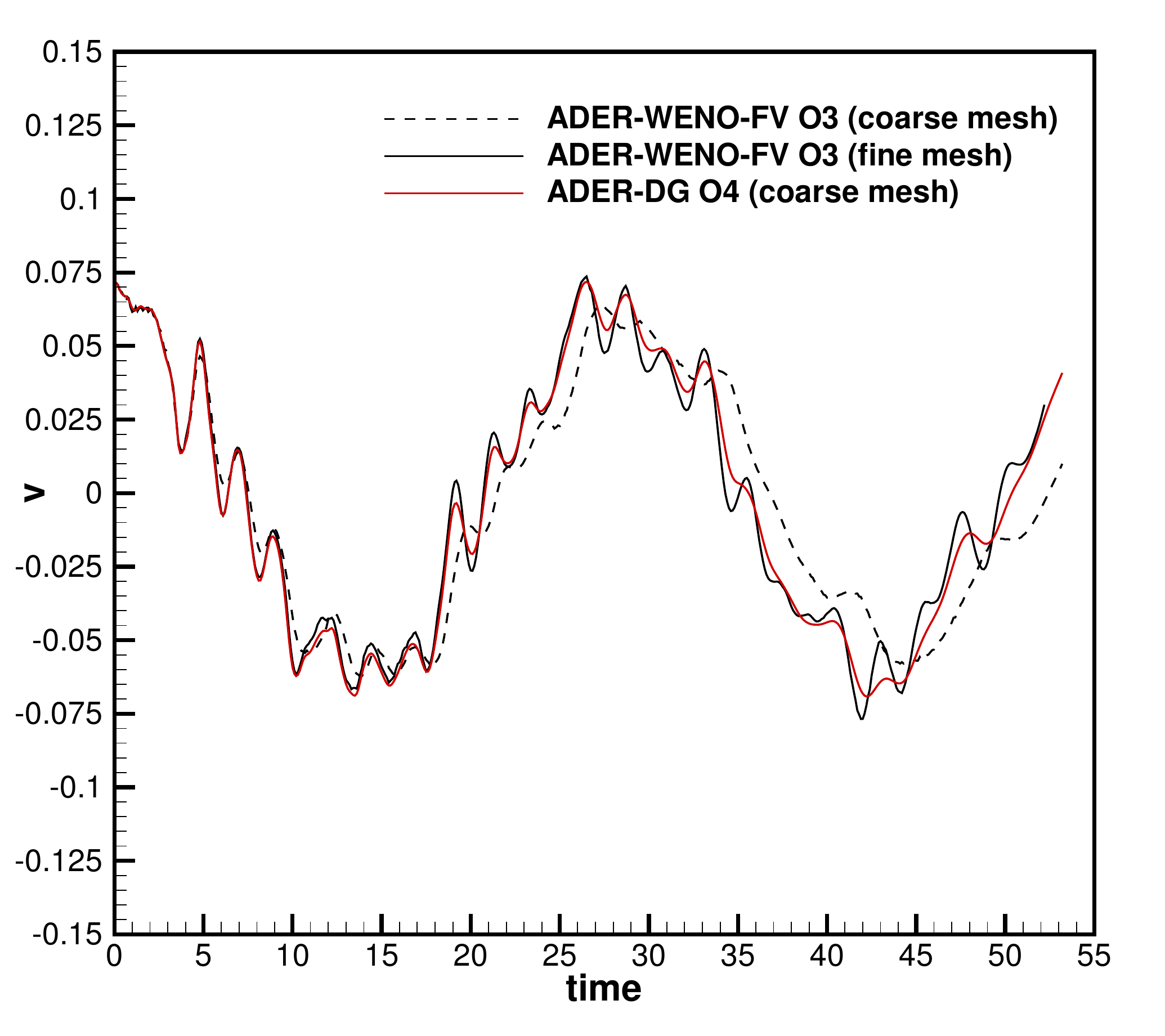}    
	\caption{Temporal evolution of the vertical velocity component $v(0,0,t)$ obtained with a third order ADER-WENO scheme applied to the diffuse interface GPR model using two different mesh resolutions of $256 \times 128$ elements (coarse mesh) and $512 \times 256$ grid cells (fine mesh). For comparison, also a fourth order ADER-DG simulation on the coarse mesh is shown (red line). }
	\label{fig.plate.velocity}
\end{figure}

For the fine grid simulation in Figure~\ref{fig.plate.alphap}, we present the temporal evolution of 
the color contour map 
of the volume fraction function $\alpha$, which represents the moving geometry of the bar. Here, dark gray color 
is used to indicate the regions with $\alpha>0.5$ and white color is used for the regions of $\alpha<0.5$. 
In the same figure, we also depict the pressure field in the region $\alpha>0.5$ at times $t=5$, 
$t=14$, $t=23$ 
and $t=28$. These time instants cover approximately one flexural period. The time evolution 
of the vertical velocity component $v(0,0,t)$ in the origin is depicted in Figure~\ref{fig.plate.velocity}. 
For comparison, in the same figure we also show the results obtained on the coarse mesh for the same test problem 
with a fourth order ADER-DG scheme with second order TVD subcell finite volume limiter (red line).  

Our computational results compare visually well against available reference solutions 
in the literature, see  \cite{Sambasivan_13,Maire_elasto_13,Burton2015,BoscheriDumbser2016,HypoHyper2}, which were all carried out with 
pure Lagrangian or Arbitrary-Lagrangian-Eulerian schemes on moving meshes, while here we use a diffuse interface
approach on a fixed Cartesian grid. 

\medskip
\subsection{Taylor bar impact problem}  
\label{sec.taylor}

So far, we have only considered \textit{ideal elastic} material, i.e. the limit case $\tau_1 \to \infty$. In this section we 
consider also nonlinear \textit{elasto-plastic} material behavior. Following \cite{BartonRomenski2010,Barton2011,LagrangeHPR,HypoHyper2} 
we choose the relaxation time $\tau_1$ as a nonlinear function of an invariant of the stress tensor as follows:  
\begin{equation}
\tau_1 = \tau_0 \left( \frac{\sigma_0}{\sigma} \right)^m, 
\label{eqn.tau.plastic} 
\end{equation}
where $\tau_0$ is a constant characteristic relaxation time, $\sigma_0$ is the yield stress of the material and the von Mises stress $\sigma$ is given by  
\begin{equation} 
\sigma = 	
\sqrt{\frac{1}{2} ((\sigma_{11} - \sigma_{22})^2 + (\sigma_{33} - \sigma_{11})^2 + (\sigma_{33} - 
	\sigma_{22})^2 + 6(\sigma_{12}^2 +\sigma_{31}^2 + \sigma_{32}^2) }  = 
\sqrt{\frac{3}{2}\mathring{\sigma}_{ij}\mathring{\sigma}_{ij}}. 
\label{eqn.vonMises} 
\end{equation} 
In the formula \eqref{eqn.vonMises} above, $\mathring{\sigma}_{ij} = \sigma_{ij} - \frac{1}{3} \sigma_{kk} \delta_{ij}$ is the stress deviator, i.e. the trace-free part of the stress tensor. The nonlinear relaxation time
\eqref{eqn.tau.plastic} tends to zero for $\sigma \gg \sigma_0$, while it tends to infinity for $\sigma \ll \sigma_0$.

The Taylor bar impact problem is a classical benchmark for an elasto-plastic aluminium projectile 
that hits a rigid solid wall, see \cite{Sambasivan_13,Maire_elasto_13,Dobrev2014,LagrangeHPR}. In 
this work the computational domain under consideration is $\Omega = [0,600] \times [-150,+150]$. 
The aluminium bar is initially located in the region $\Omega_b = [0,500] \times [-50,+50]$, where 
we set $\alpha = 1-\varepsilon$, while in the rest of the computational domain we set $\alpha = 
\varepsilon$, with $\varepsilon = 1 \cdot 10^{-2}$. 

The aluminium bar is described by the Mie-Gr\"uneisen equation of state with parameters $\rho_0 = 2.785$, 
$c_0 = 0.533$, $c_s = 0.305$, $\Gamma = 2$ and $s=1.338$. The yield stress of aluminium is set to 
$\sigma_0 = 0.003$. 

\begin{figure}[!b]
	\centering
	\includegraphics[width=0.8\textwidth]{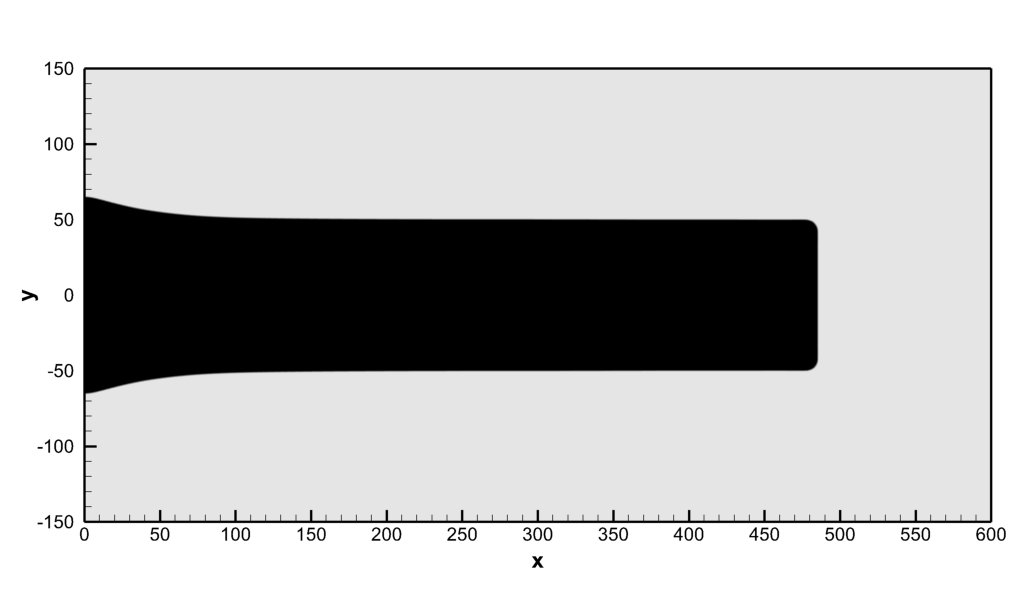}  \\ 
	\includegraphics[width=0.8\textwidth]{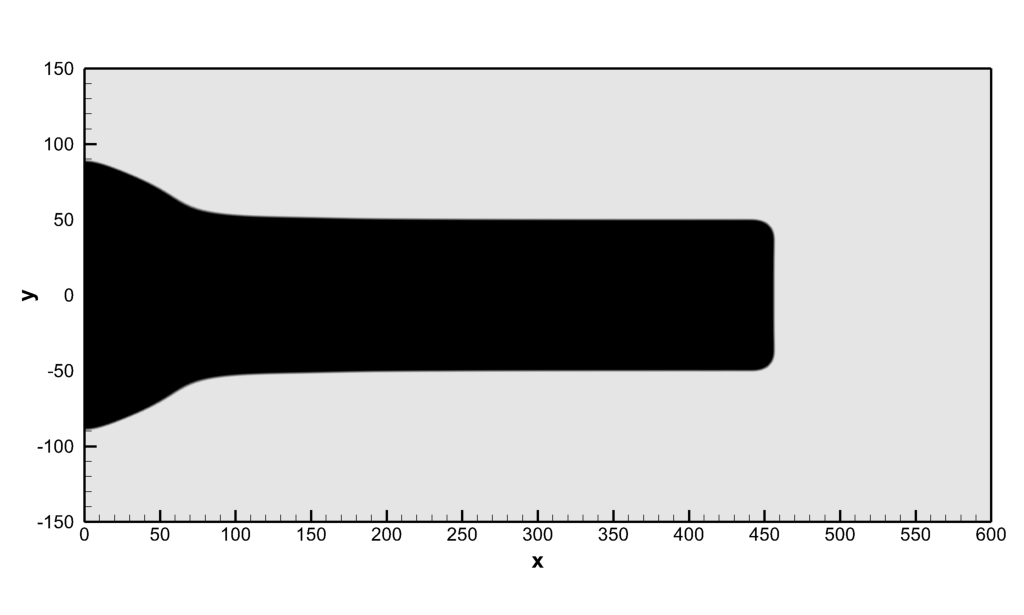}   	
	\caption{Geometry of the Taylor bar at time $t=1000$ (top) and at the final time $t=5000$ (bottom) obtained with a third order ADER-WENO finite volume scheme applied to the diffuse interface GPR model. We plot the contour colors of the volume fraction function $\alpha$, where black regions denote $\alpha>0.5$ and white regions $\alpha<0.5$. }
	\label{fig.Taylor}
\end{figure}

The projectile is initially moving with velocity $\vv=(-0.015,0)$ towards a wall located at $x=0$. 
This velocity field is imposed within the subregion $\Omega_b$, while in the rest of the domain we 
set $\vv=\mathbf{0}$. The remaining initial conditions are chosen as $\rho=\rho_0$, $p=p_0$, 
$\AA=\Id$, 
$\JJ=\mathbf{0}$ and with the parameters $\tau_0=1$ and $m=20$ for the computation of the 
relaxation time 
\eqref{eqn.tau.plastic}. Unlike in Lagrangian schemes, we do not need to set any boundary 
conditions on the free surface of the moving bar. We only apply reflective slip wall boundary 
conditions on the wall in $x=0$. According to \cite{Maire_elasto_13,Dobrev2014,LagrangeHPR} the 
final time of the simulation is $t=5000$. The computational domain is discretized on a regular 
Cartesian grid composed of $512 \times 256$ elements using a third order ADER-WENO finite volume 
scheme. As in \cite{LagrangeHPR} we employ a classical source splitting for the treatment of the 
stiff sources that arise in the regions of plastic deformations, i.e. when $\sigma \gg \sigma_0$. 
In Figure  \ref{fig.Taylor}, we show the computational results at $t=1000$ and at the final time 
$t=5000$. The obtained solution is in agreement with the results presented in 
\cite{Maire_elasto_13,LagrangeHPR,HypoHyper2}. At time $t=5000$, we measure a final 
length of the projectile of $L_f=456$, which fits the results achieved in \cite{Maire_elasto_13,LagrangeHPR} up to $2\%$.

\medskip
\subsection{Dambreak problem}  
\label{sec.dambreak}

\begin{figure}[!bp]
	\centering
	\includegraphics[trim= 0 0 10 0,clip,width=0.8\textwidth]{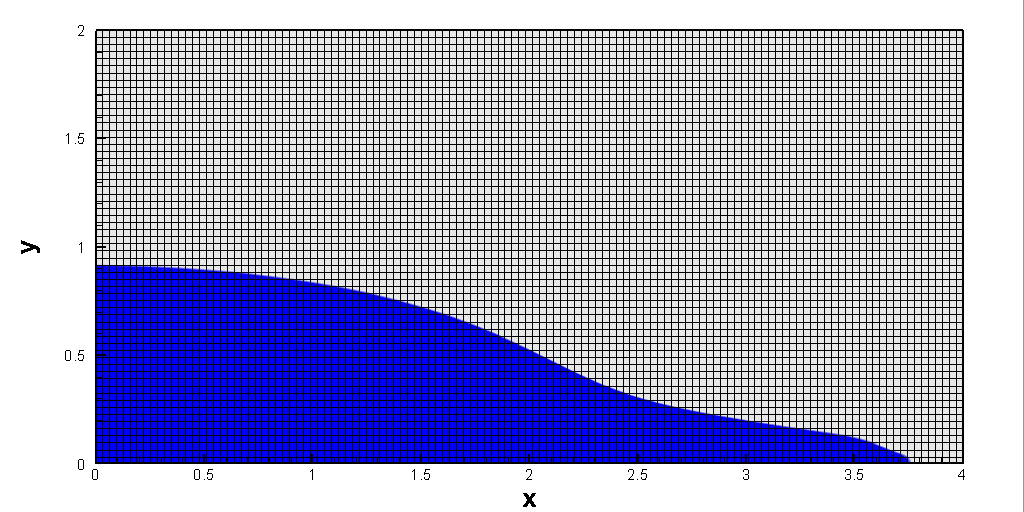}   
	\\   
	\includegraphics[trim= 0 0 10 0,clip,width=0.8\textwidth]{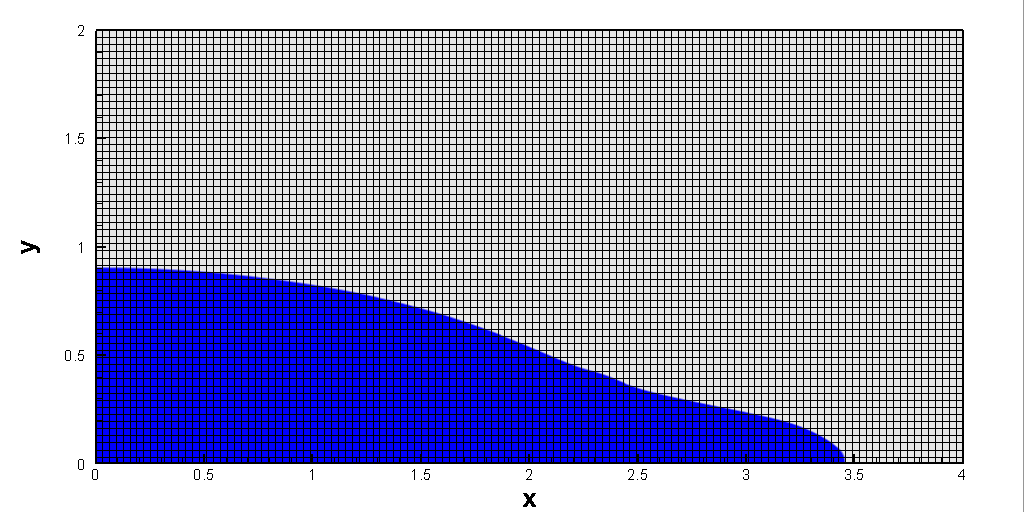}  
	\\
	\includegraphics[trim= 0 0 10 0,clip,width=0.8\textwidth]{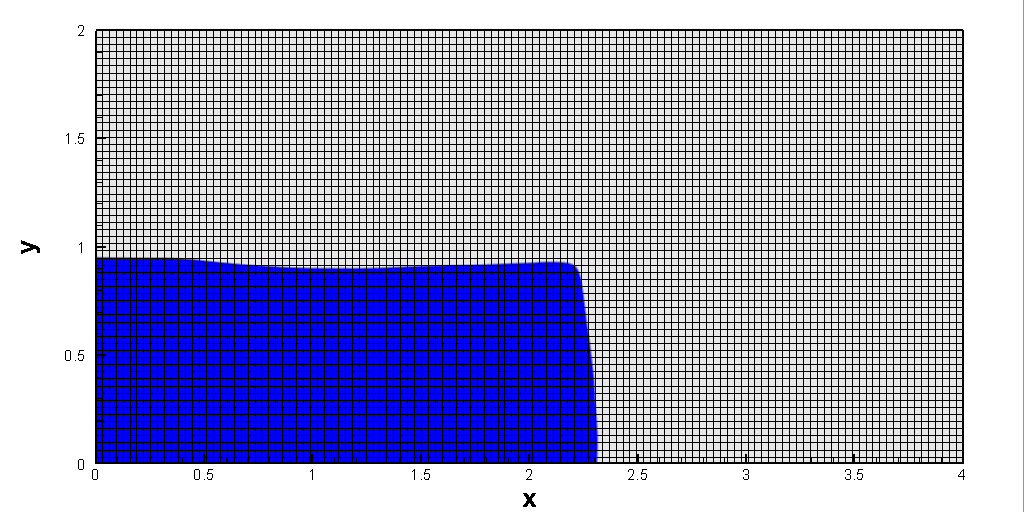}      
	\caption{Dambreak problem at $t=0.5$, simulated with a fourth order ADER-DG scheme using 
		different relaxation times. Top: low viscosity fluid (stiff relaxation limit) with $\nu = 
		10^{-3}$. Center: high viscosity fluid with $\nu = 10^{-1}$. Bottom: ideal elastic solid 
		($\tau_1 \to \infty$) with low shear resistance. }
	\label{fig.dambreak}
\end{figure}

\begin{figure}[!bp]
	\centering
	\includegraphics[trim= 0 0 10 0,clip,width=0.8\textwidth]{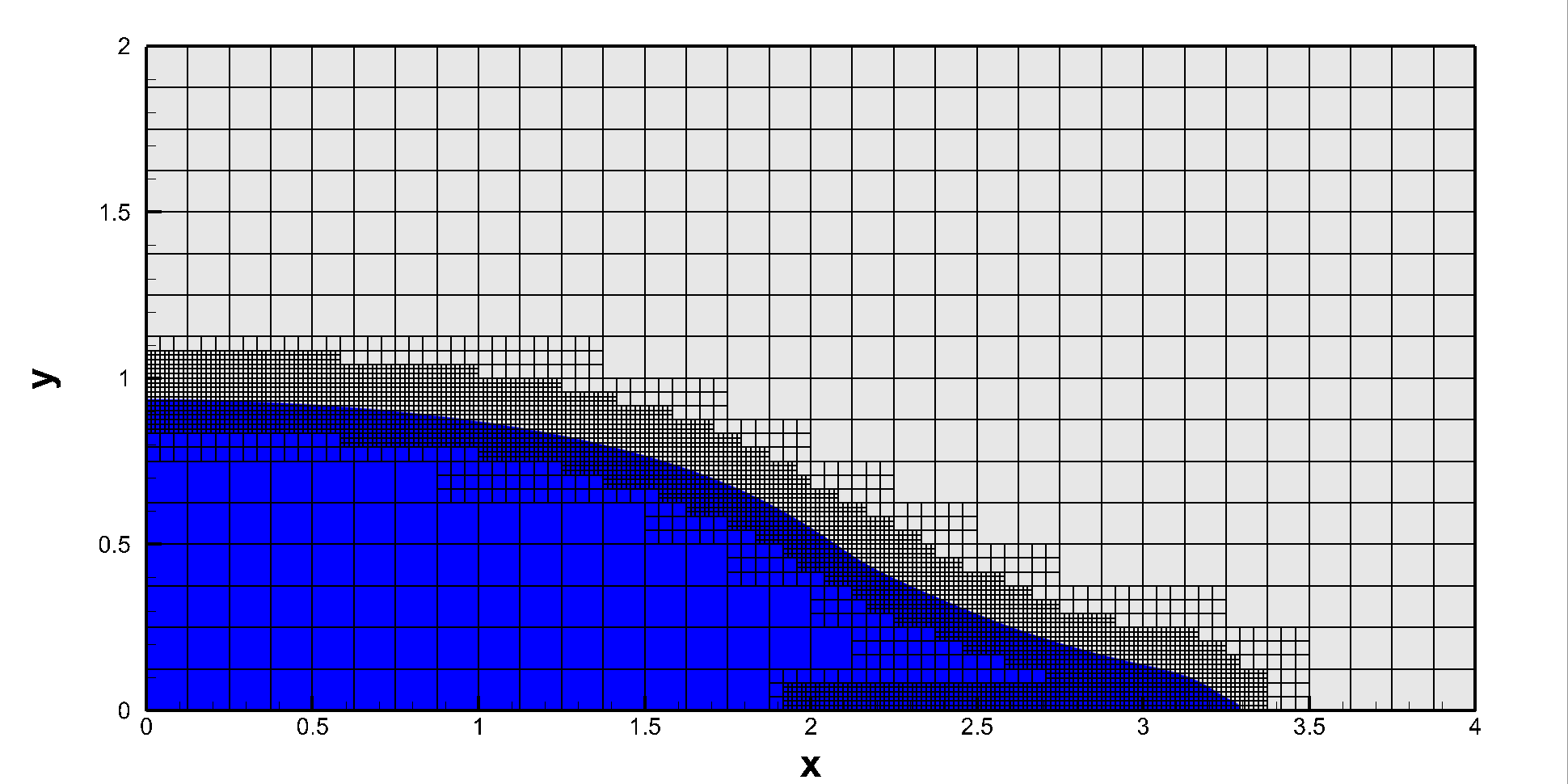}   	\\   
	\includegraphics[trim= 0 0 10 0,clip,width=0.8\textwidth]{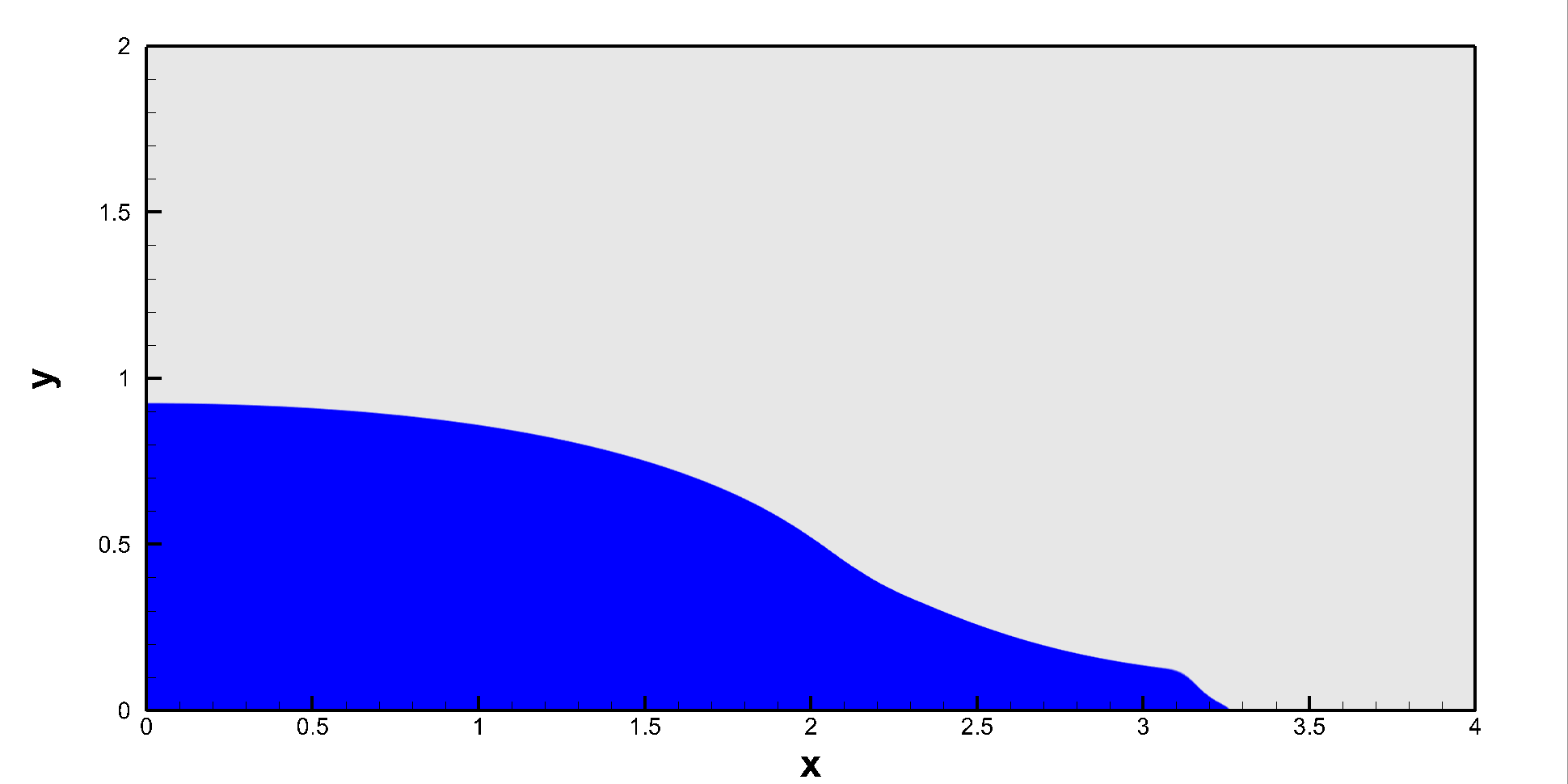}  
	\caption{Dambreak problem at $t=0.4$, simulated with a fourth order ADER-DG scheme 
		using  a space-time adaptive Cartesian AMR mesh applied to the GPR model with with $\nu=10^{-3}$ 
		(top panel), and reference solution, computed with a third order ADER-WENO finite volume scheme on a very fine uniform Cartesian grid, solving the inviscid and barotropic reduced Baer-Nunziato 
		approach presented in \cite{DIM2D,DIMWB} (bottom panel). } 
	\label{fig.dambreak.amr}
\end{figure}

In this last section on numerical test problems, we solve a two-dimensional dambreak problem with 
different relaxation times
in order to show the entire range of potential applications of the GPR model. For this purpose, we also activate the gravity source term, setting the gravity vector to $\mathbf{g}=(0,-g)$ with $g=9.81$. The computational domain is chosen as $\Omega = [0,4] \times [0,2]$ and is discretized with a fourth order ADER discontinuous Galerkin finite element scheme with polynomial 
approximation degree $N=3$ and \textit{a posteriori} subcell TVD finite volume limiter. Computations are run on a uniform 
Cartesian mesh composed of $128 \times 64$ elements until the final time $t=0.5$. 
The initial condition is chosen as follows: we set $\rho = \rho_0$, $\vv=\mathbf{0}$, $\AA=\Id$ and 
$\JJ=\mathbf{0}$ in the entire 
computational domain. We impose the slip boundary condition on the bottom. In the subdomain 
$\Omega_d = [0,2] \times [0,1]$, we set $\alpha = 1 - \varepsilon$, and $p= \rho_0 g (y-1)$, while in 
the rest of the 
domain we set $\alpha=\varepsilon$ and $p=0$. In this test problem we set $\varepsilon = 10^{-2}$ and use a stiffened gas equation of state with parameters $\rho_0 = 1000$, $p_0 = 5 \times 10^4$, $\gamma = 2$, $c_h=0$ and a shear sound speed $c_s=6$. Simulations are run in three different regimes, 
only characterized by a different choice of the strain relaxation time $\tau_1$. In the first 
simulation, we set $\tau_1$ so 
that a kinematic viscosity $\nu = \mu / \rho_0 = 10^{-3}$ is reached in the stiff relaxation limit, i.e. the GPR model in this case describes an almost inviscid fluid. In the second simulation we choose $\tau_1$ so that $\nu = 0.1$, i.e. a high
viscosity Newtonian fluid behavior is reached. In the last simulation we set $\tau_1 \to \infty$, 
i.e. the strain relaxation term is switched off so that an ideal elastic solid with low shear 
resistance is described, similar to a jelly-type medium. In all cases, we apply solid slip wall 
boundary conditions on the left and on the right of the computational domain, while on the right 
and upper boundary, transmissive boundary conditions are set. The temporal evolution of the volume 
fraction function $\alpha$, together with the coarse mesh used in this simulation, are 
depicted in Figure~\ref{fig.dambreak}. 
The results for the almost inviscid fluid agree qualitatively well with those shown in 
\cite{SPH3D,DIM2D,DIMWB} for nonhydrostatic dambreak problems. In order to corroborate this statement quantitatively,
we now repeat the simulation with $\nu=10^{-3}$ using a fourth order ADER-DG scheme on a coarse AMR grid 
composed of only $32 \times 16$ elements on the level zero grid. We then apply two levels of AMR refinement with refinement
factor $\mathfrak{r}=3$, i.e. we employ a general space-tree, rather than a simple quad-tree. We note that the simulations on the AMR grid are run in combination with time-accurate local time stepping (LTS), which is trivial to implement in high order ADER-DG and ADER-FV schemes, due to their fully-discrete one-step nature. For details on LTS, see \cite{LTS,AMR3DCL,ALELTS1D,ALELTS2D}. As a reference solution of this almost inviscid flow problem, we solve the reduced barotropic and inviscid Baer-Nunziato model introduced in \cite{DIM2D} and \cite{DIMWB}, using a third 
order ADER-WENO finite volume scheme on a very fine uniform Cartesian grid composed of $1024 \times 512$ elements. The direct comparison of the two simulations at time $t=0.4$ is shown in Figure \ref{fig.dambreak.amr}. Overall we can indeed note an excellent agreement between the behaviour of the diffuse interface GPR model in the stiff relaxation limit and the weakly compressible inviscid non-hydrostatic free surface flow model of \cite{DIM2D,DIMWB}.

\section{Conclusions and Outlook}
\label{sec.concl} 

In the first part of this paper we have provided a review of the ADER approach, whose development started about 20 
years ago with the seminal works of \cite{toro1,mill} and \cite{toro3,toro4} in the context of 
approximate solvers for the generalized Riemann problem (GPR). The ADER method provides \textit{fully discrete} 
explicit one-step schemes that are in principle arbitrary high order accurate in both space and time. The most recent 
developments include ADER schemes for stiff source terms, as well as ADER finite volume and discontinuous Galerkin 
finite element schemes on fixed and moving meshes, which are all based on a space-time predictor-corrector approach. 
The fact that ADER schemes are fully discrete makes the implementation of time accurate local time stepping (LTS) 
particularly simple, both on adaptive Cartesian AMR meshes \cite{AMR3DCL}, as well as in the context of Lagrangian
schemes on moving grids \cite{ALELTS1D,ALELTS2D}. The fully discrete space-time formulation also allows the treatment
of topology changes during one time step in a very natural way \cite{GBCKSD2019}. 
In the second part of the paper we have then shown several applications of high order ADER finite volume and discontinuous Galerkin finite element schemes to the novel unified hyperbolic model of continuum mechanics (GPR model) proposed 
by Godunov, Peshkov and Romenski \cite{GodunovRomenski72,PeshRom2014,GPRmodel}. The presented test problems
cover the entire range of continuum mechanics, from ideal elastic solids over plastic solids to viscous fluids. 
The use of a diffuse interface approach allows also to simulate moving boundary problems on fixed Cartesian meshes.  
Future developments will concern the extension of the mathematical model to non-Newtonian fluids 
\cite{Jackson2019a} and to free surface flows with surface tension, see 
\cite{HypSurfTension,SHTCSurfaceTension}, as well as to the conservative multi-phase model of  
\cite{RomenskiTwoPhase2007,RomenskiTwoPhase2010}. In future work we will also consider the use of novel 
all speed schemes \cite{AbateIolloPuppo} and semi-implicit space-time discontinuous Galerkin finite element schemes 
\cite{TavelliDumbser2017,IoriattiDumbser,SIDGGravity} for the diffuse interface version of the GPR model used in 
this paper.

\section*{Conflict of Interest Statement}
The authors declare that the research was conducted in the absence of any commercial or financial relationships that could be construed as a potential conflict of interest.

\section*{Author Contributions}

The governing PDE system was developed by IP. The numerical method and the computer codes were developed by MD, EG and SC. The test problems were computed by MD, EG and SC. The analysis of the method was performed by SB. All authors discussed the results and contributed to the final manuscript.     

\section*{Funding}
The research presented in this paper has been financed by the European Union's Horizon 2020 Research and  
Innovation Programme under the project \textit{ExaHyPE}, grant agreement number no. 671698 (call 
FET-HPC-1-2014). 

S.B. has also received funding by INdAM (\textit{Istituto Nazionale di Alta Matematica}, Italy) 
under a Post-doctoral grant of the research project \textit{Progetto premiale FOE 2014-SIES}.

S.C. acknowledges the financial support received by
the Deutsche Forschungsgemeinschaft (DFG) under the project 
\textit{Droplet Interaction Technologies (DROPIT)}, grant no. GRK 2160/1.

M.D. also acknowledges the financial support received from the Italian Ministry of Education, University and Research (MIUR) 
in the frame of the Departments of Excellence Initiative 2018--2022 attributed to DICAM of the University of Trento 
(grant L. 232/2016) and in the frame of the PRIN 2017 project. M.D. has also received funding from the University of 
Trento via the  \textit{Strategic Initiative Modeling and Simulation}. 

E.G. has also been financed by a national mobility grant for young researchers in Italy, funded by GNCS-INdAM and acknowledges the support given by the University of Trento through the \textit{UniTN Starting Grant} initiative.

\bibliographystyle{abbrv}
\bibliography{./biblio}

\end{document}